# A compilation of solar atlases (from Delbouille, Kurucz, Gandorfer, Stenflo) at disk centre and at limb from λ 3000 Å to λ 8800 Å

J.-M. Malherbe, Observatoire de Paris, Université de Recherche PSL, CNRS, LESIA, Meudon, France

Email : Jean-Marie.Malherbe@obspm.fr

ORCID : https://orcid.org/0000-0002-4180-3729

## Abstract

We present in this paper a compilation of solar atlases from λ 3000 Å to λ 8800 Å with spectral lines identified by the Moore table and with the corresponding equivalent Lande factors g*. We used two spectra at disk centre (µ = 1.0), from Delbouille and Kurucz, and two spectra at the limb from Stenflo and Gandorfer, respectively at µ = 0.145 and µ = 0.10.

## Description of the spectra shown in annex

We used the solar atlases of Delbouille *et al* (1973), from λ 3000 Å to λ 10000 Å (step 2 mÅ) and from Kurucz et al (1984) from λ 4080 Å to λ 9950 Å (step 5 mÅ) at disk centre (µ = 1.0), taken respectively at the Jungfraujoch station (Switzerland, a very high and dry site) and Kitt Peak (USA). We restricted the wavelength range to λ 3000 - 8800 Å for which we have the line identification provided by the Moore table (Moore *et al*, 1966). The equivalent Lande factors g* were also provided in the range λ 3700 – 8800 Å, as this is important to characterize the sensitivity to the Zeeman effect (Stokes V and the circular polarization rate are proportional to g*). The spectra are displayed in annex, by pages of 100 Å displaying bands of 10 Å. Delbouille's spectrum is shown in **black**, while Kurucz's spectrum is plotted in green colour. The blue numbers are equivalent Lande factors g*. The line identification follows the convention of the Moore *et al* (1966) table (slash for main contributors in case of a blend, dash for blends, parenthesis for masked lines, p for predicted line, see page XVIII of their book for details). The equivalent width w of lines (in fact 0.2*log(w), with w in mÅ) is represented by the violet dashed vertical bars (for instance, CaII K 3933.68 Å line has w = 20253 mÅ, which gives a bar length of 0.2*log(20253) = 0.86). Limb spectra obtained by Stenflo (2014, 2015) with the FTS at Kitt Peak (USA) and Gandorfer (2000, 2002, 2005) with ZIMPOL at IRSOL (Switzerland) are superimposed, respectively for µ = 0.145 (blue dotted) and µ = 0.10 (red dotted). All spectra ($I/I_c$) are normalized to the adjacent continuum $I_c$.

The PDF document is based on 2280 x 3324 pixel GIF images which are available here:

https://www.lesia.obspm.fr/perso/jean-marie-malherbe/spectrevisible/spectreCL/index.html

**Atlases at 5" and 10'' from the limb (µ = 0.1, IRSOL, µ = 0.145, FTS Kitt Peak)**

*The data for this analysis have been provided in electronic form by IRSOL as a compilation by Stenflo (2014), based on the atlases of Stenflo (2015) and Gandorfer (2000, 2002, 2005), which are on line at https://www.irsol.usi.ch/it/data-archive/second-solar-spectrum-ss2-atlas*

**Line identification table**

# ANNEX : THE SPECTRA

The spectra from 3000 Å to 8800 Å, 58 pages of 100 Å with 10 bands of 10 Å per page.

For wavelength intervals [λ1, λ2] (in Å), see the following **page numbers** :

3000-3100 : **3** / 3100-3200 : **4** / 3200-3300 : **5** / 3300-3400 : **6** / 3400-3500 : **7**

3500-3600 : **8** / 3600-3700 : **9** / 3700-3800 : **10** / 3800-3900 : **11** / 3900-4000 : **12**

4000-4100: **13** / 4100-4200 : **14** / 4200-4300 : **15** / 4300-4400 : **16** / 4400-4500 : **17**

4500-4600 : **18** / 4600-4700 : **19** / 4700-4800 : **20** / 4800-4900 : **21** / 4900-5000 : **22**

5000-5100: **23** / 5100-5200 : **24** / 5200-5300 : **25** / 5300-5400 : **26** / 5400-5500 : **27**

5500-5600 : **28** / 5600-4700 : **29** / 5700-5800 : **30** / 5800-5900 : **31** / 5900-6000 : **32**

6000-6100 : **33** / 6100-6200 : **34** / 6200-6300 : **35** / 6300-6400 : **36** / 6400-6500 : **37**

6500-6600 : **38** / 6600-6700 : **39** / 6700-6800 : **40** / 6800-6900 : **41** / 6900-7000 : **42**

7000-7100 : **43** / 7100-7200 : **44** / 7200-7300 : **45** / 7300-7400 : **46** / 7400-7500 : **47**

7500-7600 : **48** / 7600-7700 : **49** / 7700-7800 : **50** / 7800-7900 : **51** / 7900-8000 : **52**

8000-8100 : **53** / 8100-8200 : **54** / 8200-8300 : **55** / 8300-8400 : **56** / 8400-8500 : **57**

8500-8600 : **88** / 8600-8700 : **59** / 8700-8800 : **60**

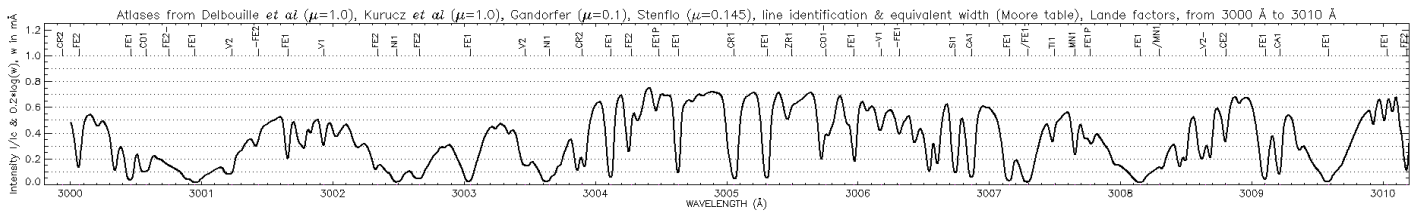
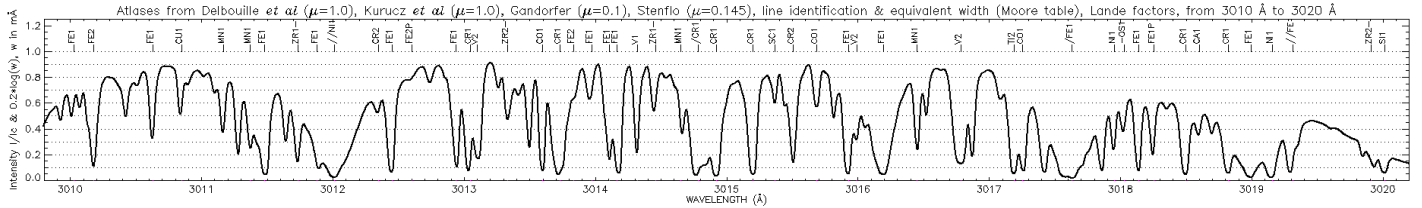
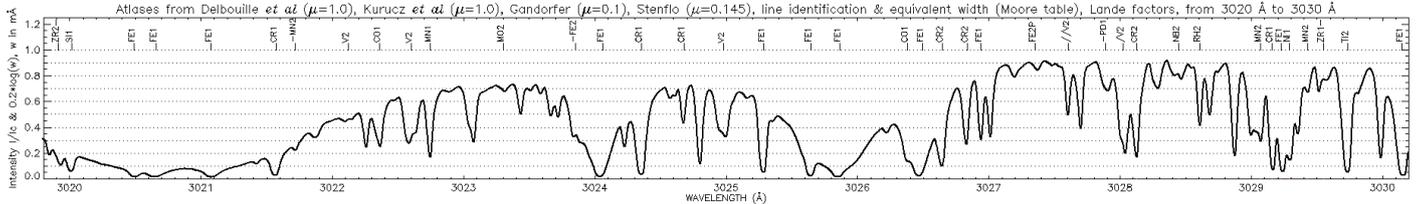
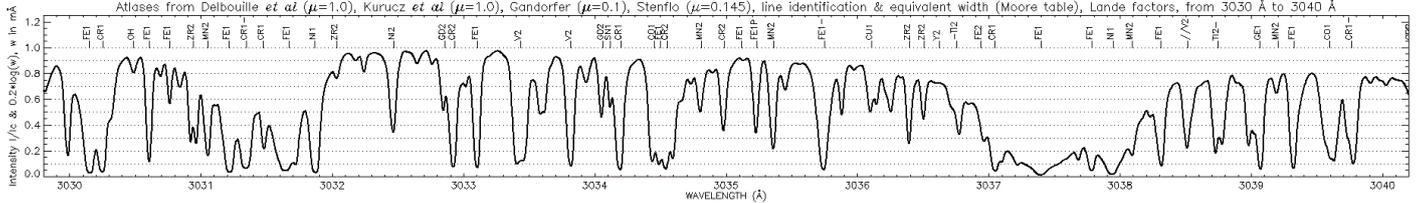
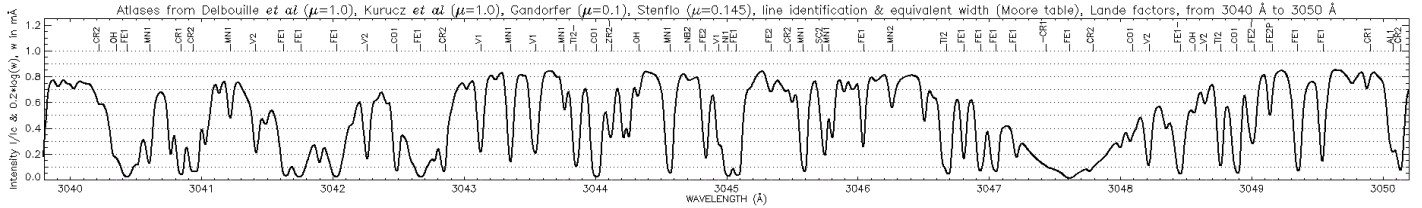
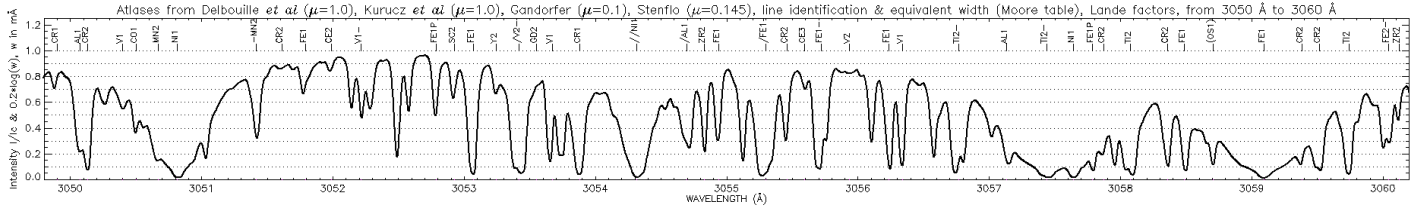
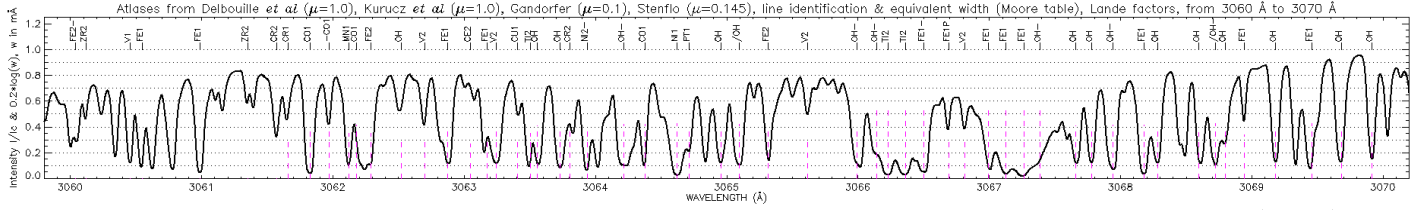
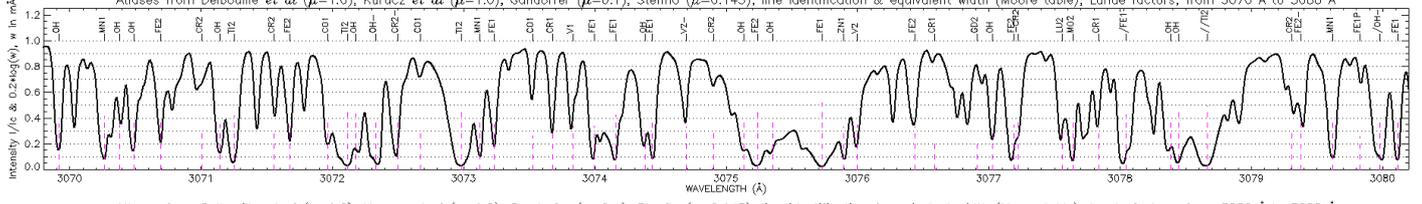
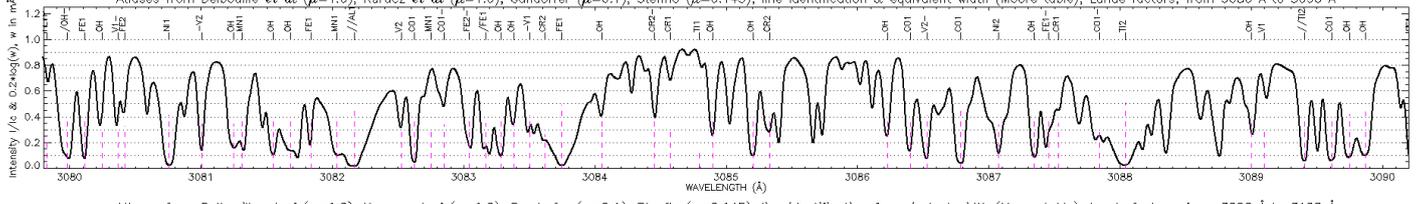
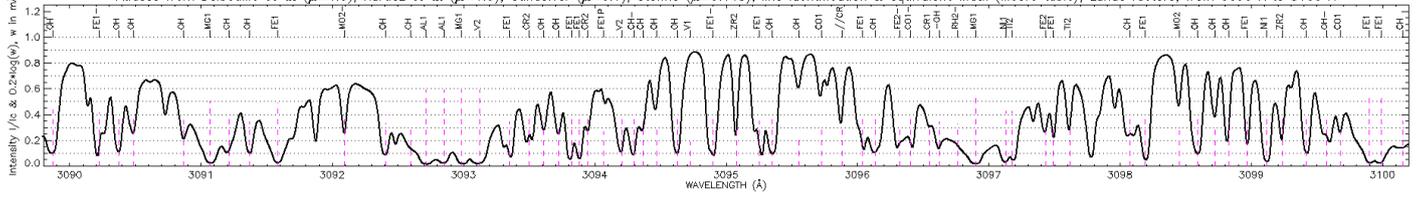

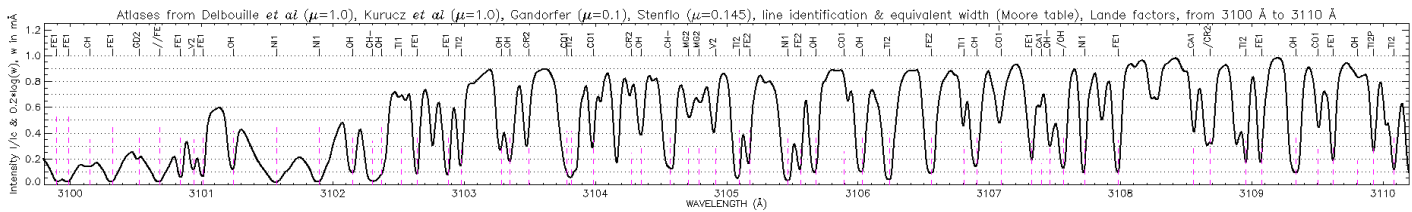
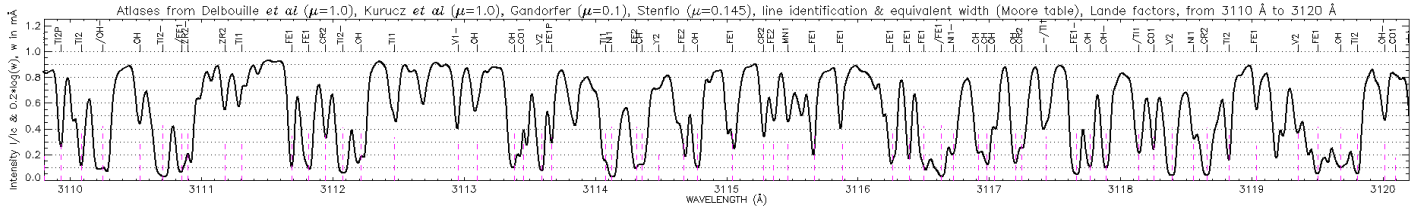
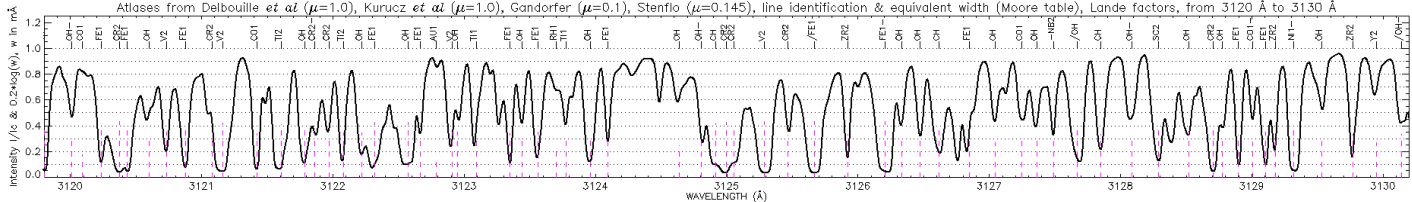
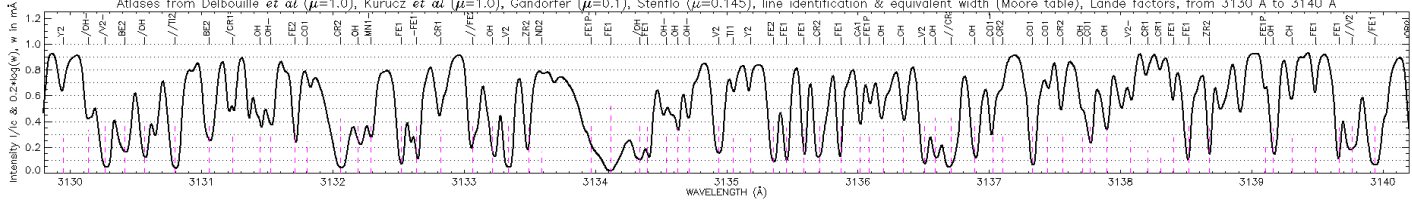
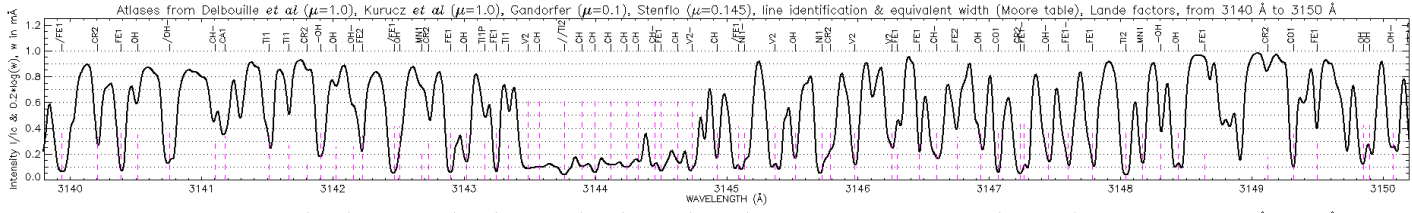
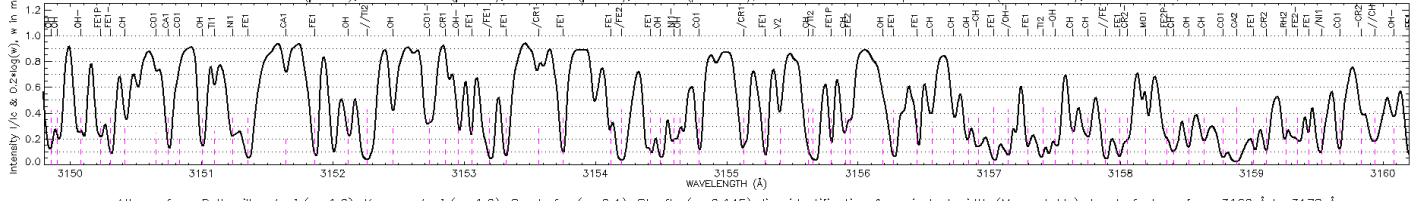
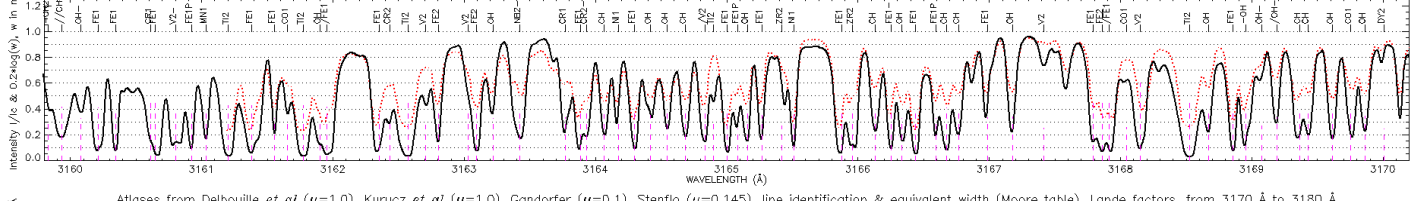
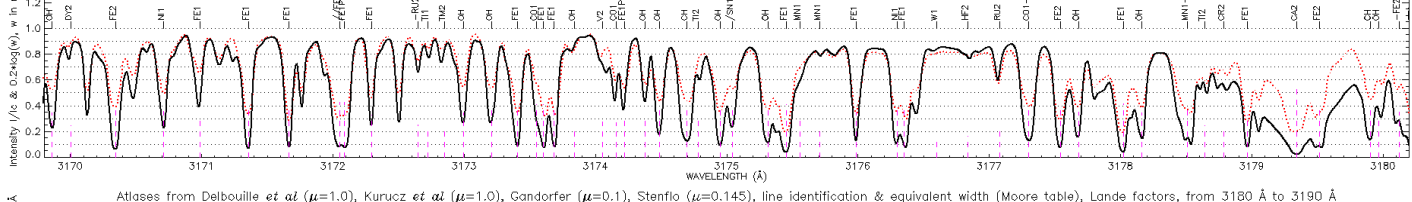
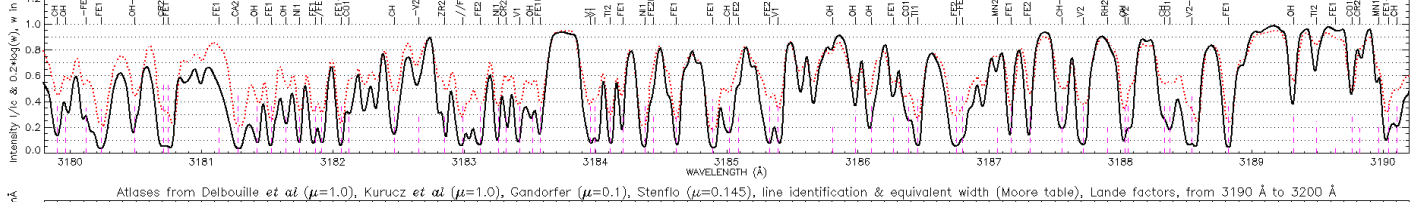
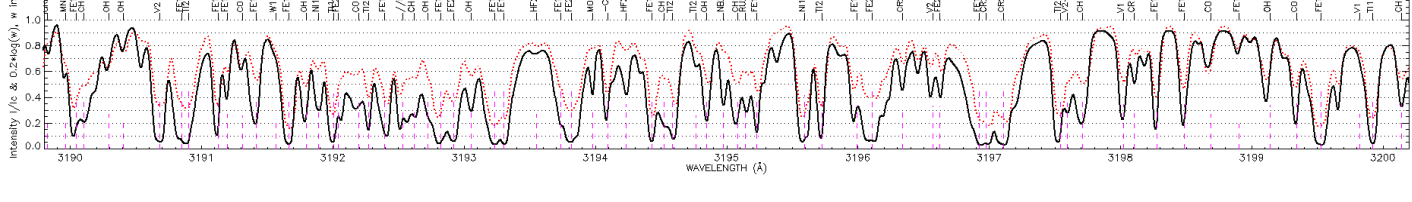

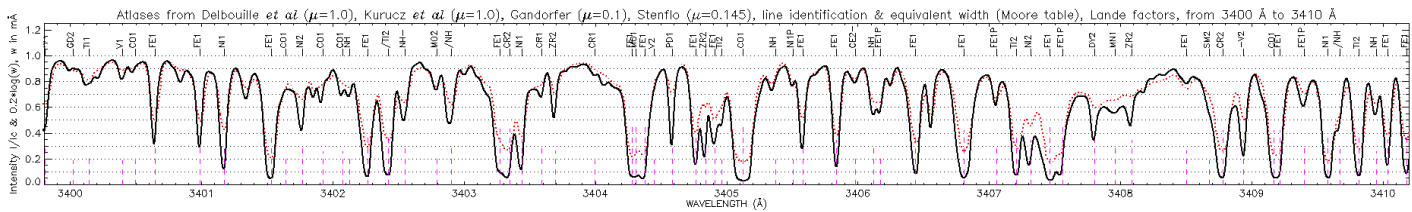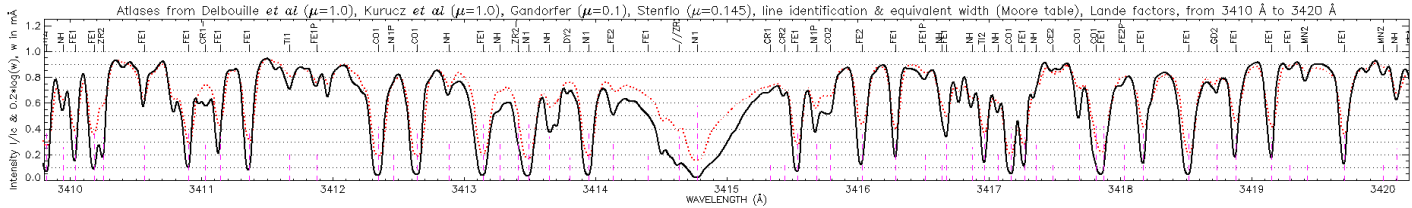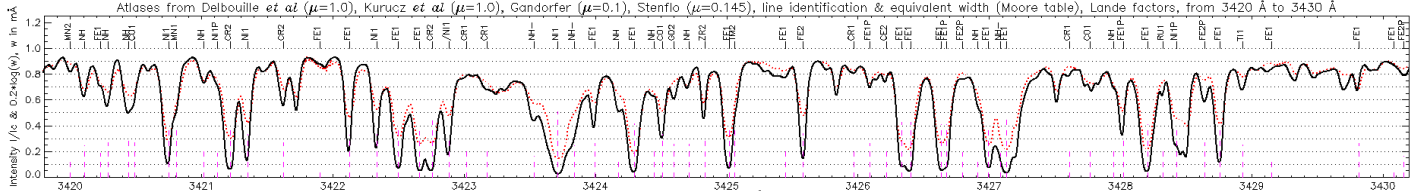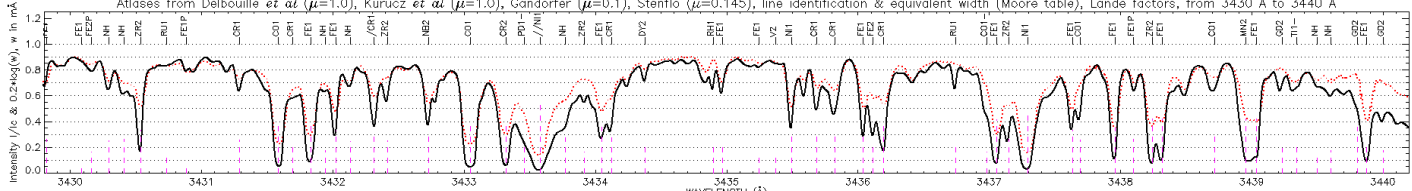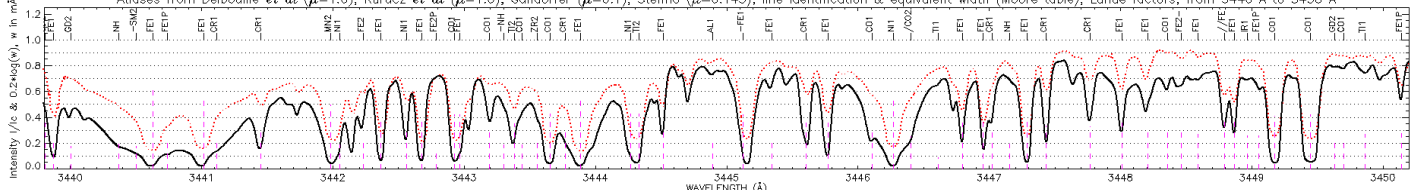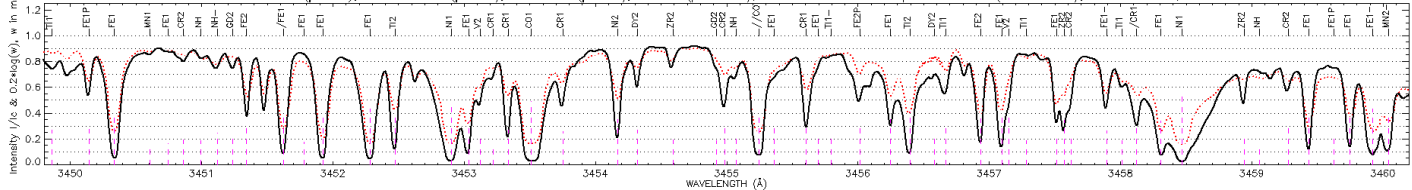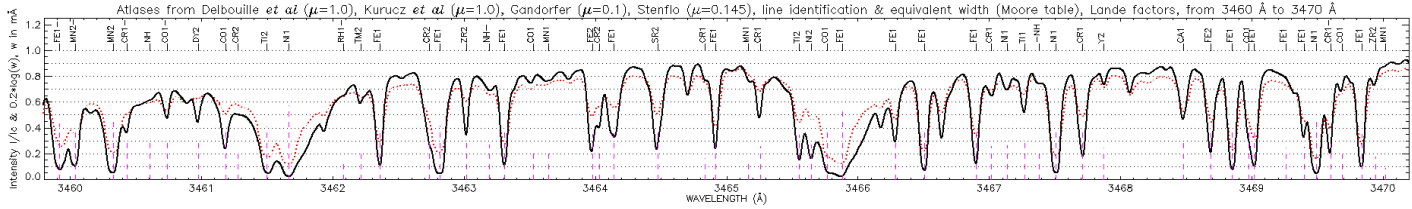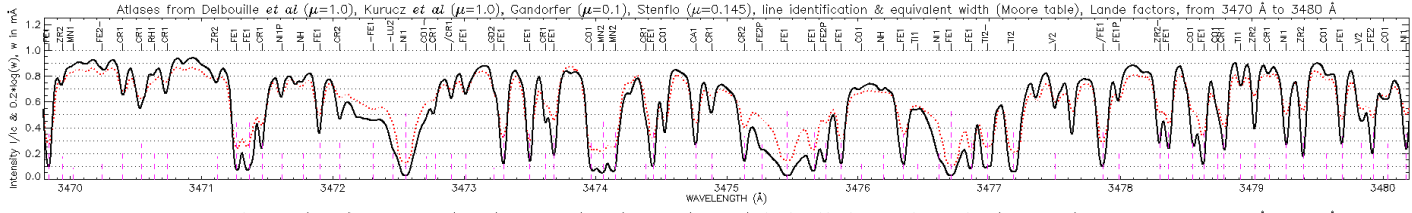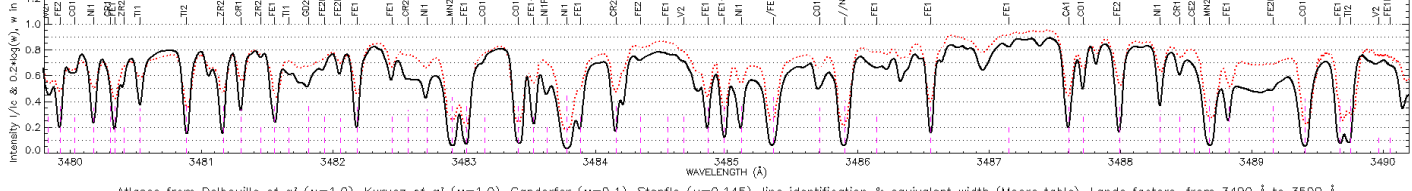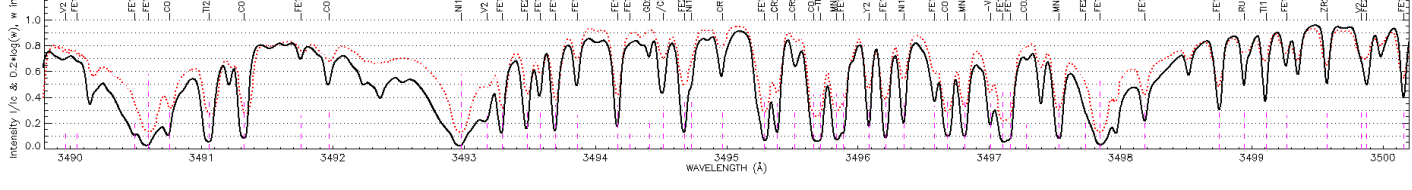

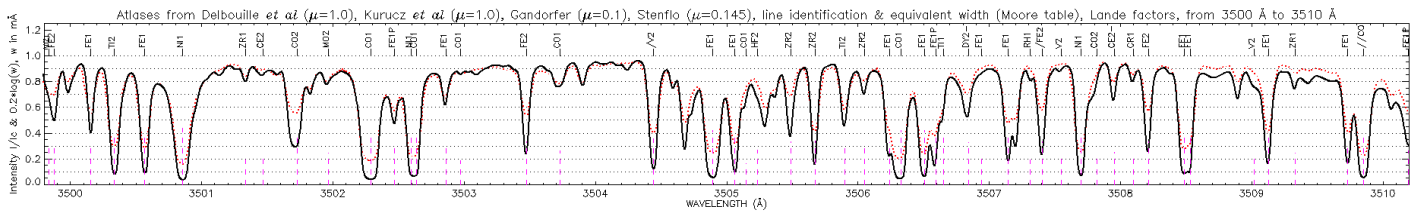
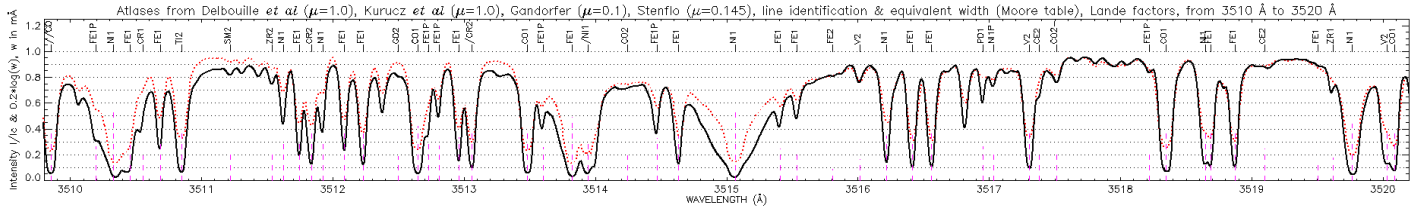
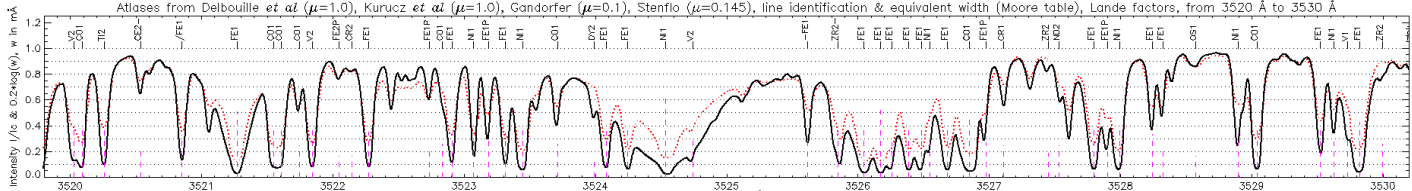
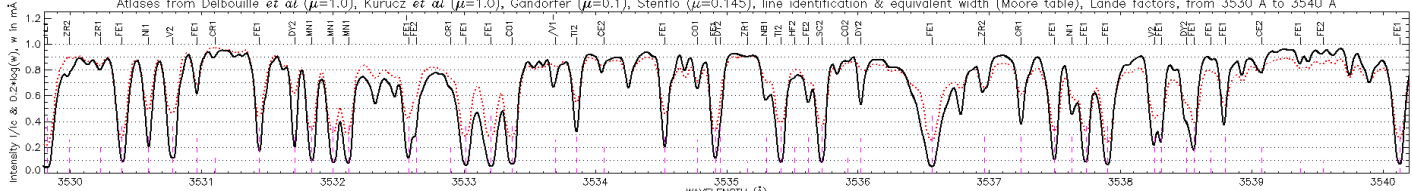
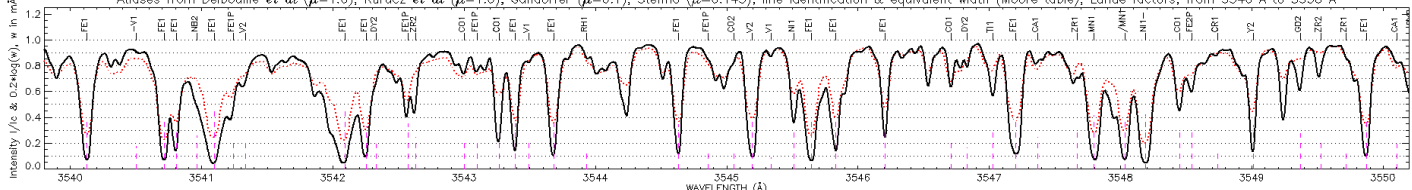
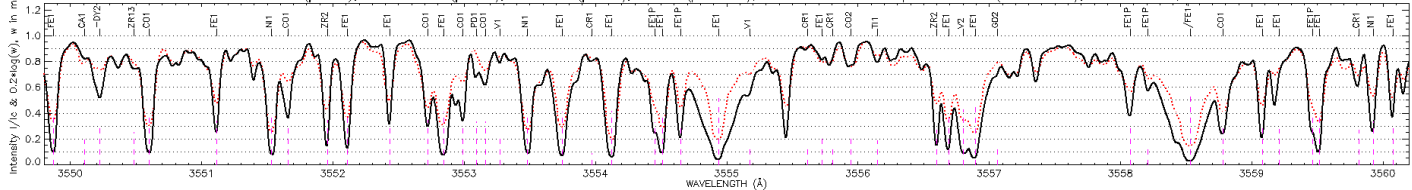
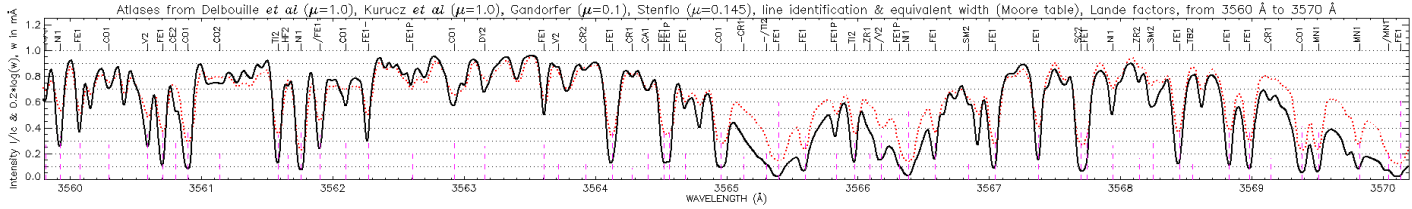
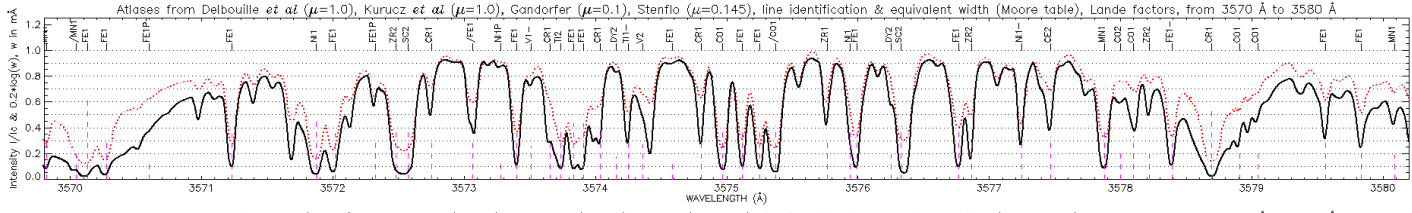
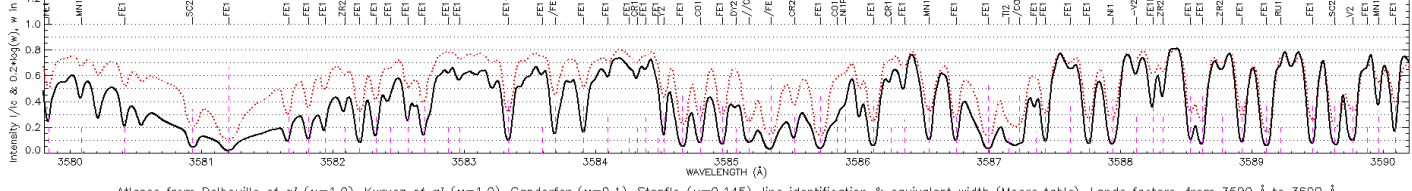
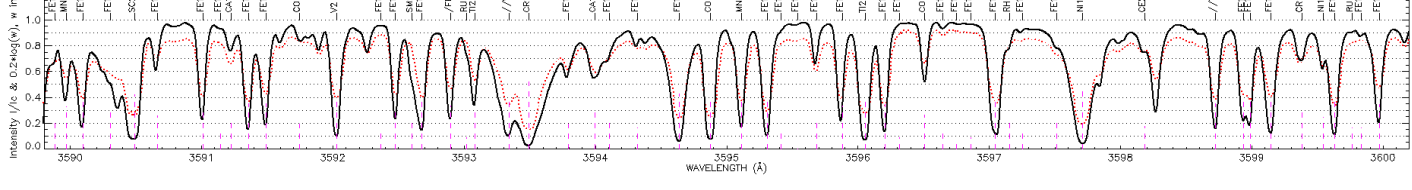

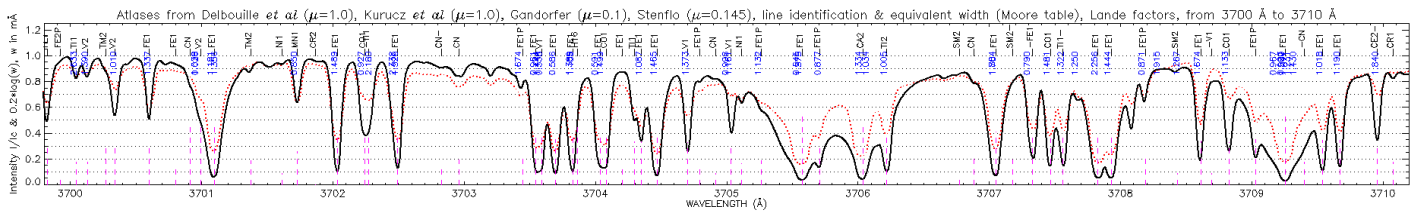
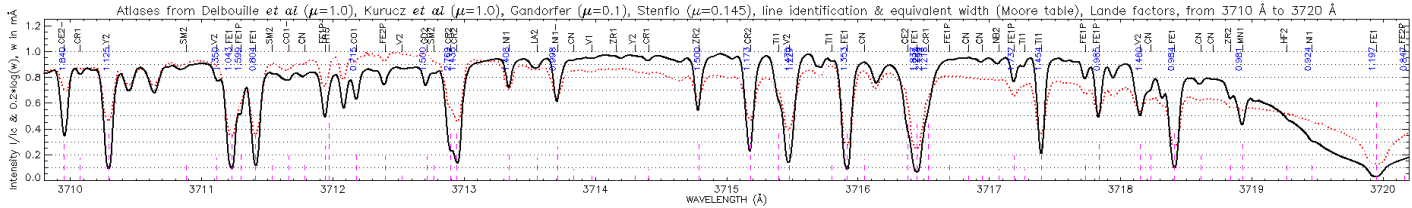
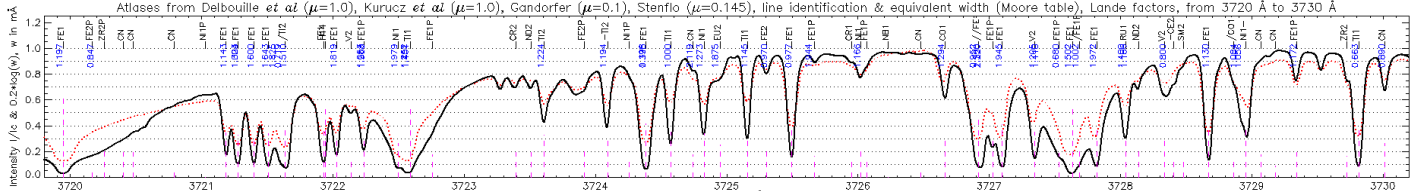
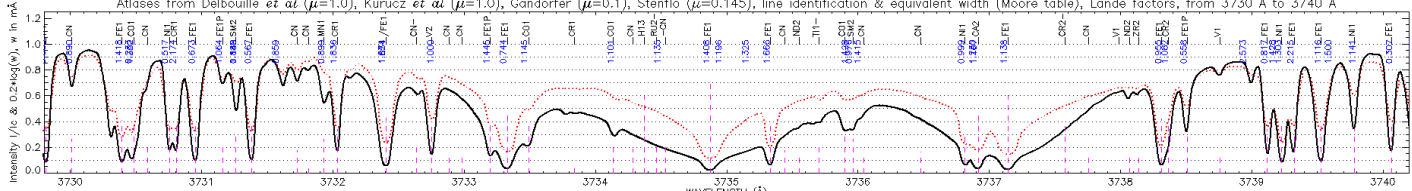
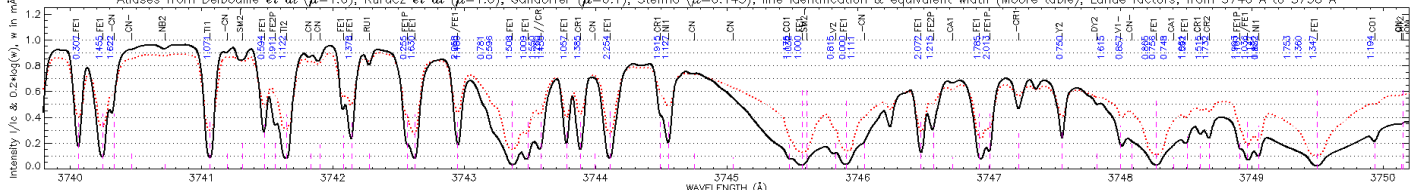
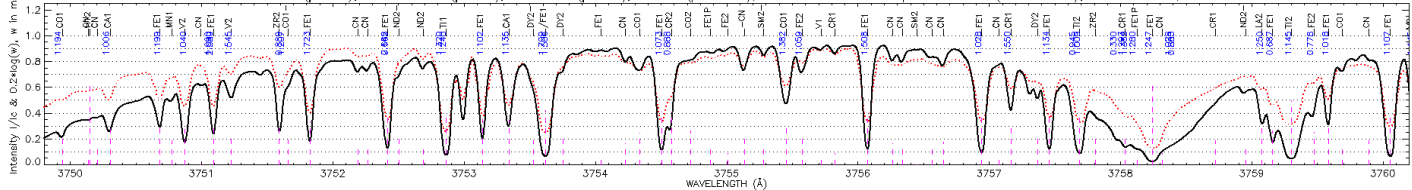
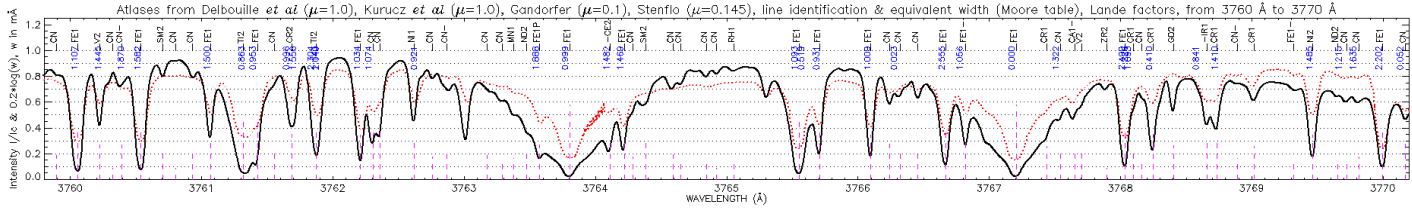
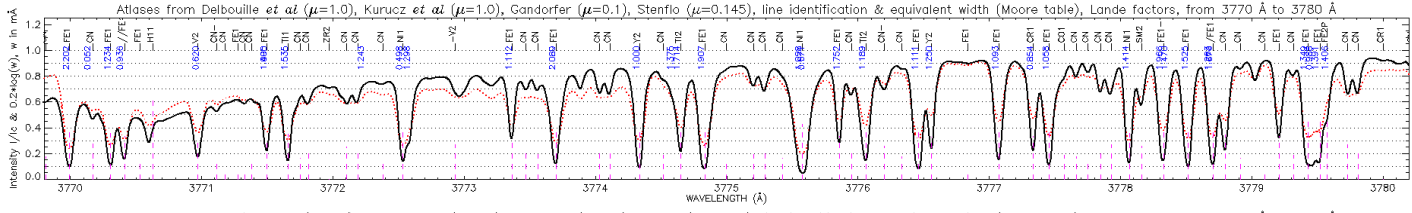
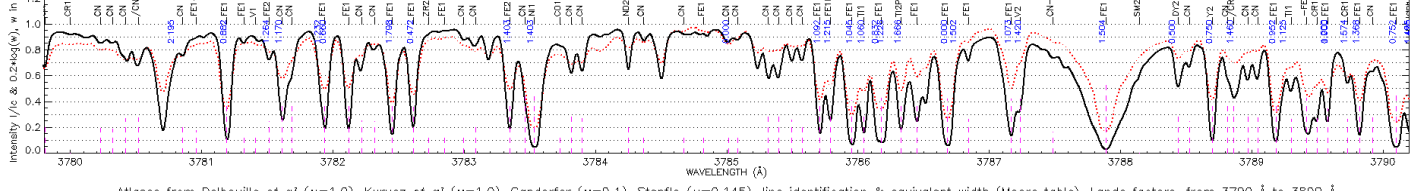
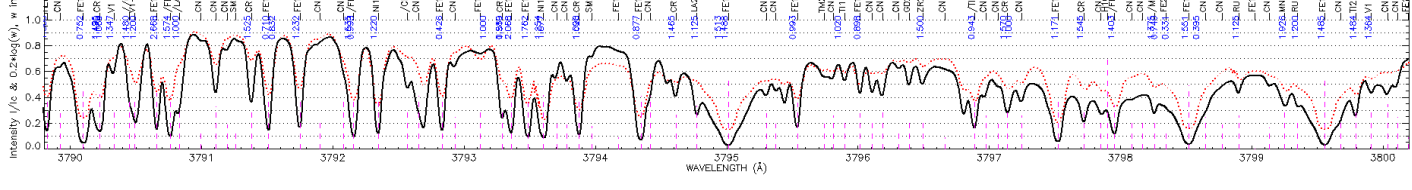

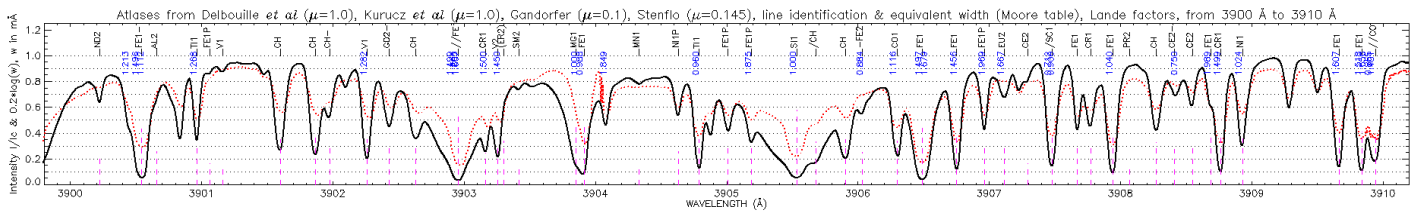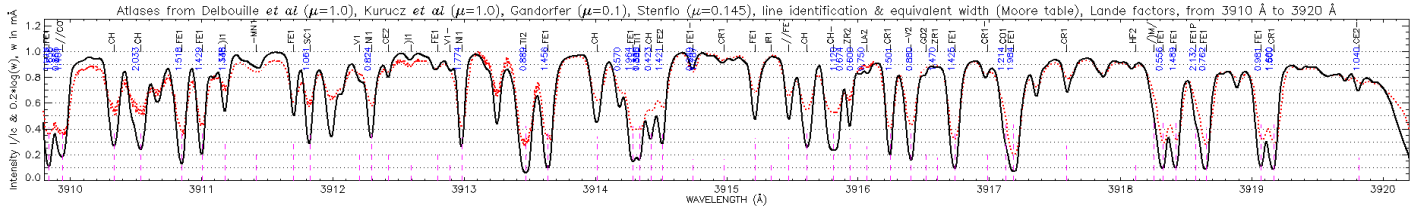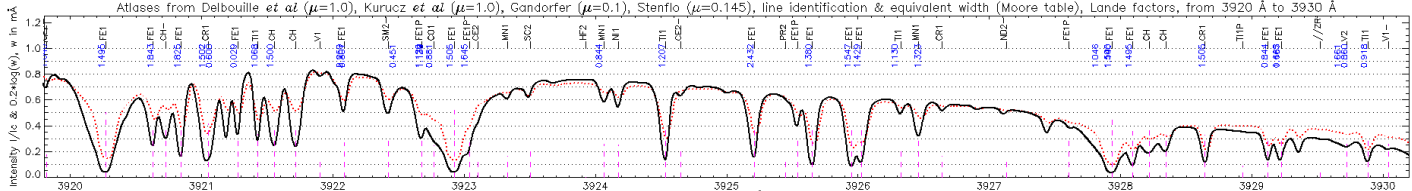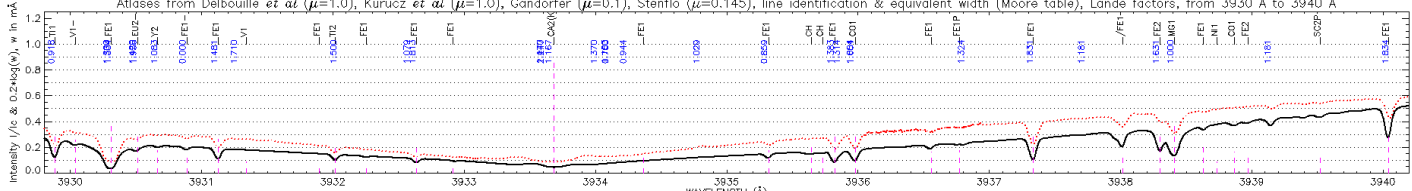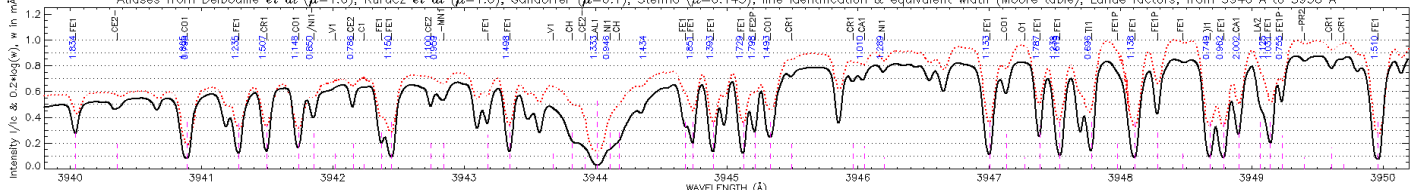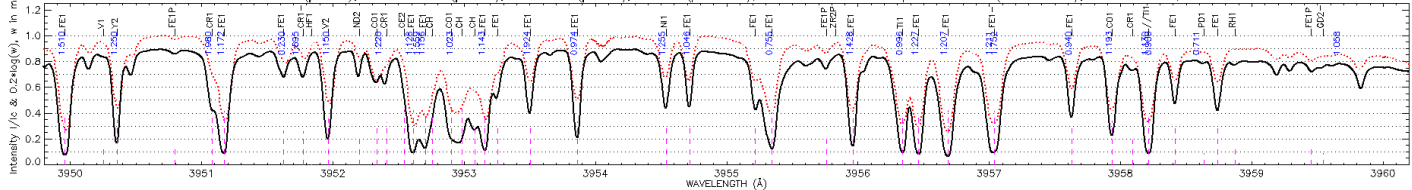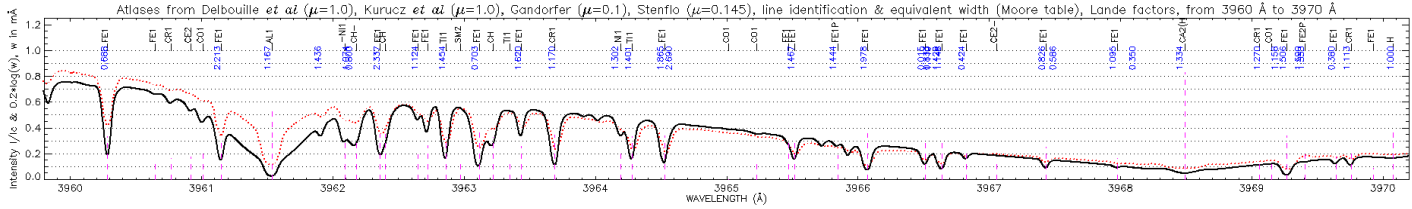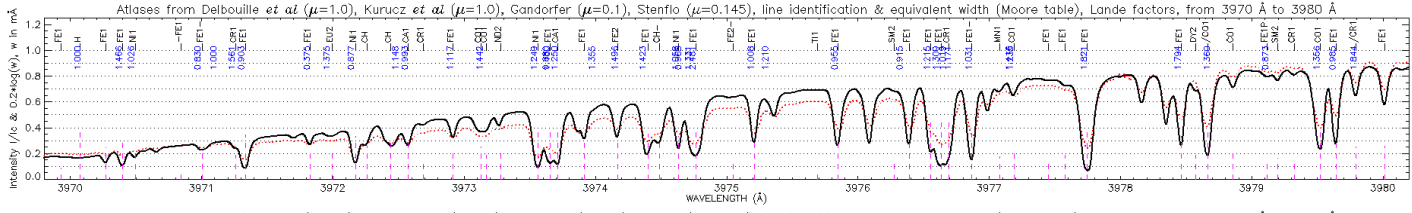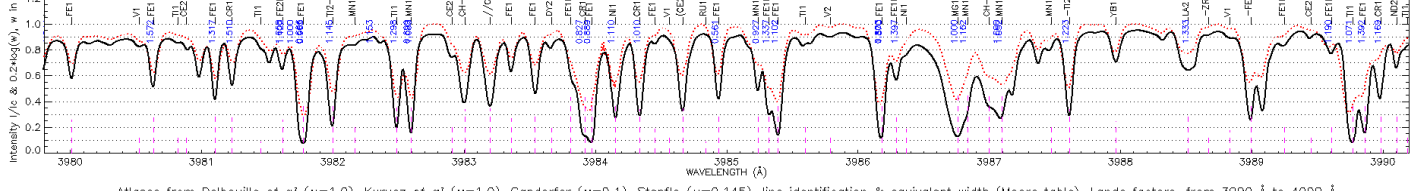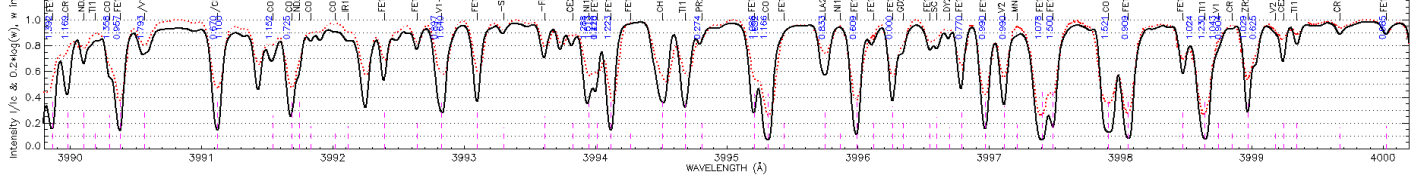

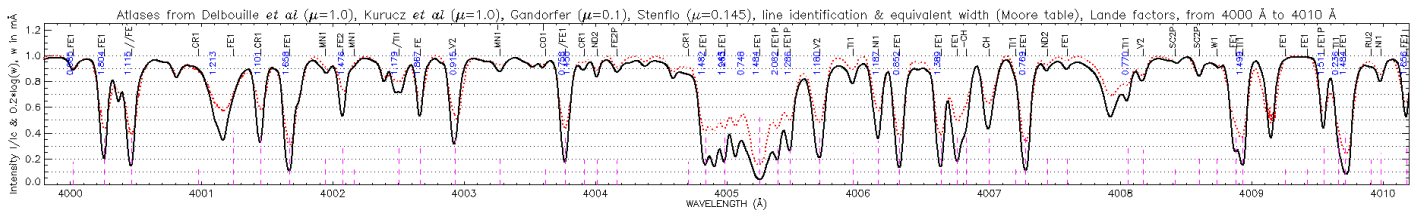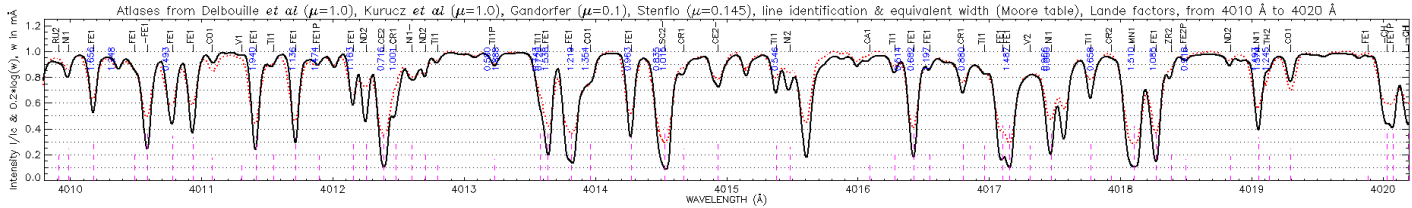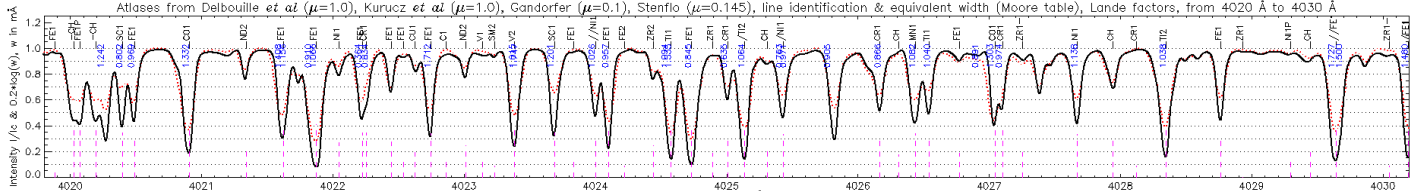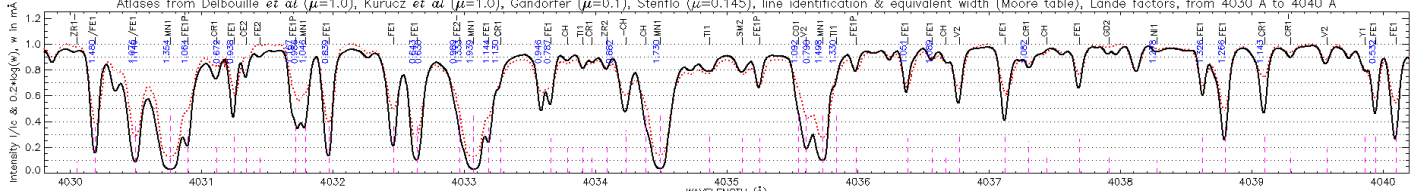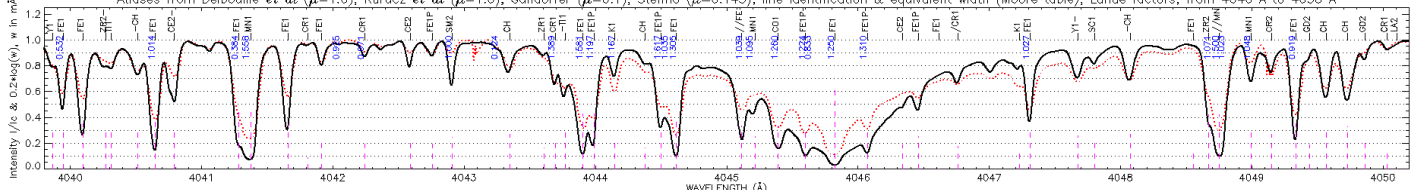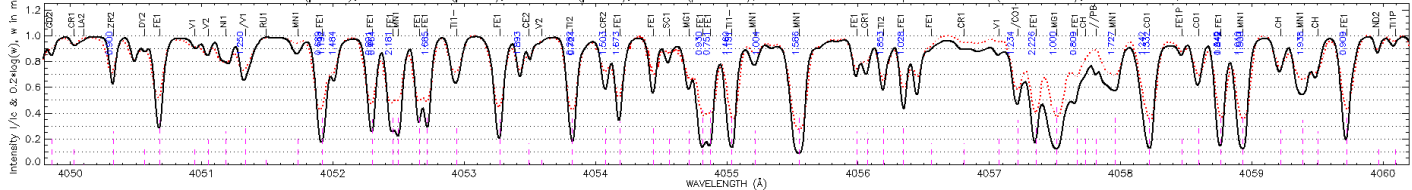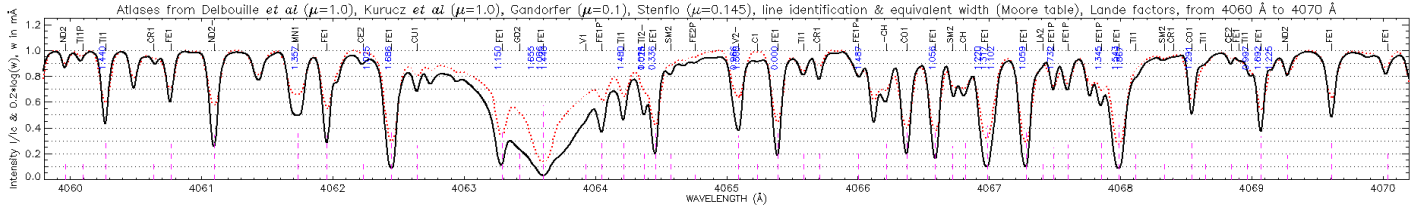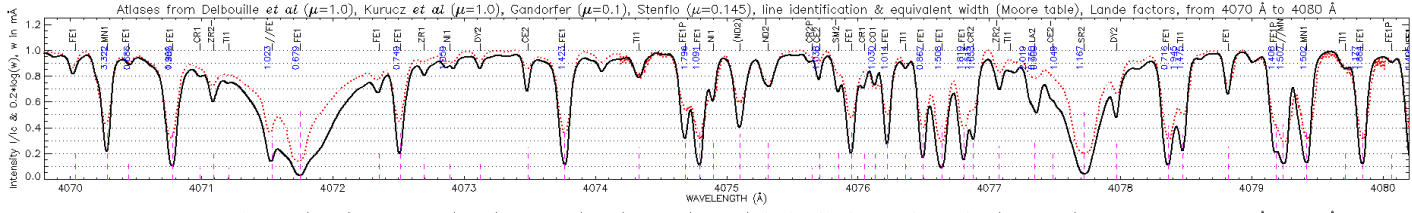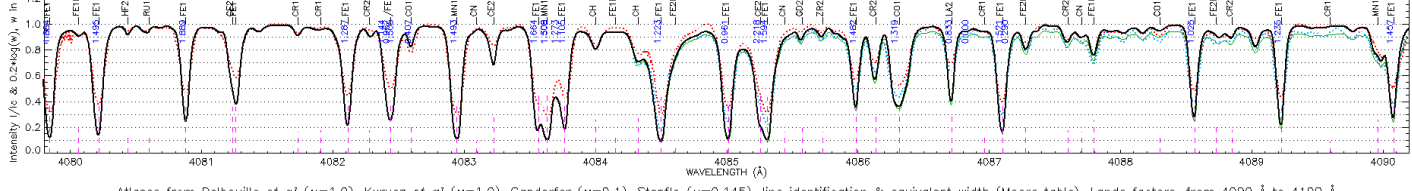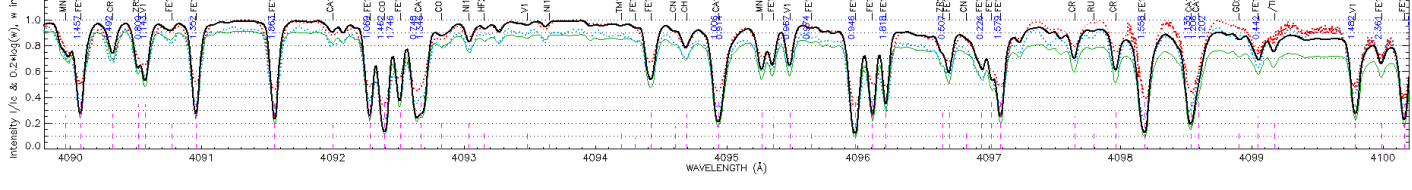

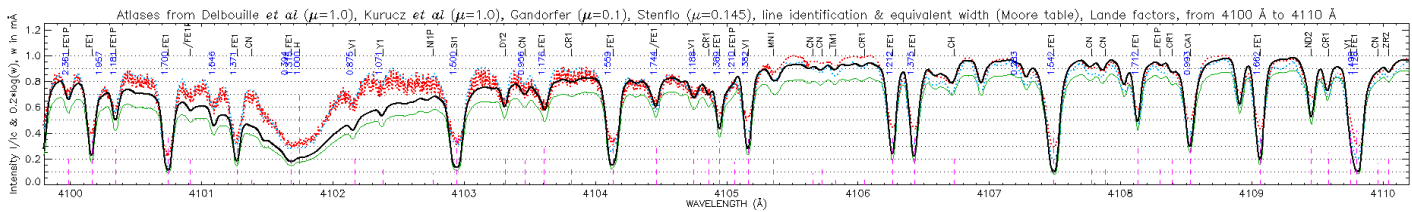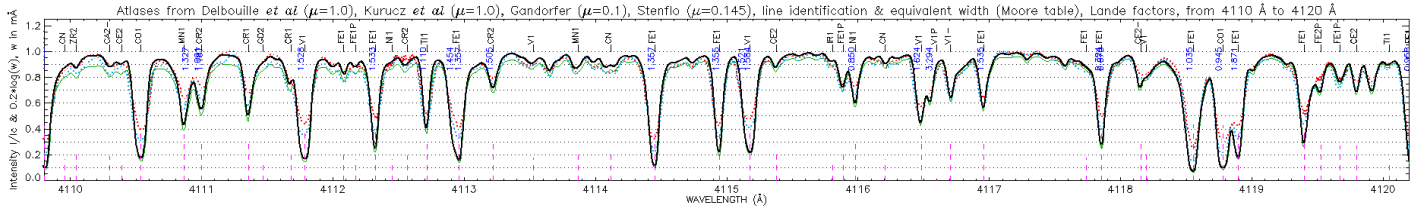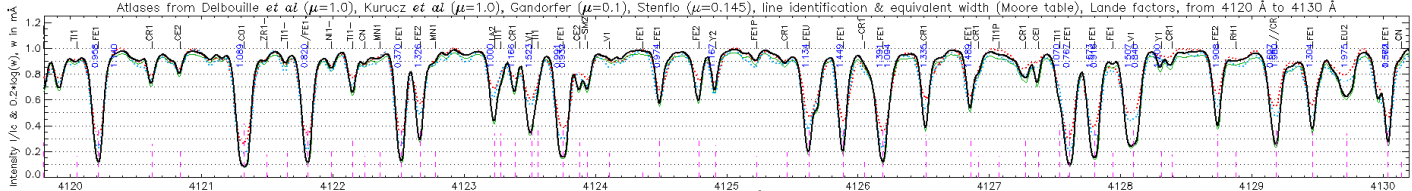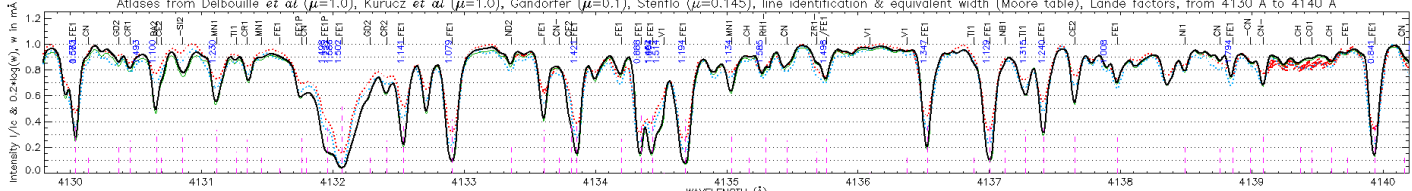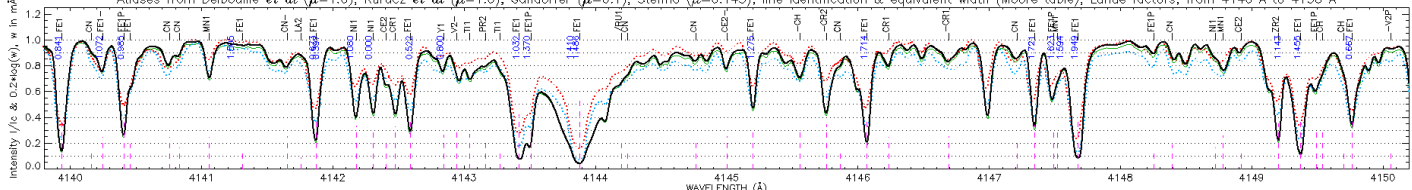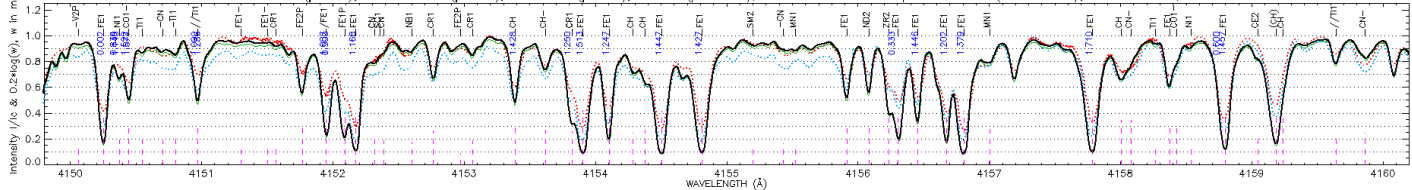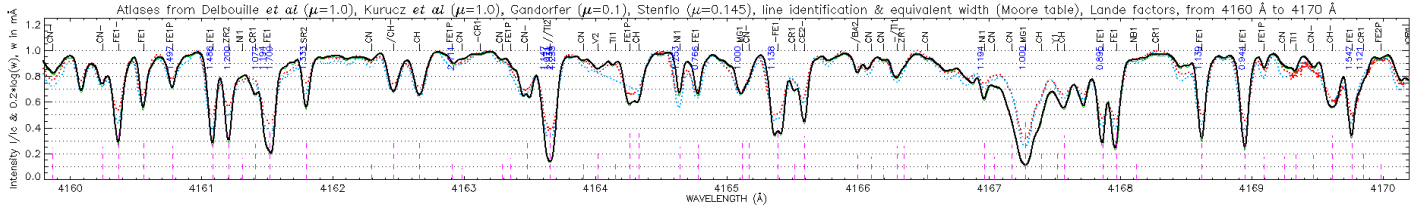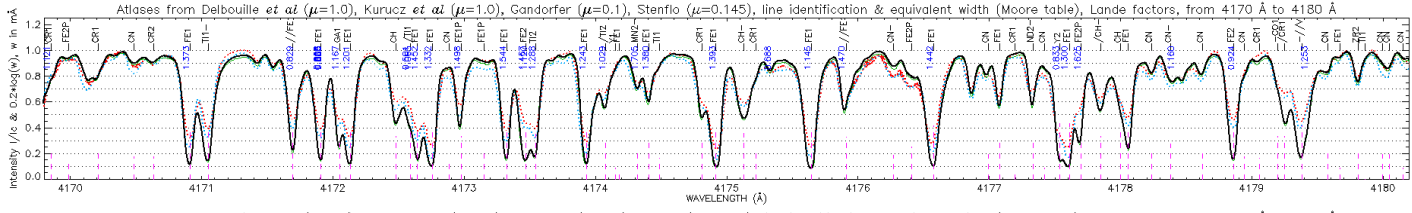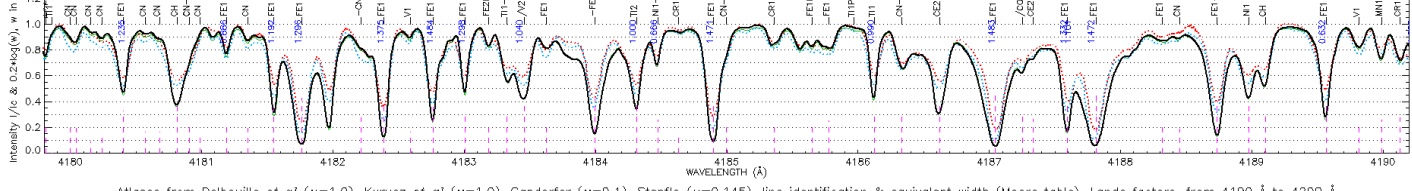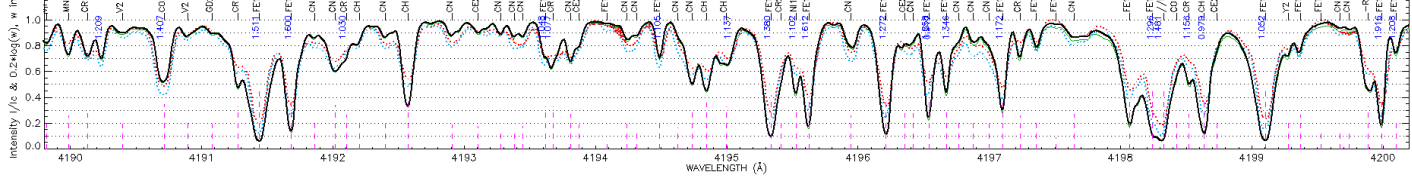

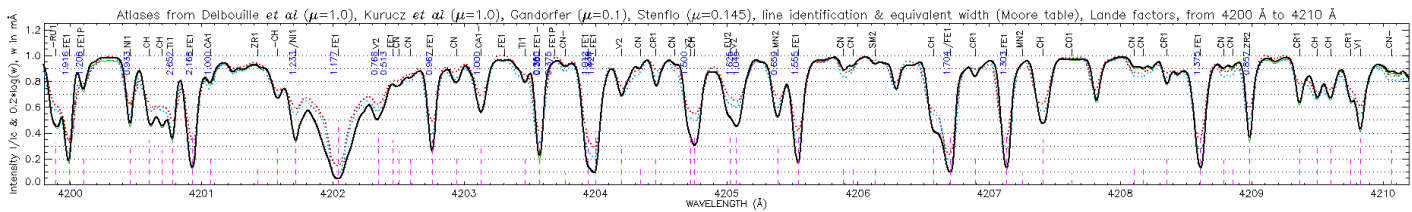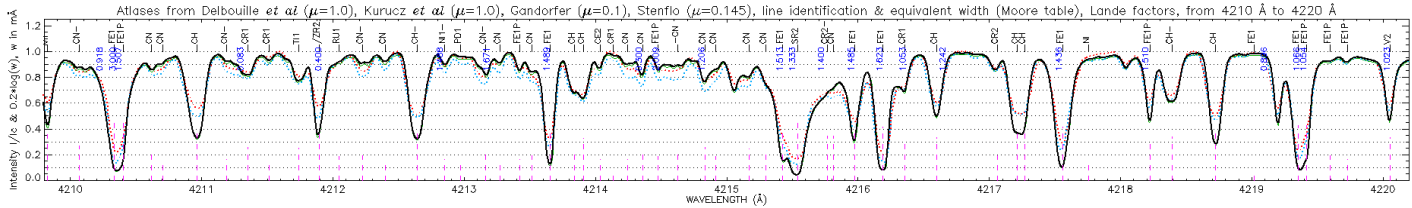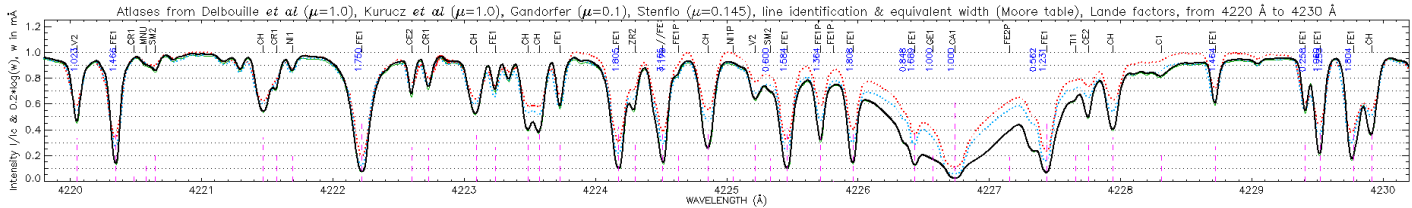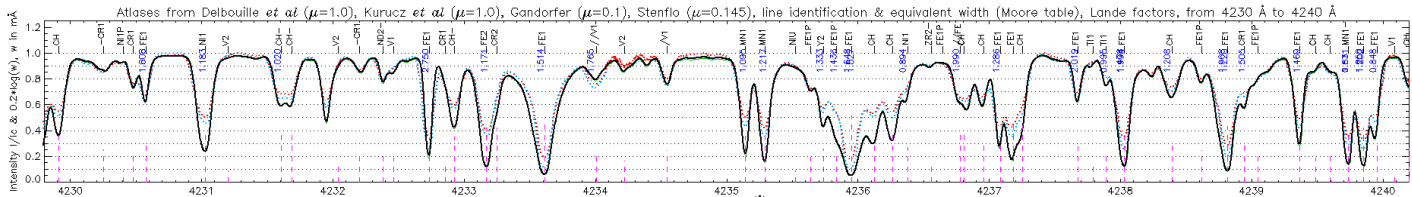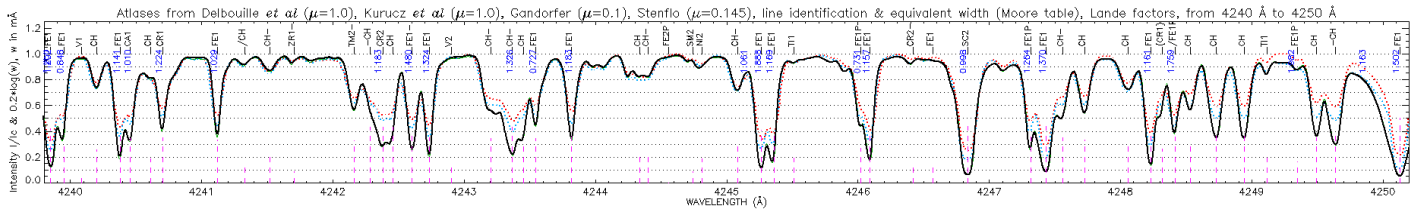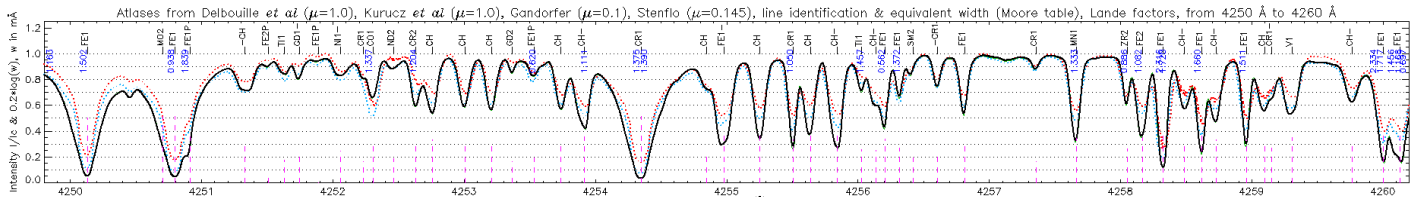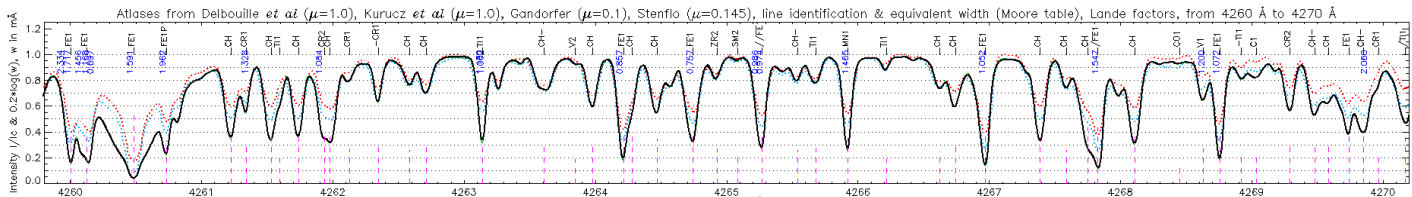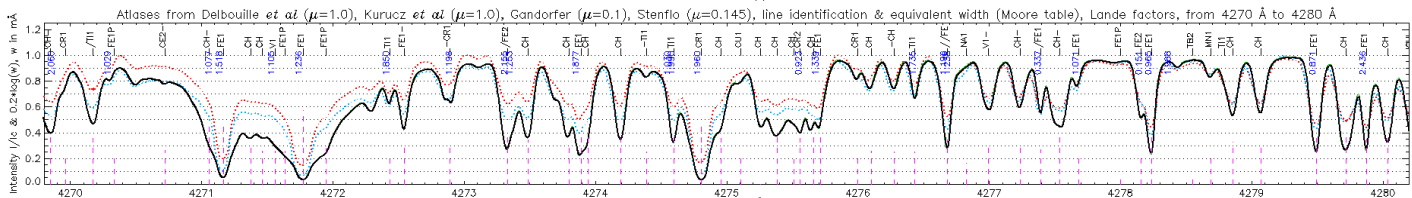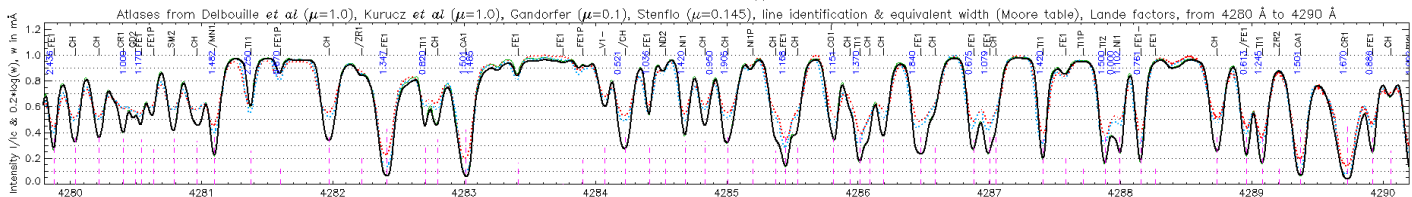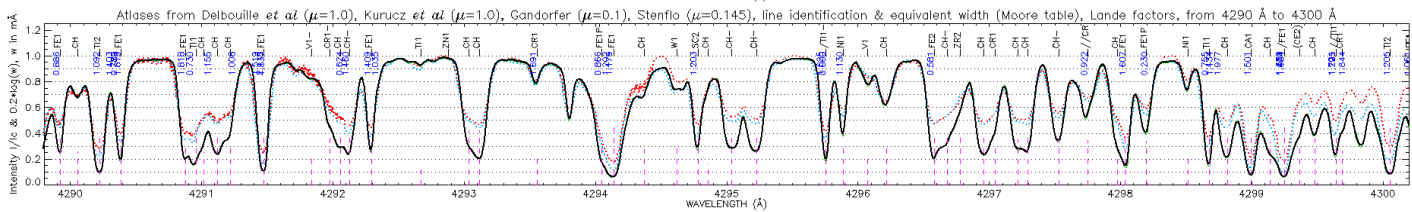

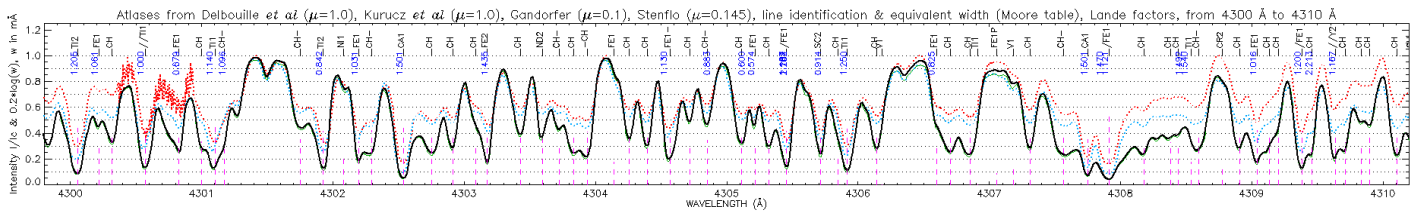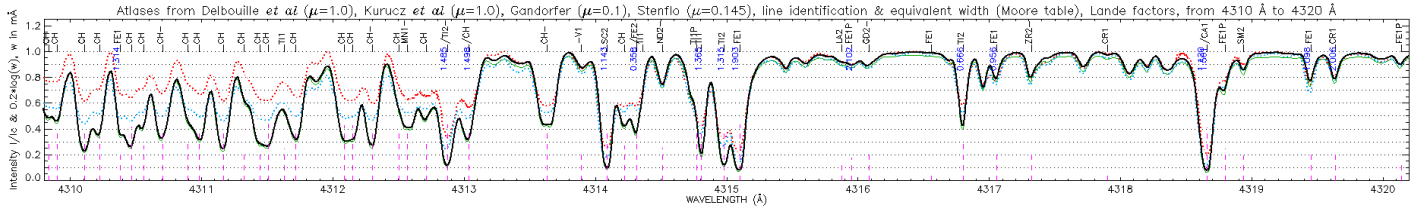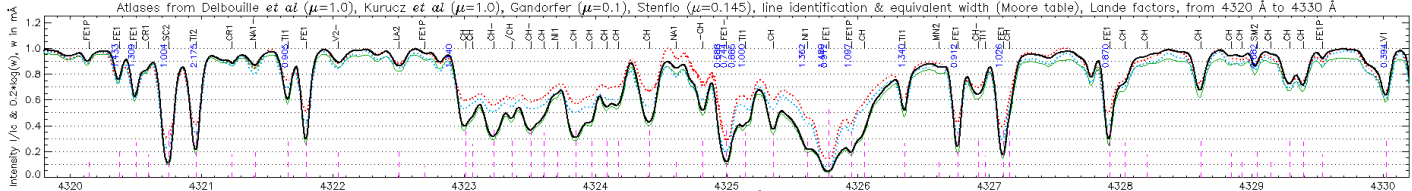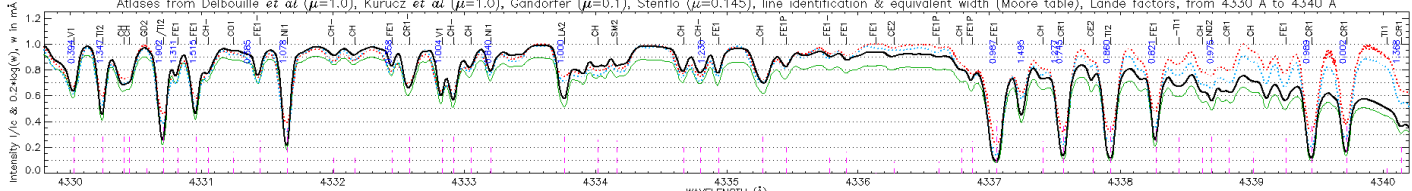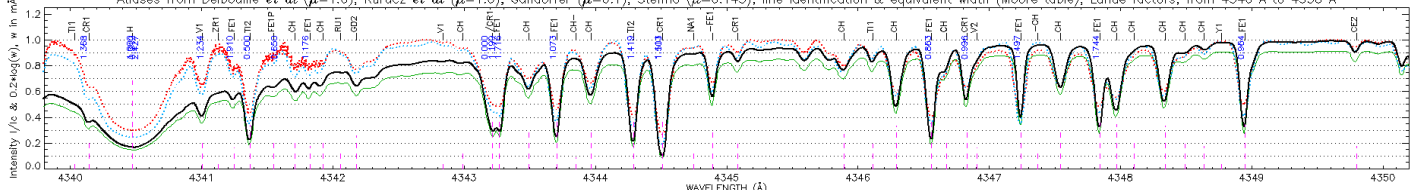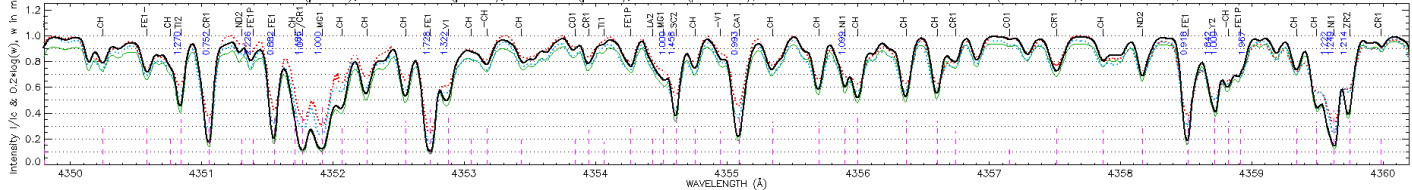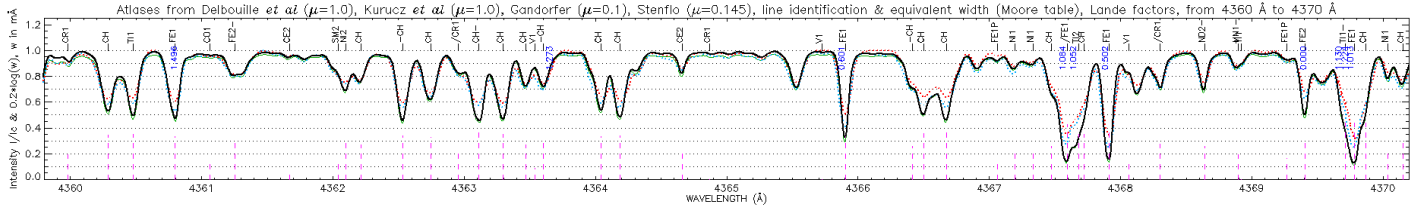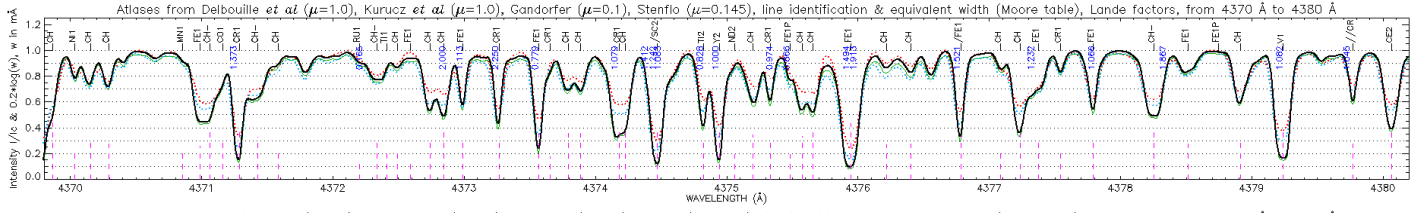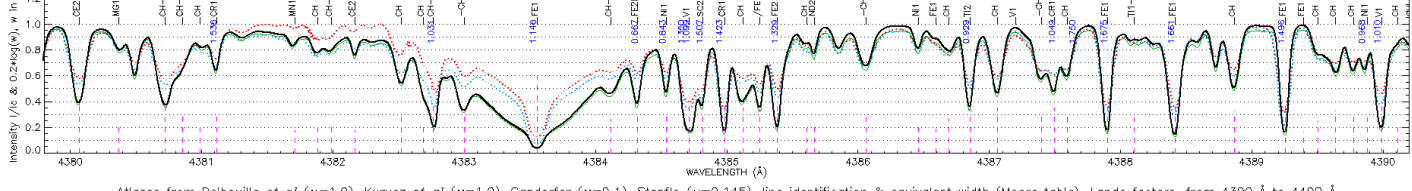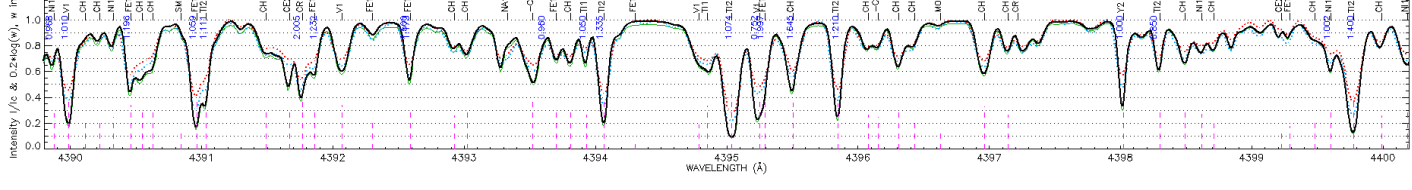

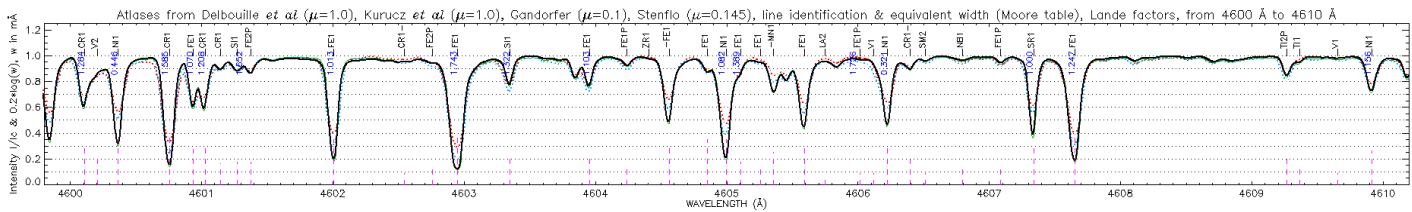
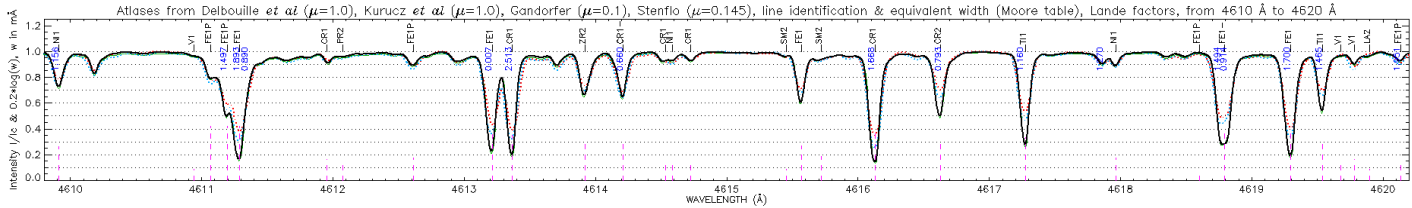
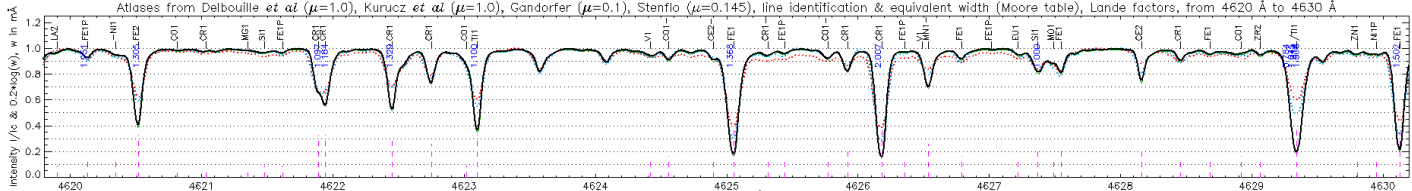
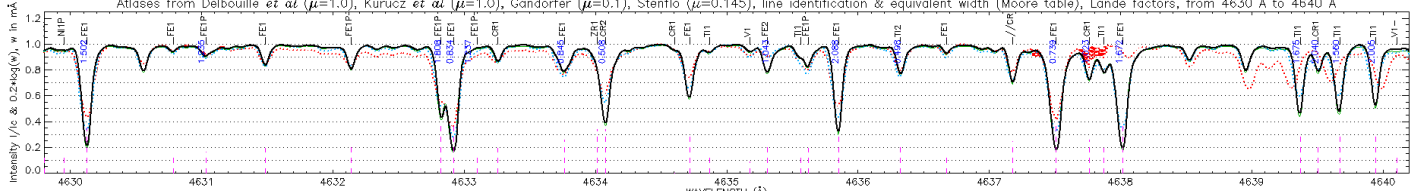
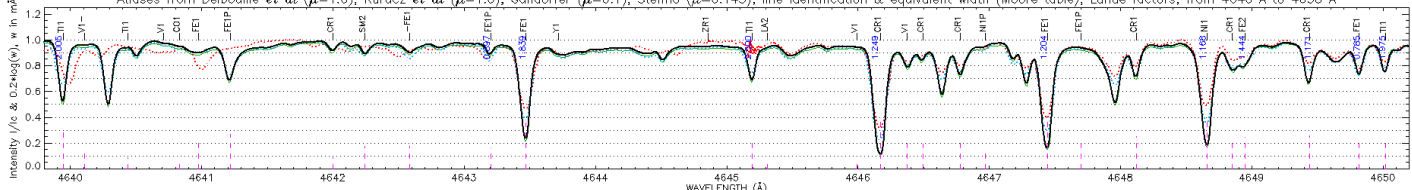
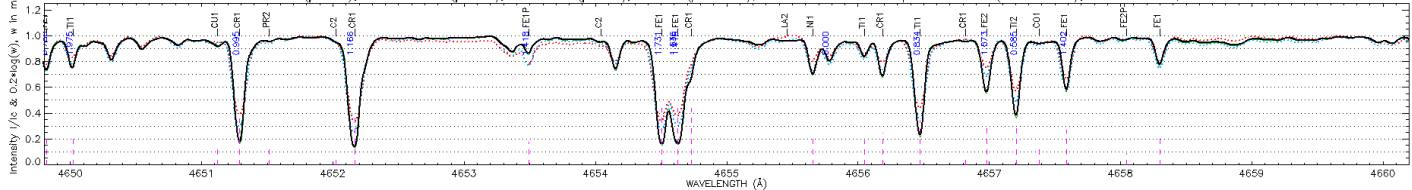
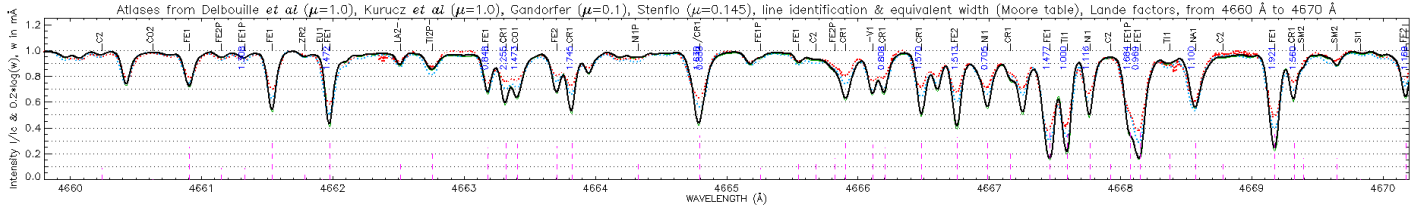
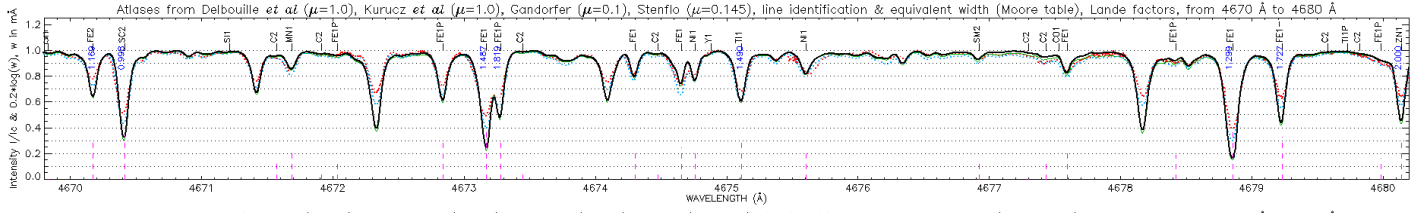
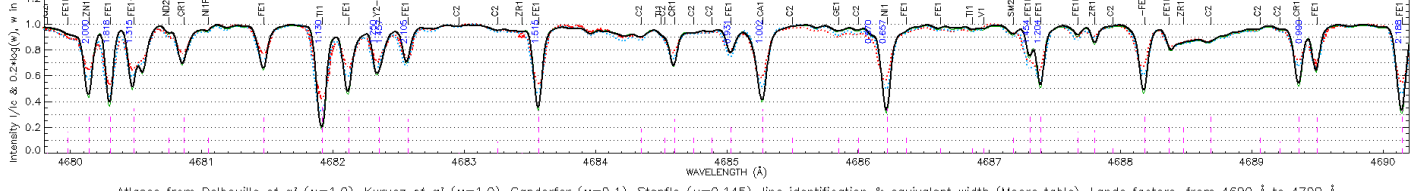
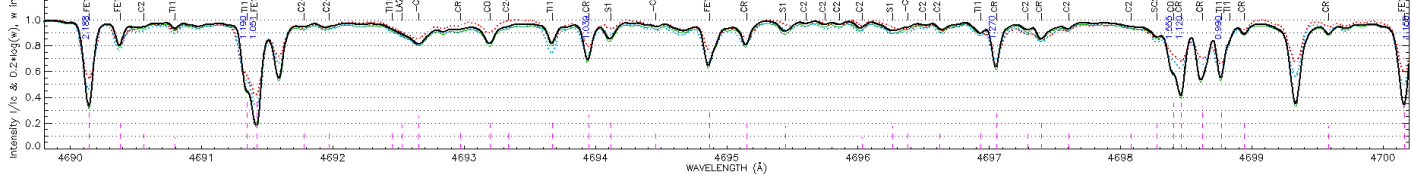

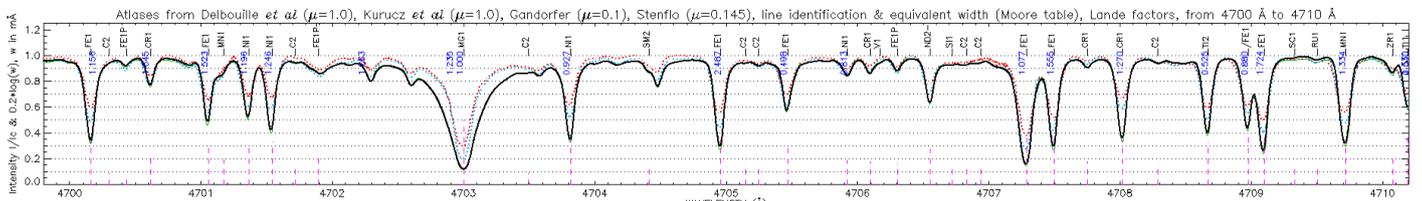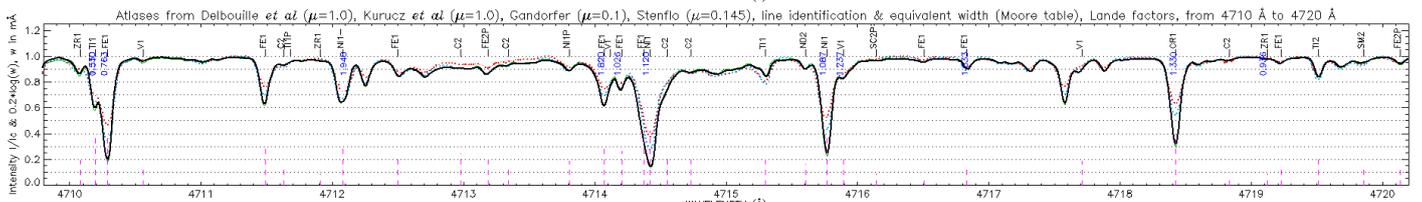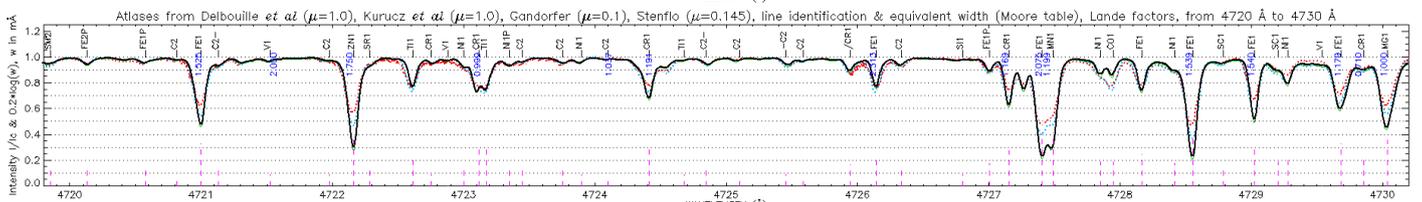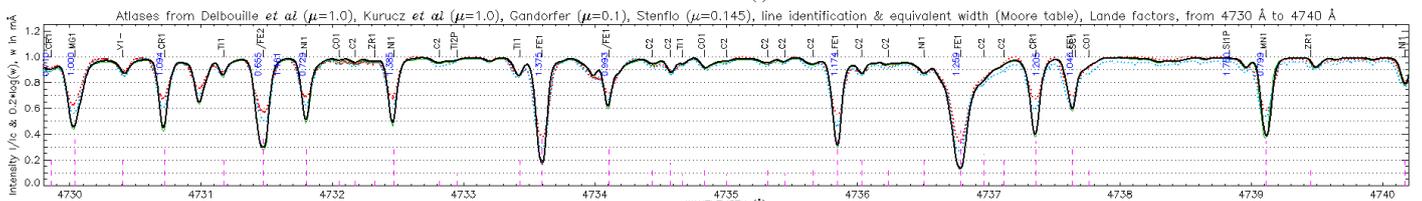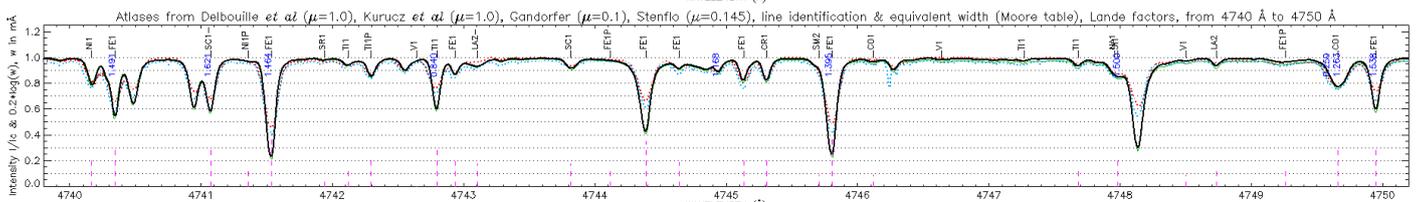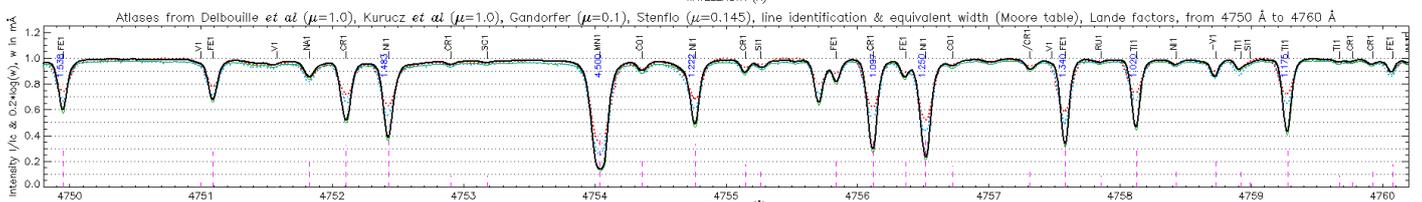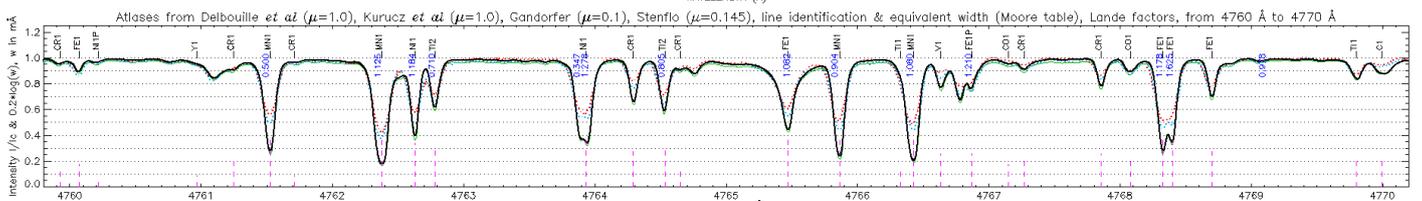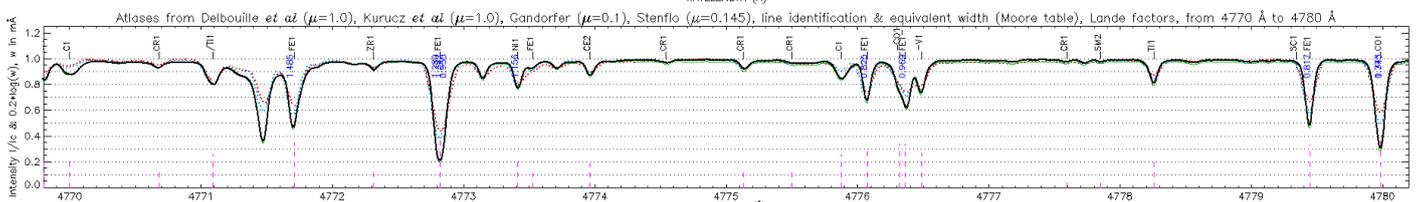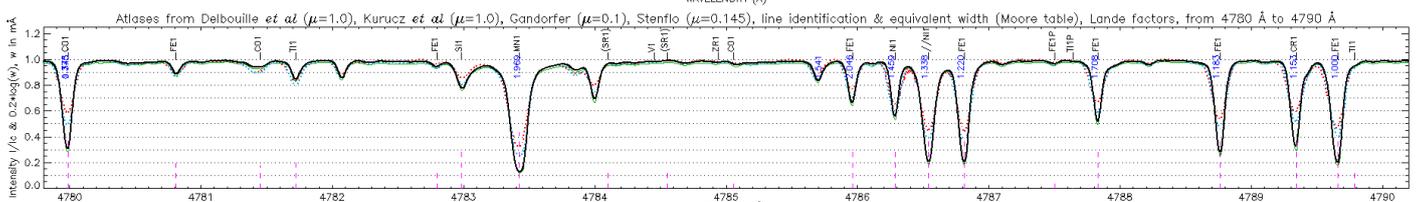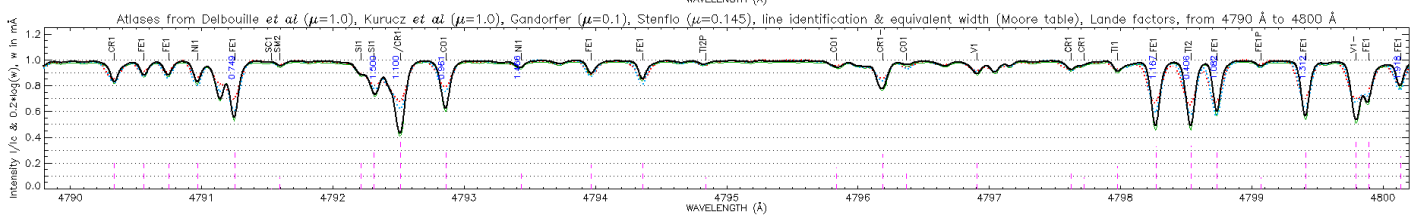

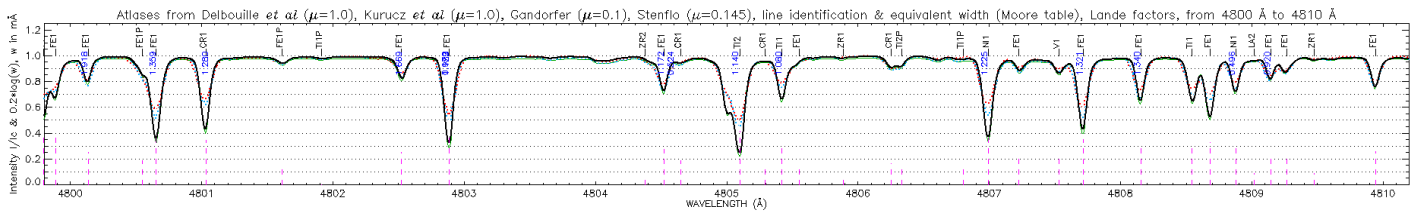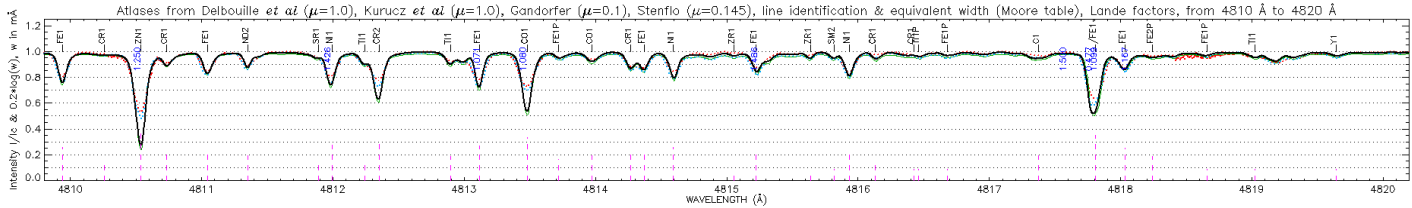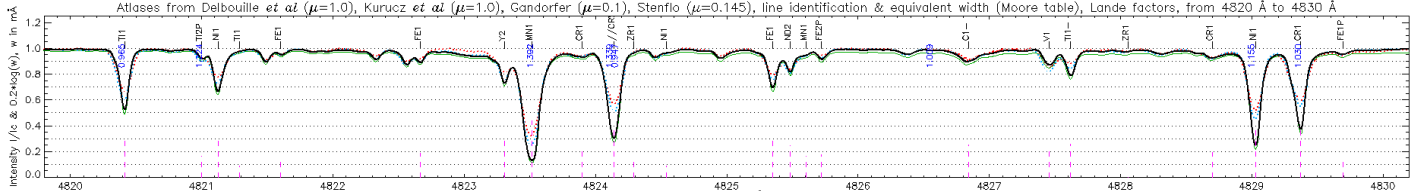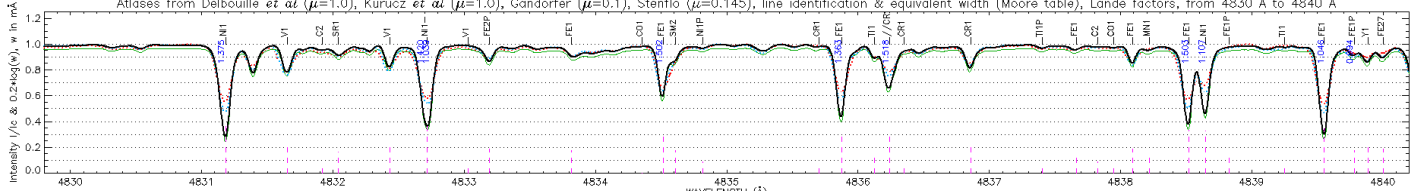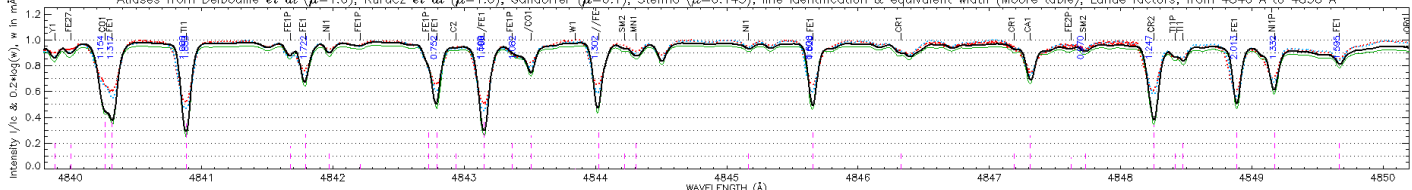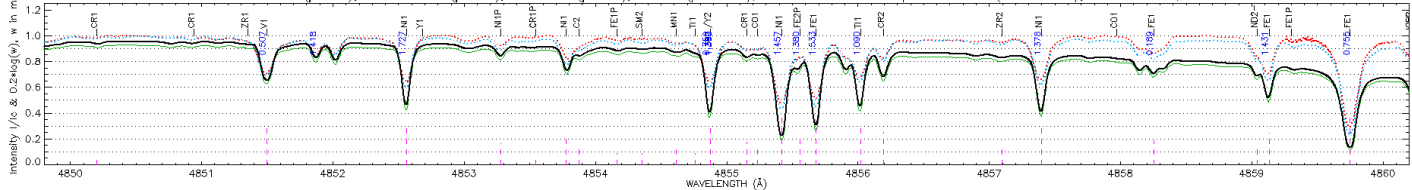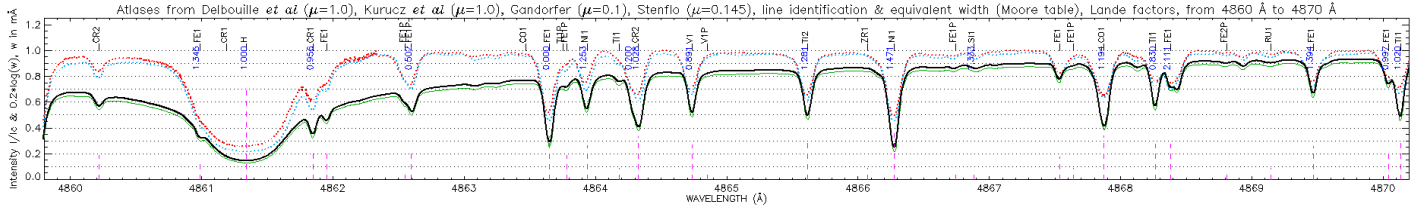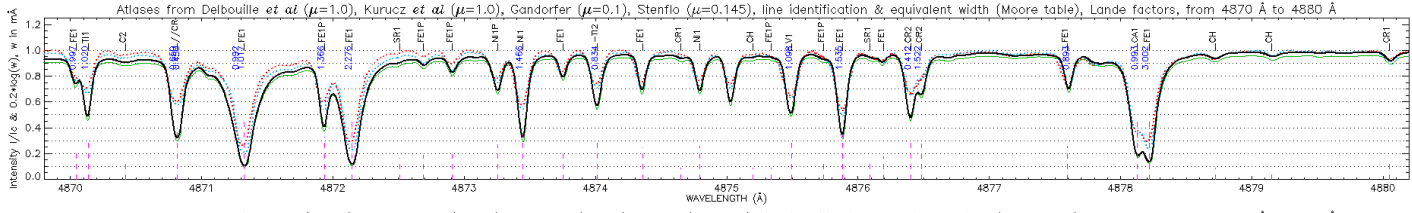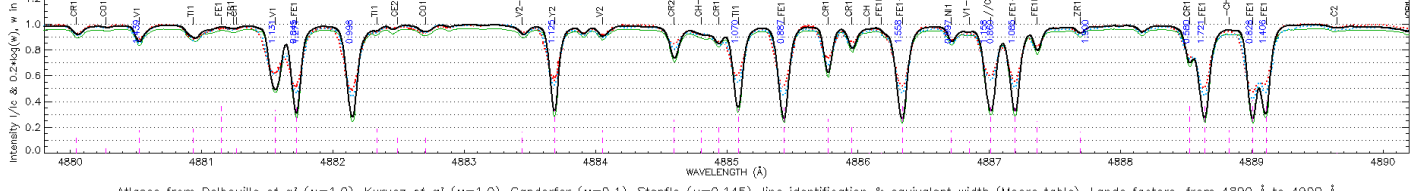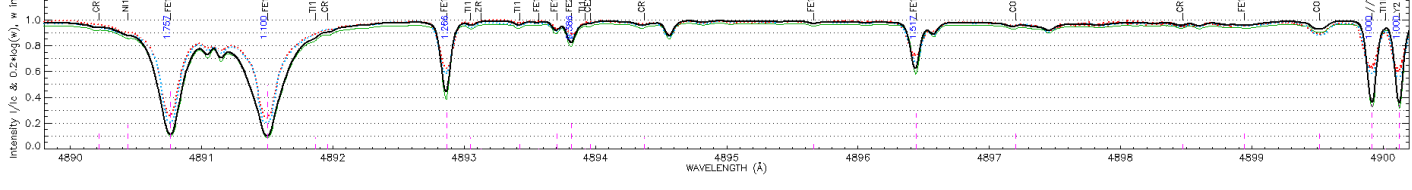

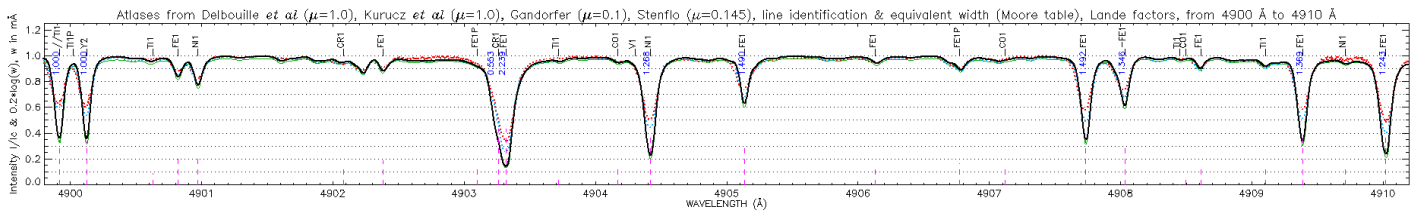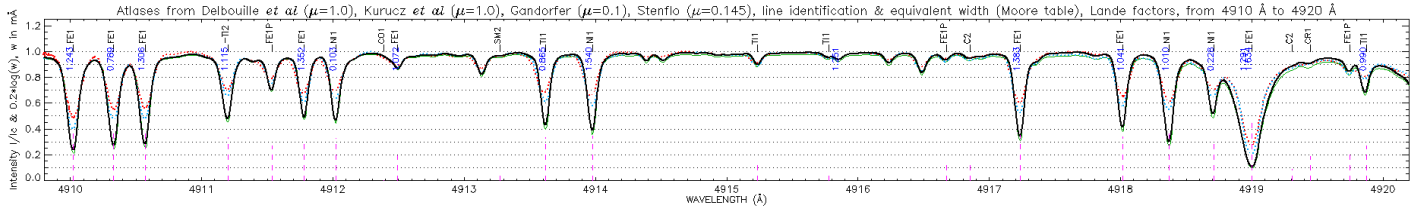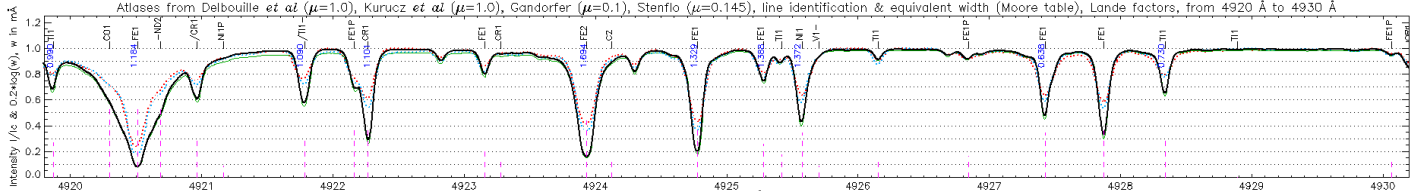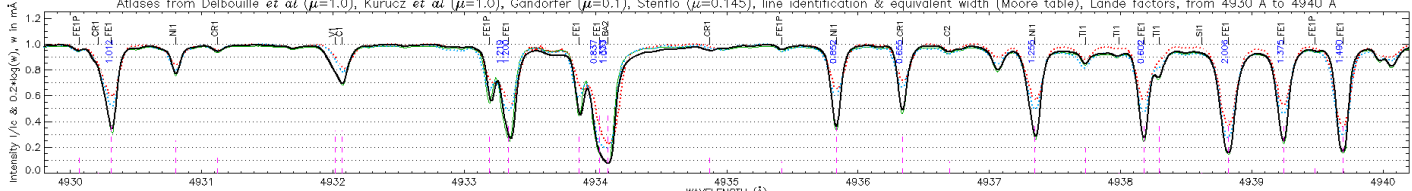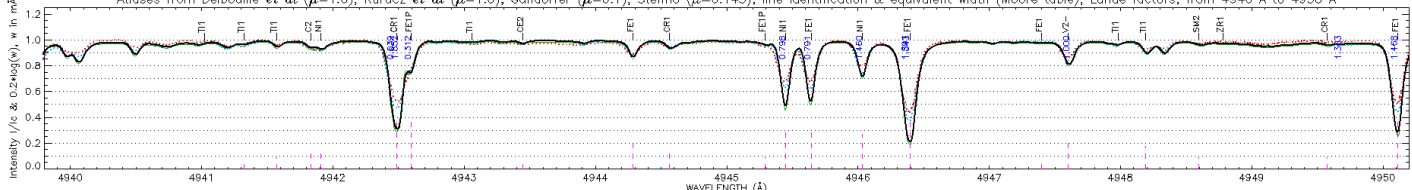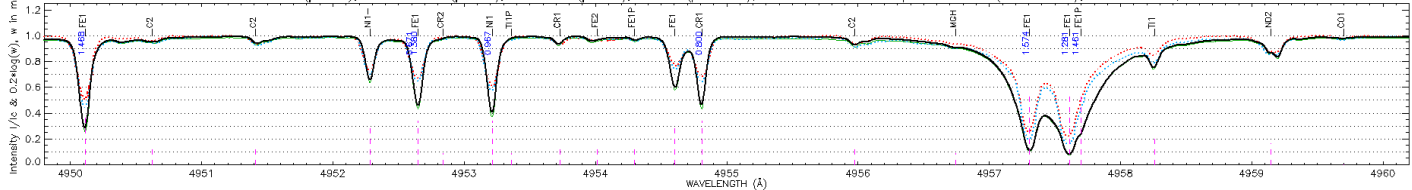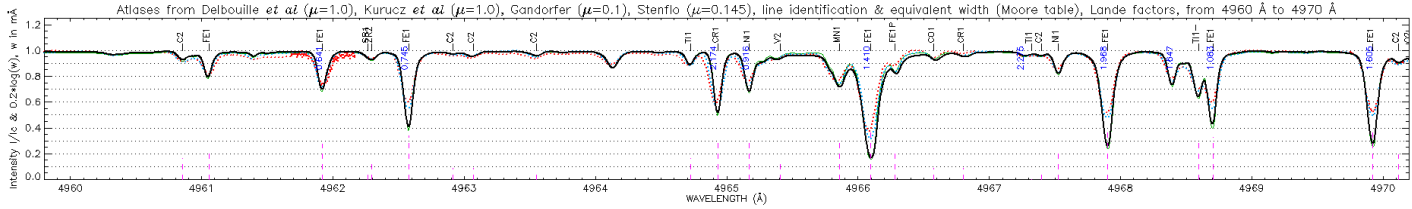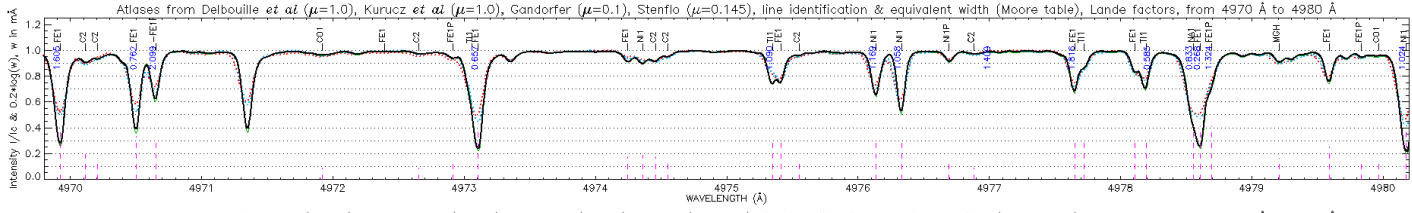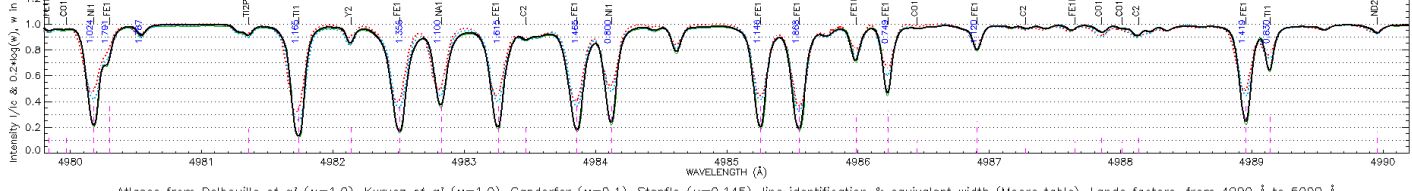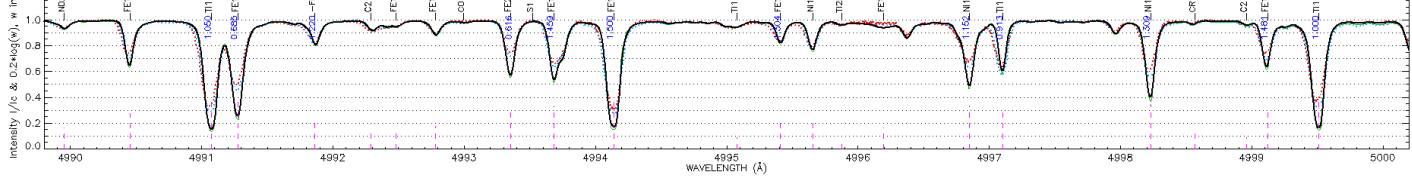

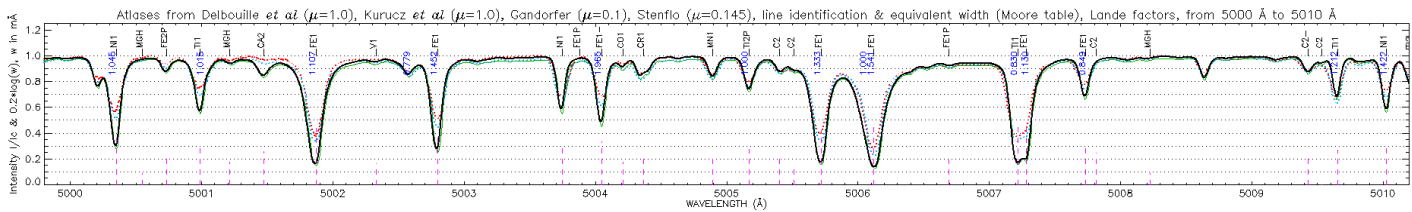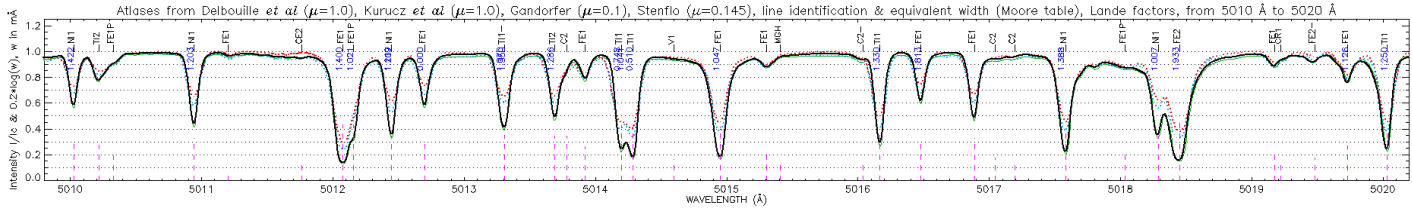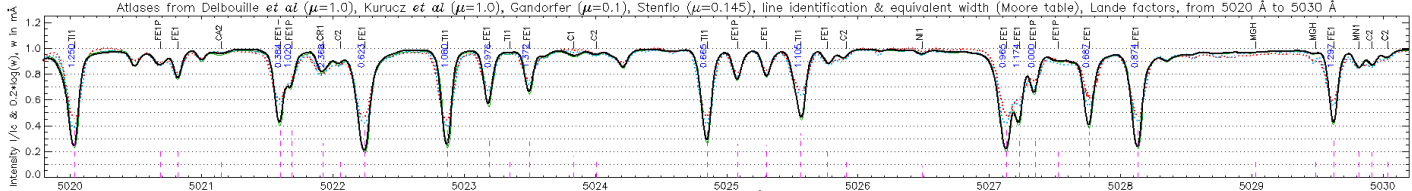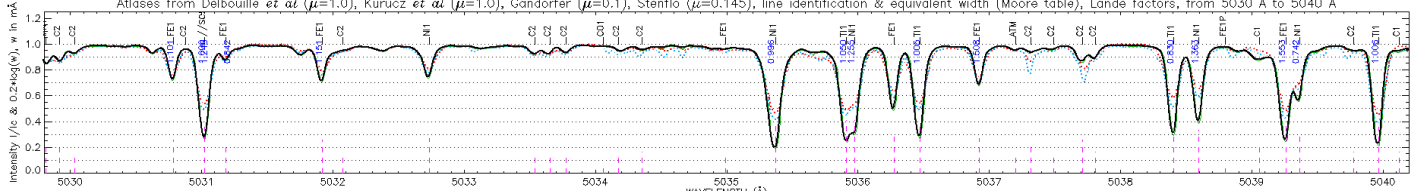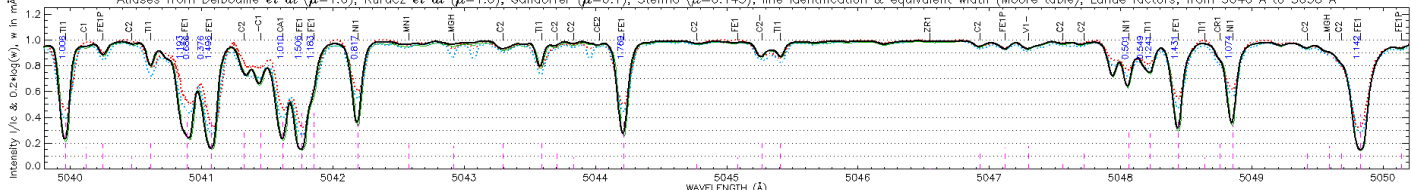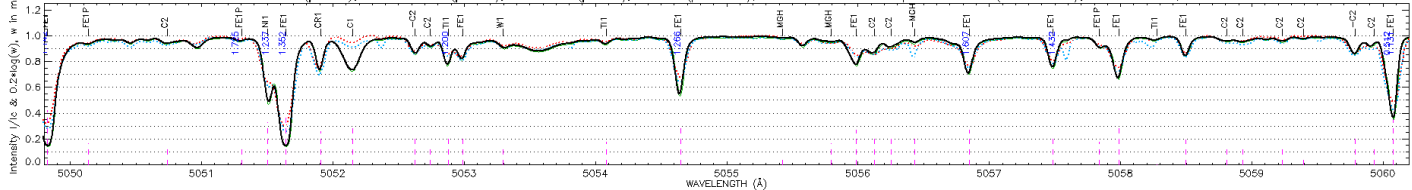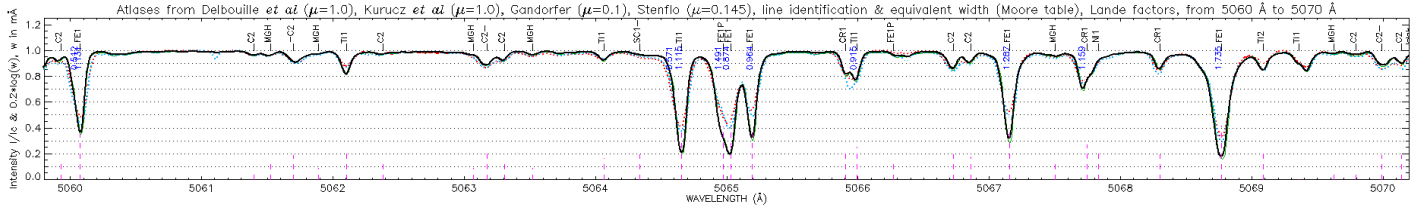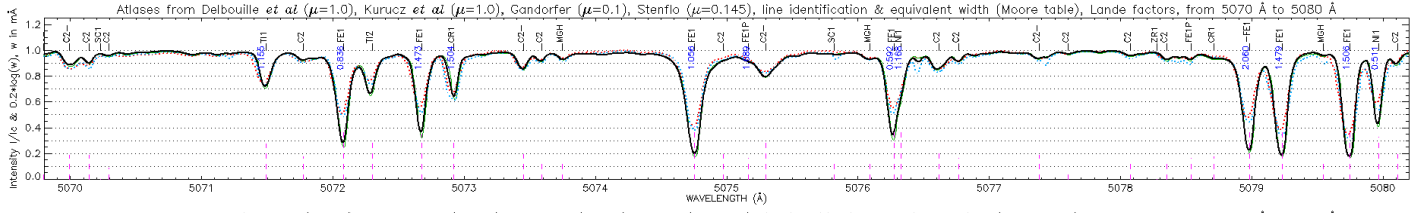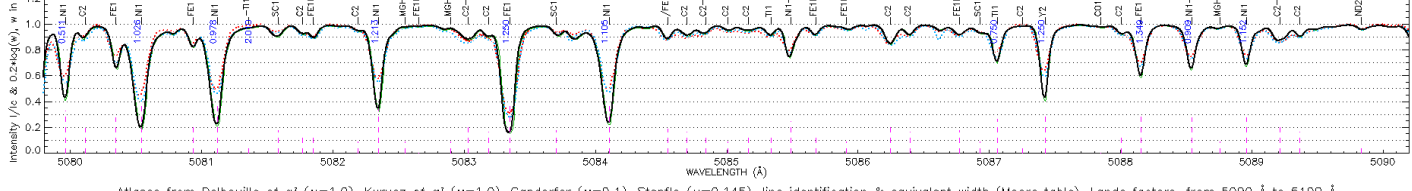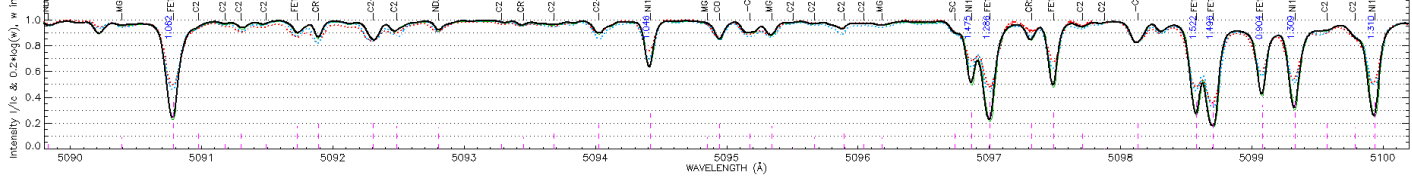

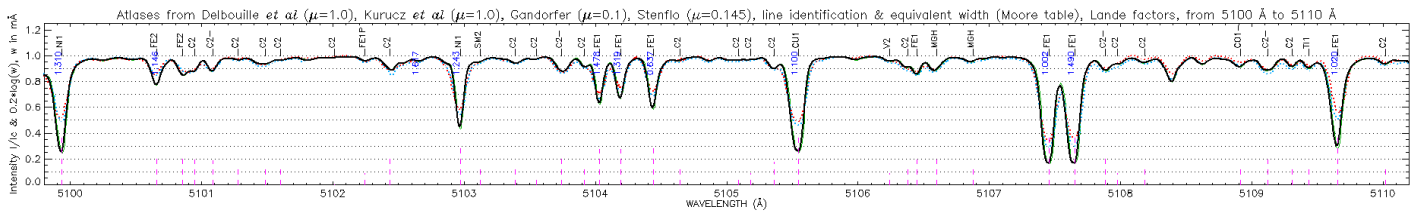
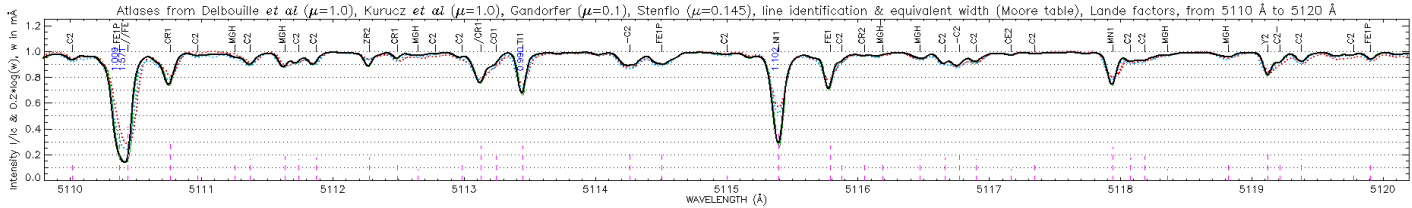
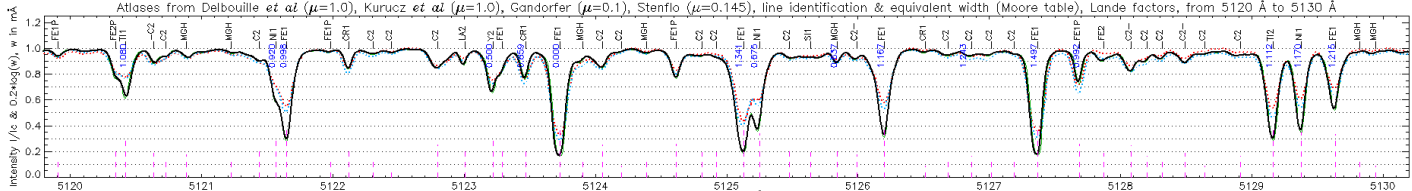
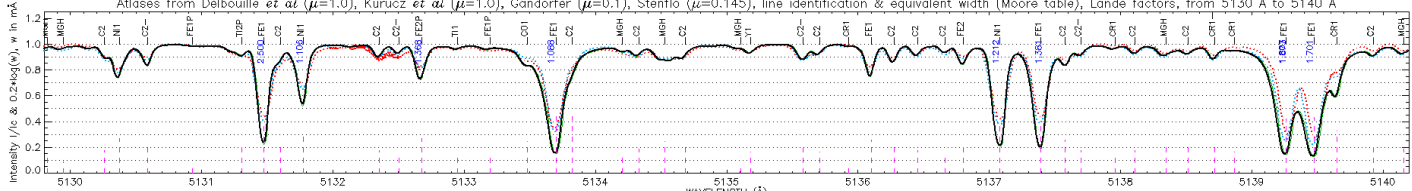
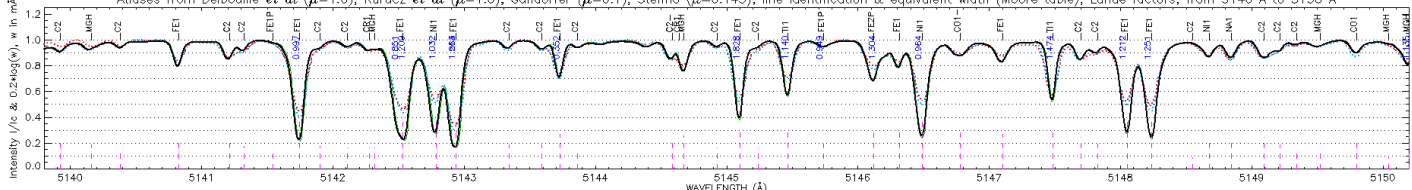
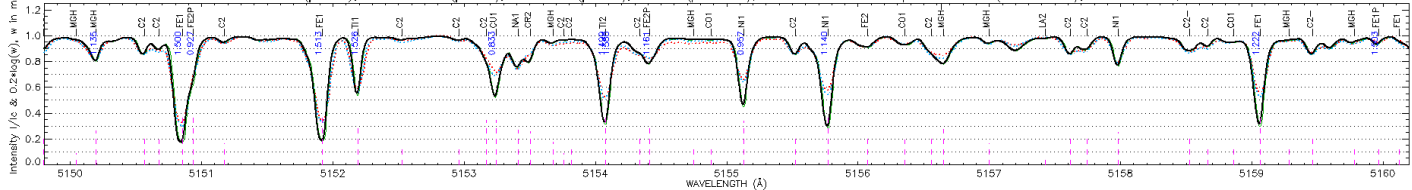
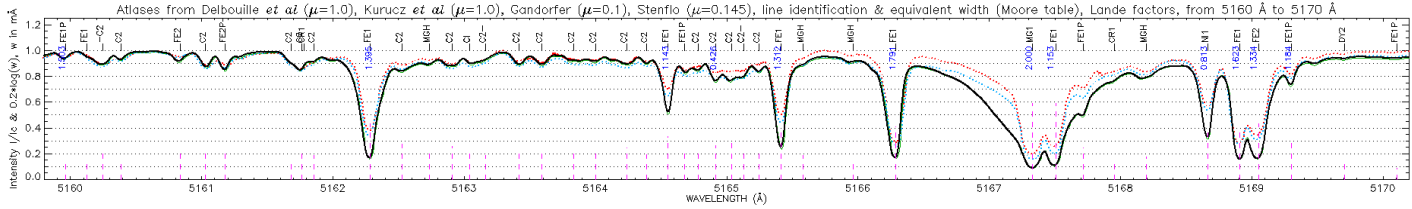
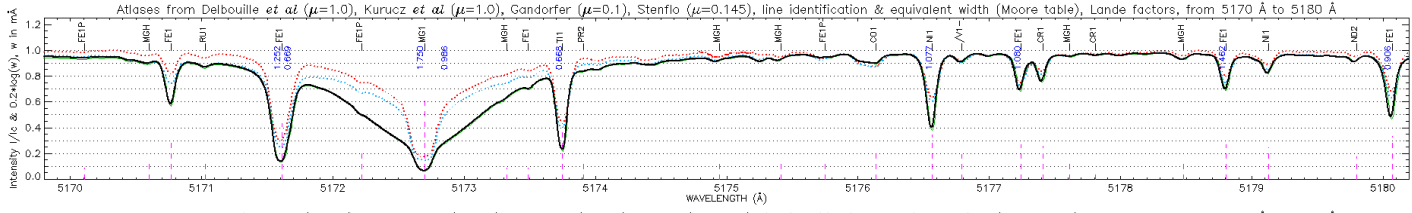
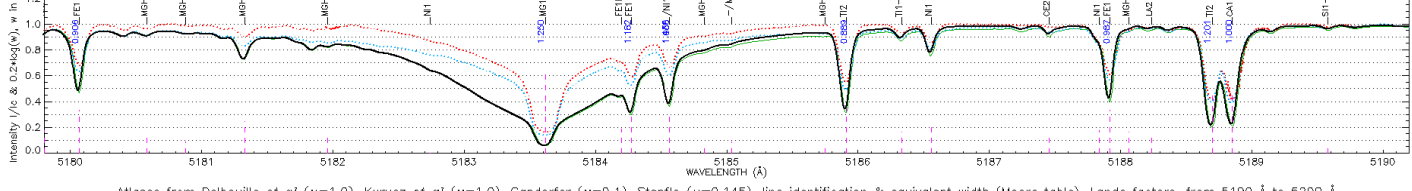
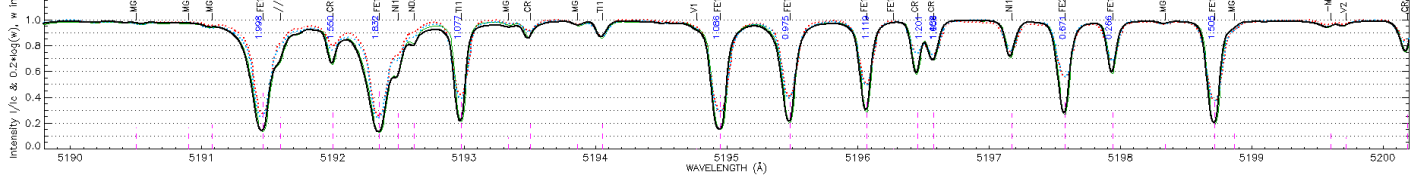

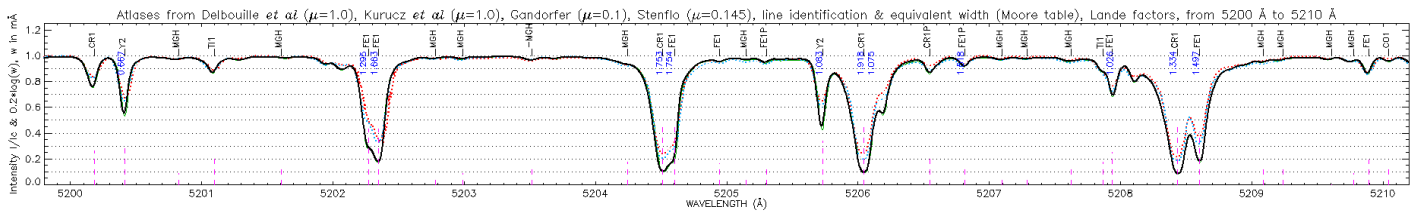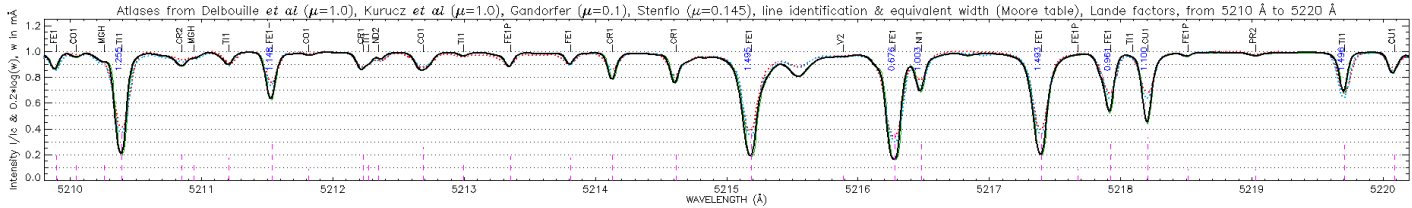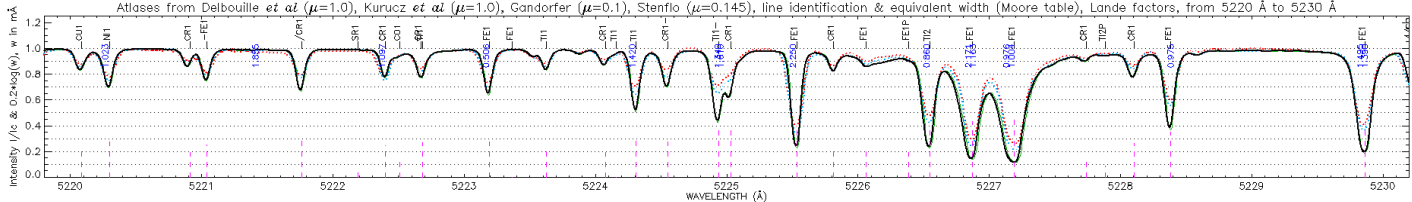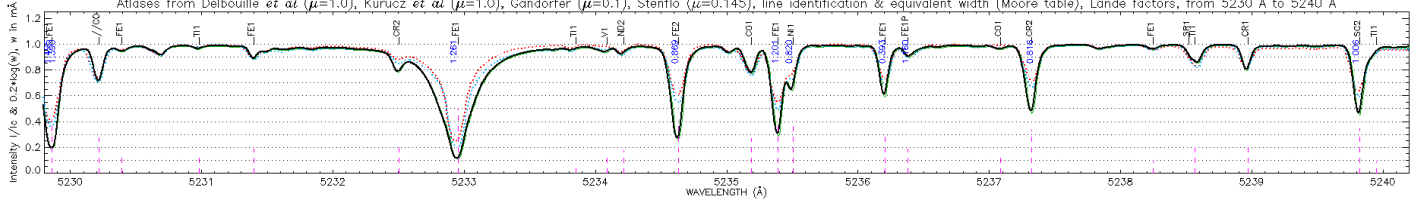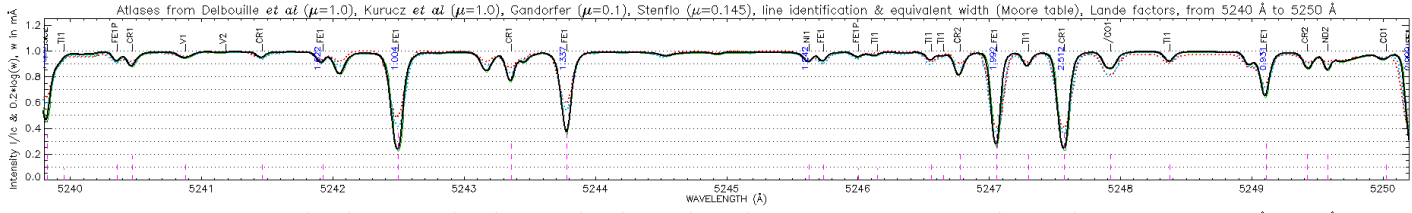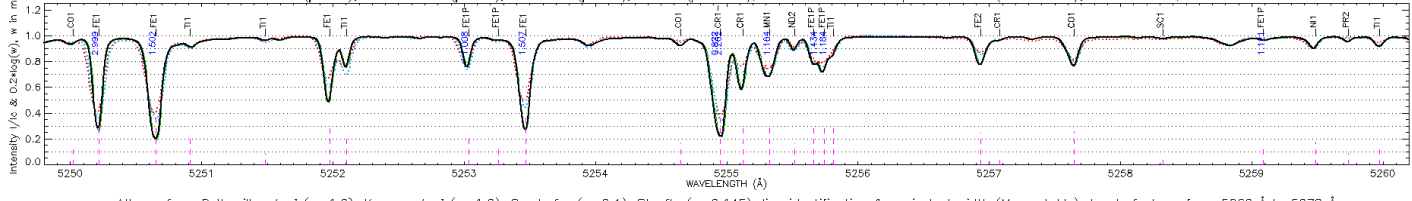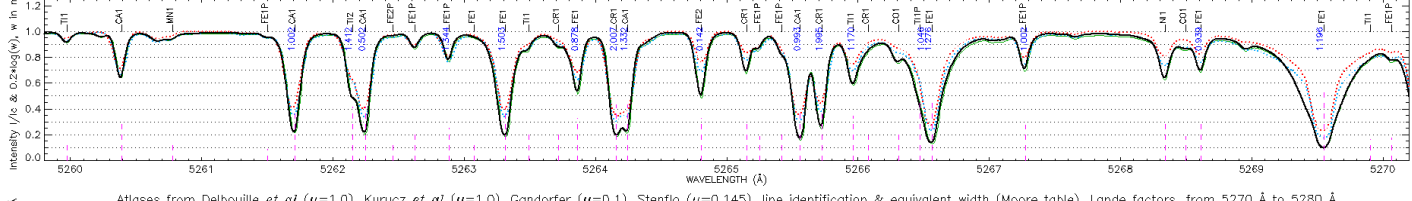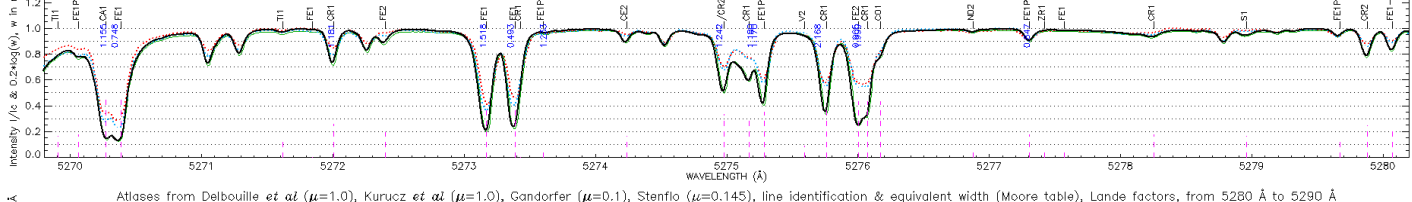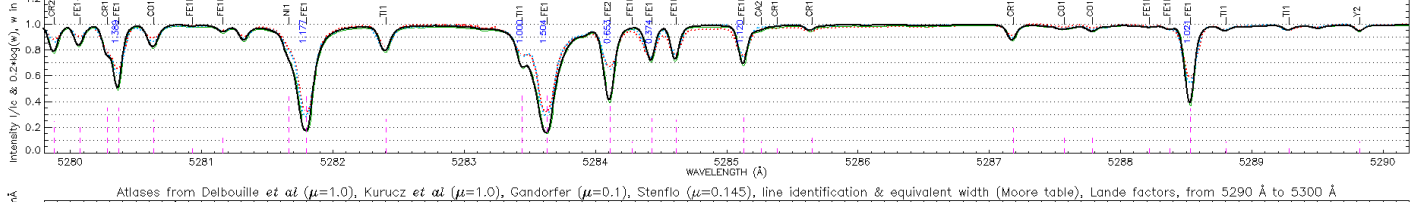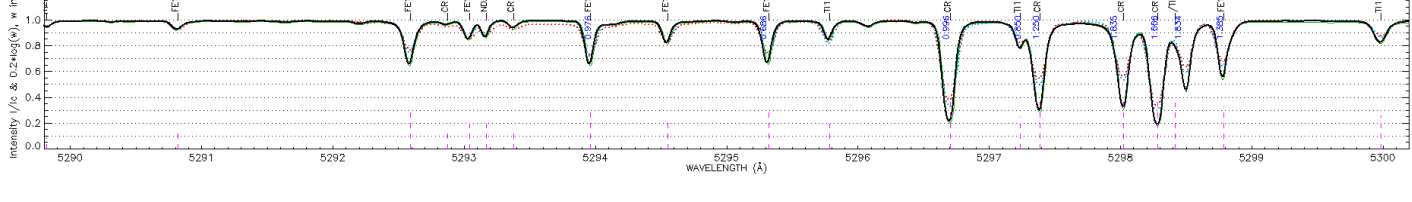

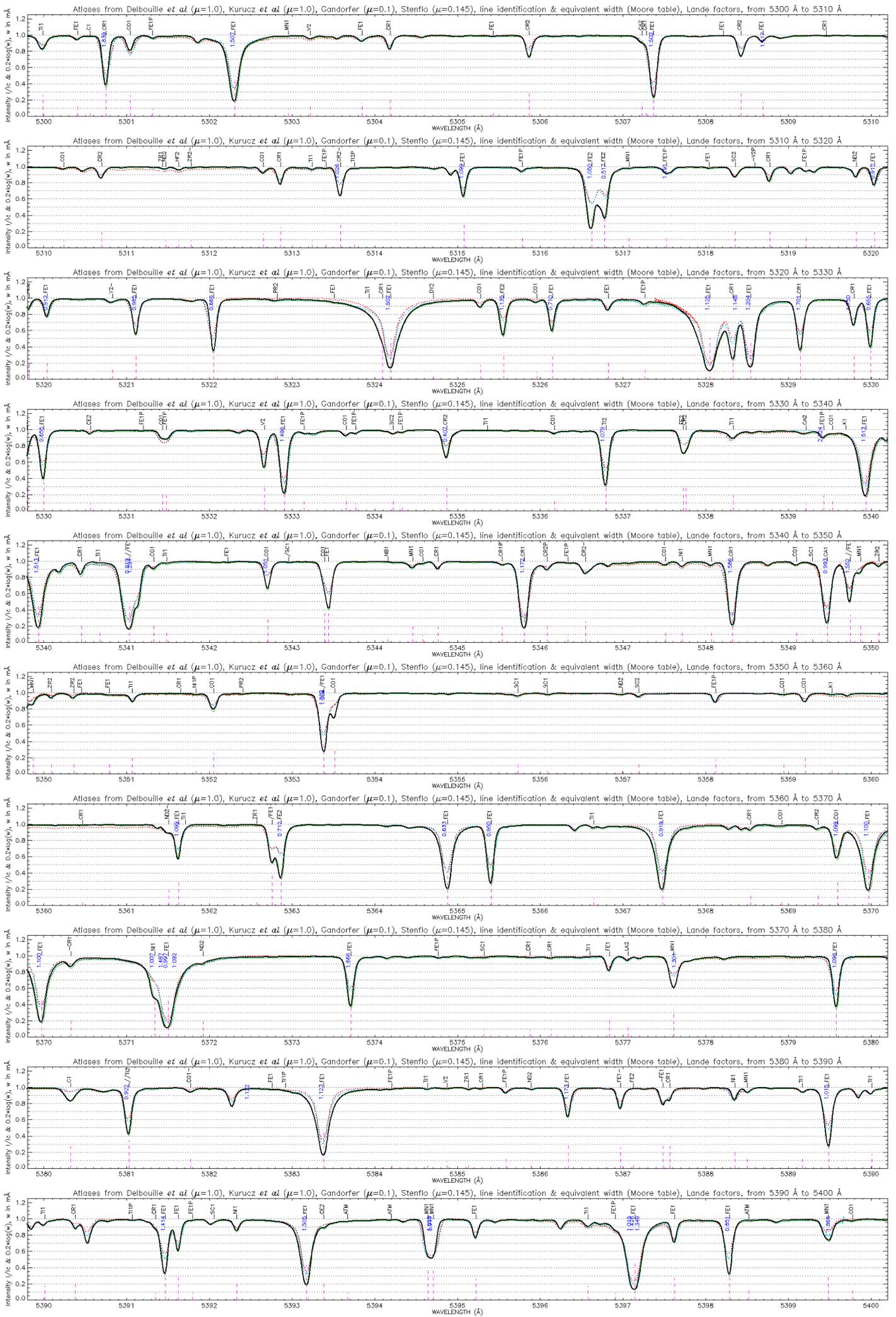

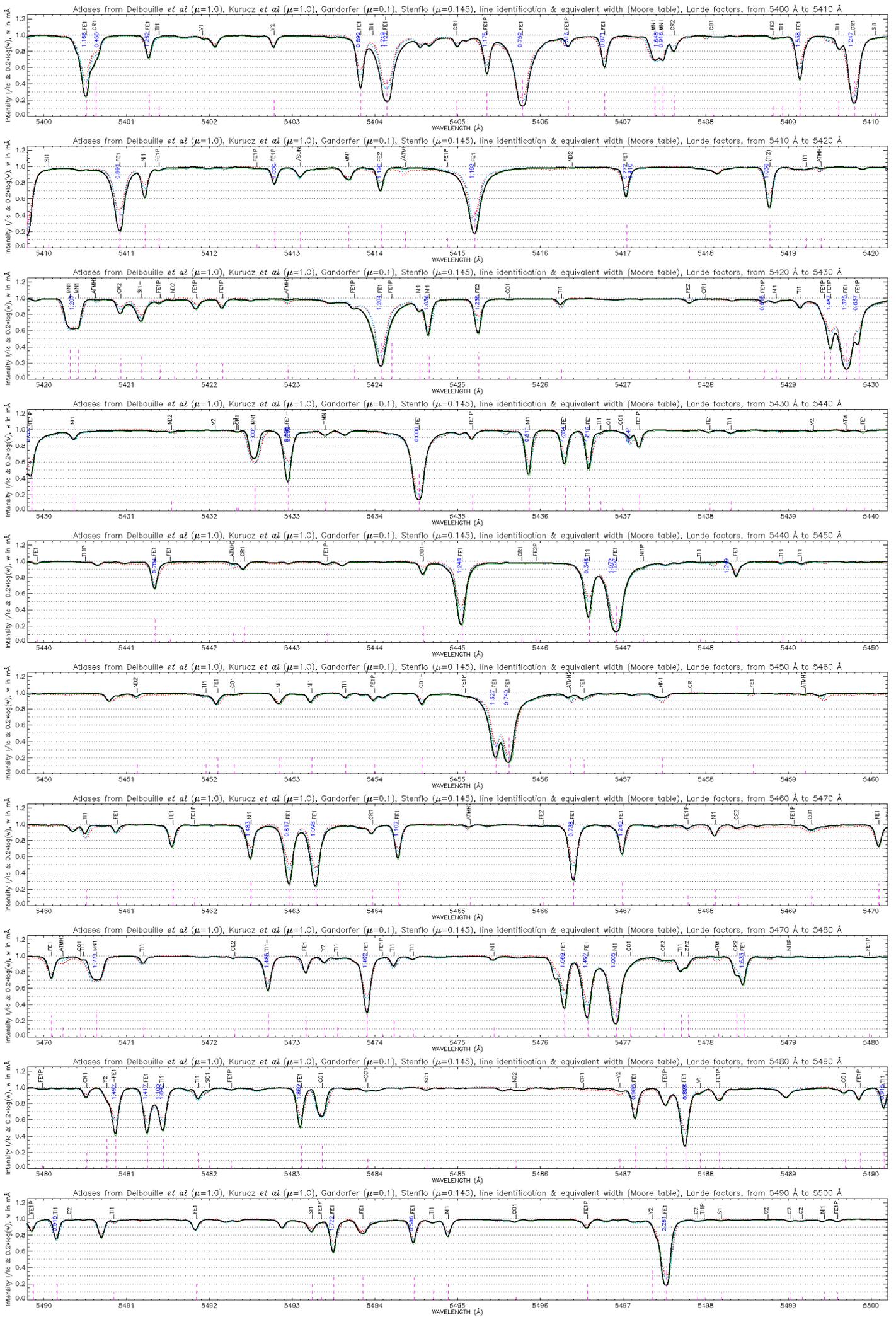

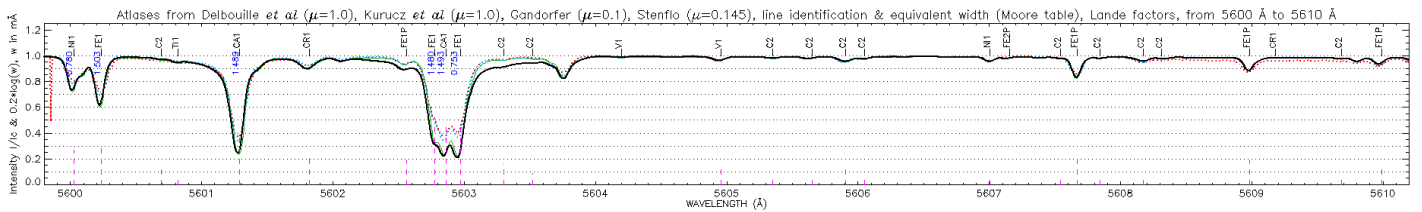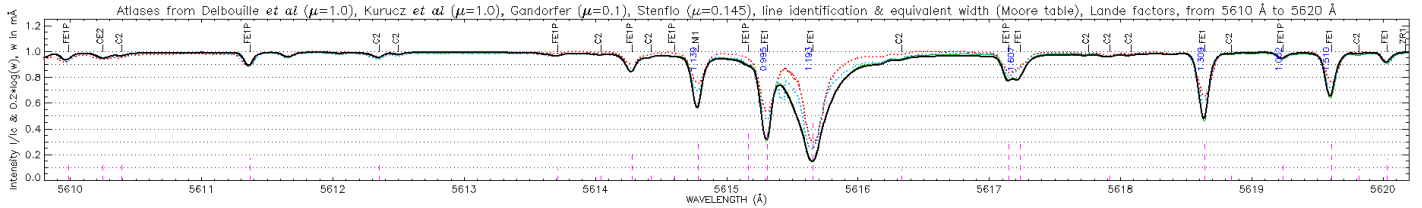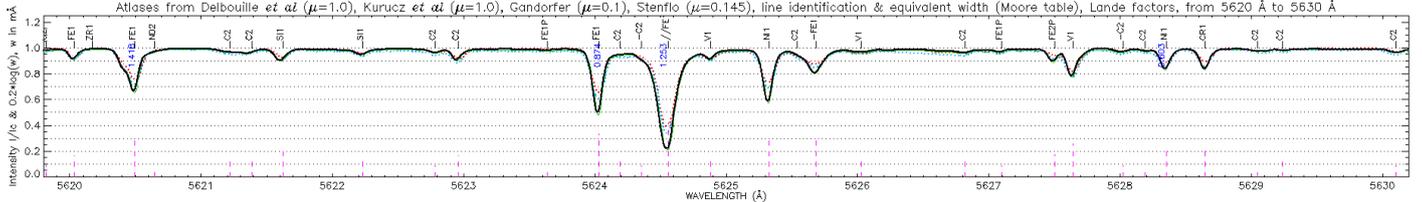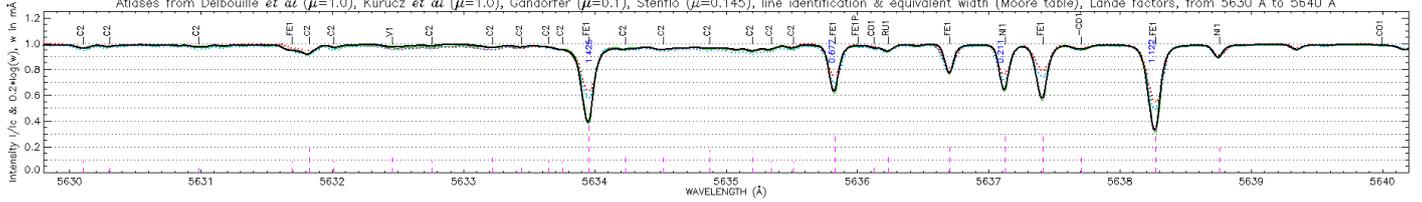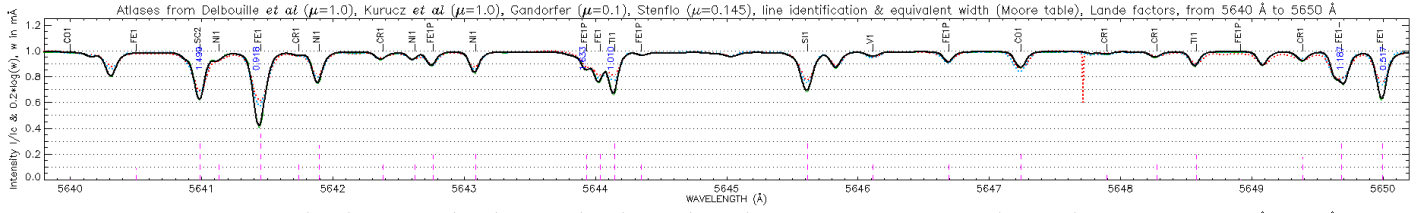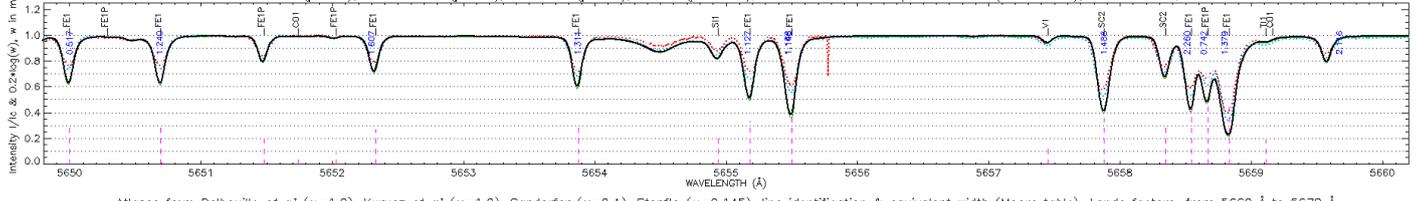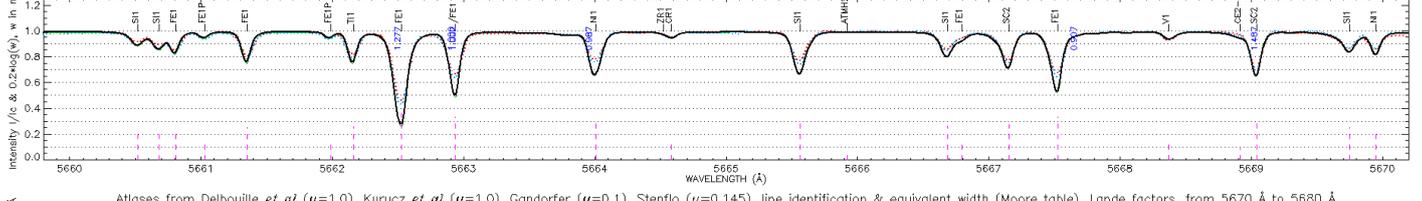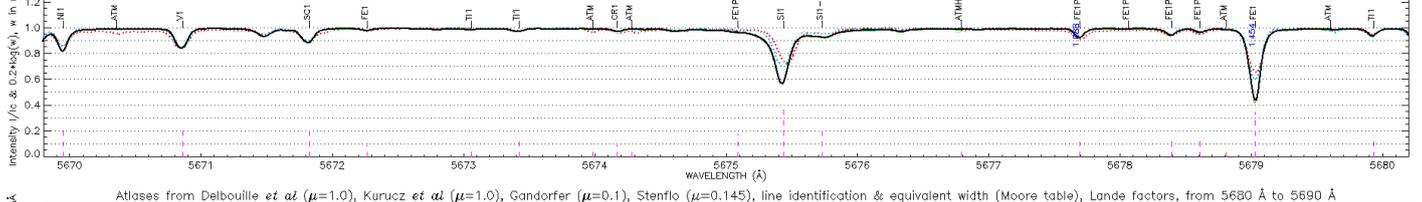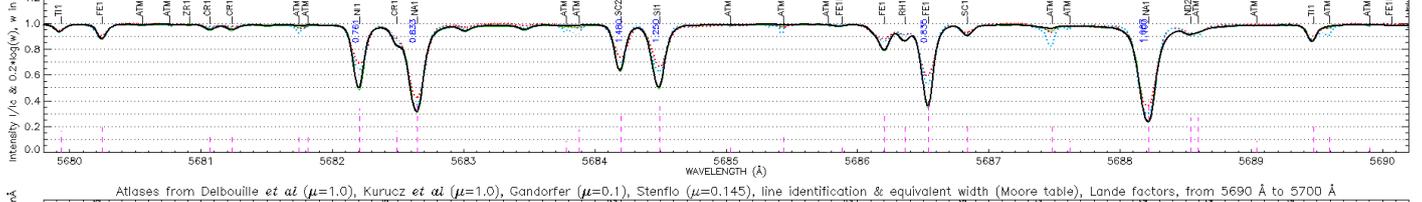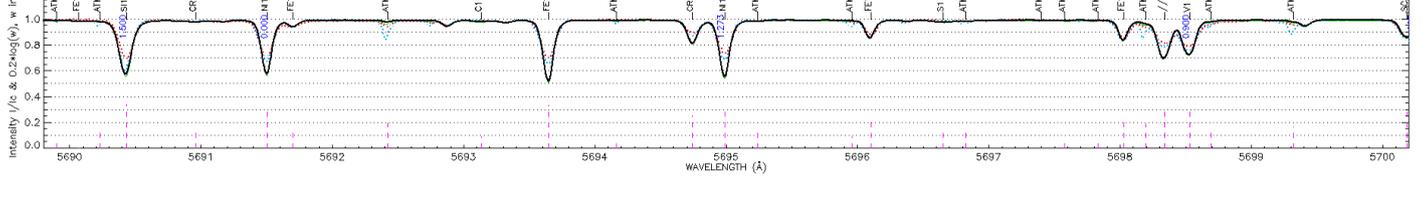

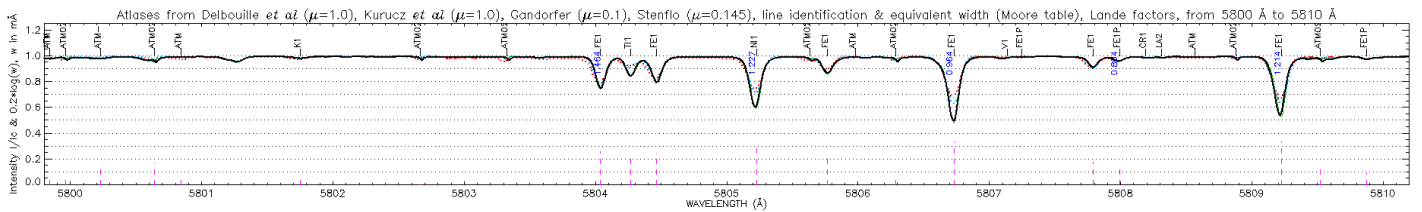
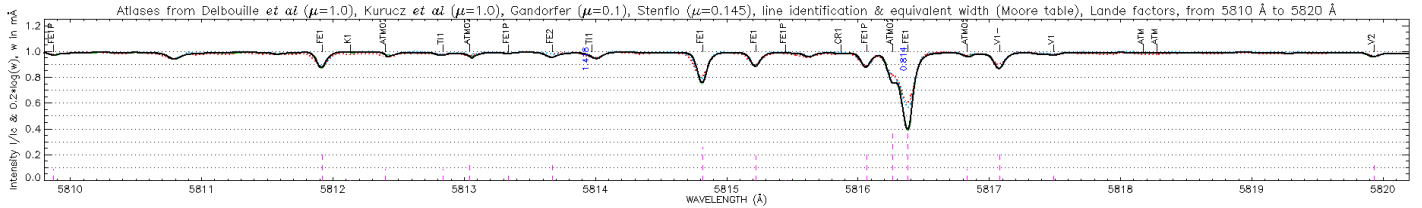
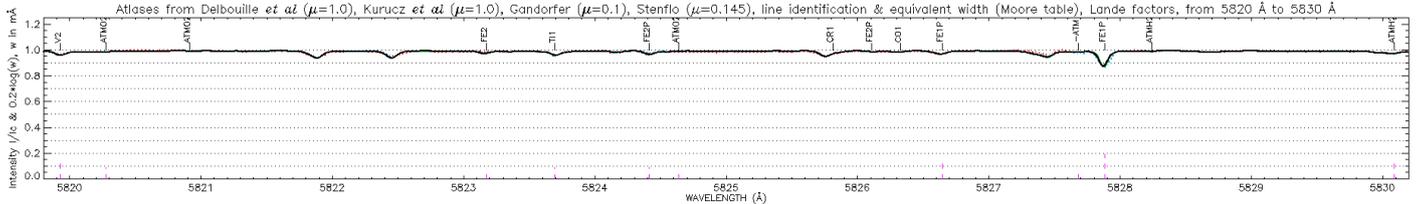
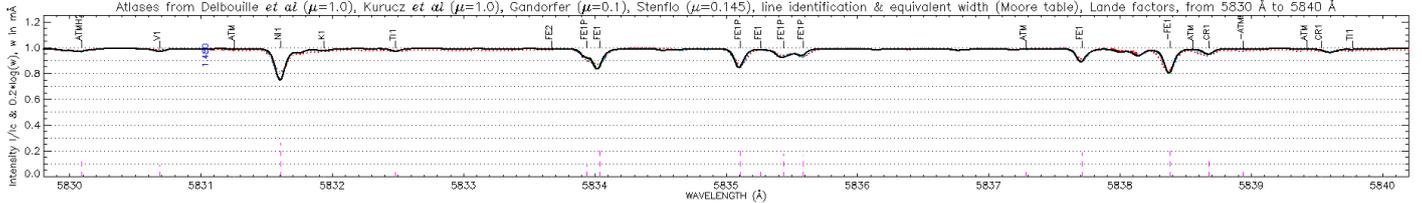
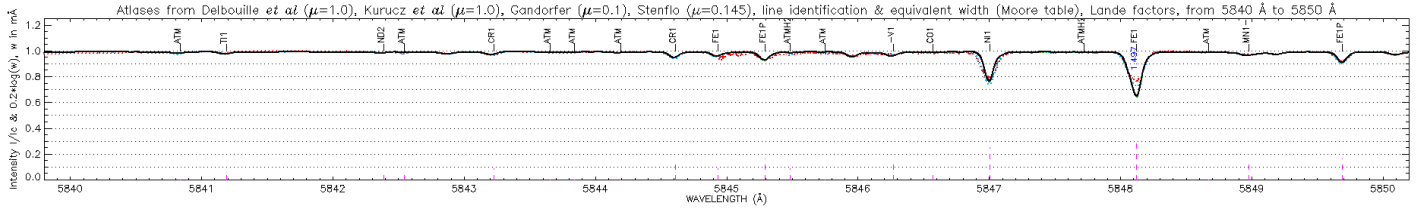
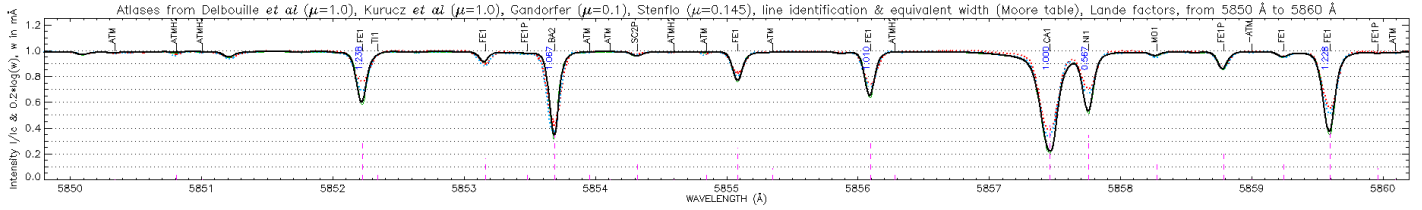
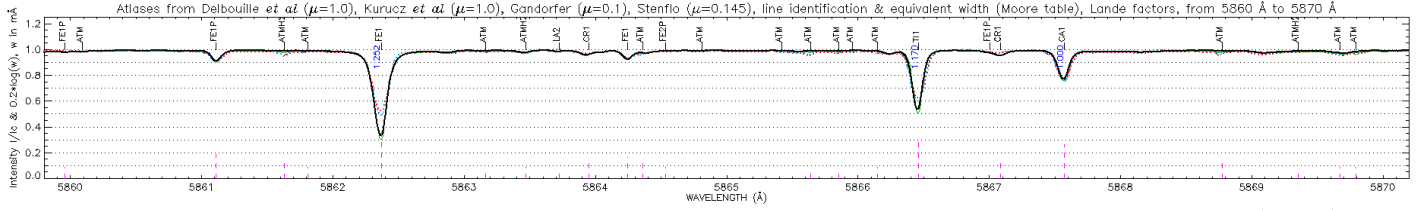
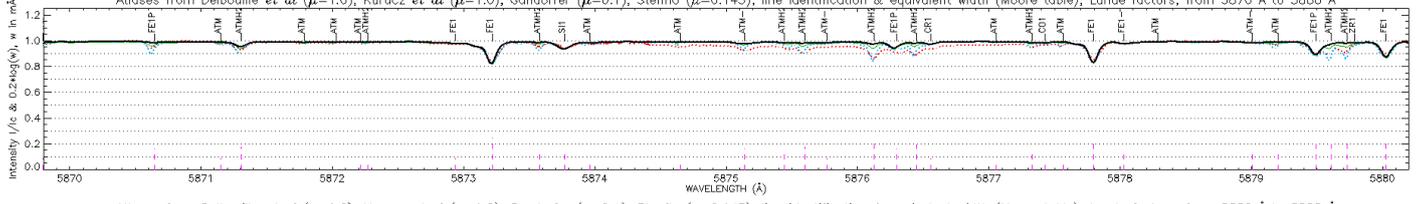
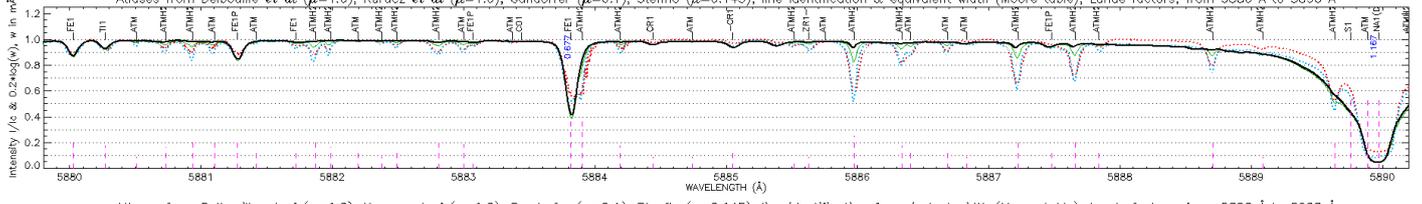
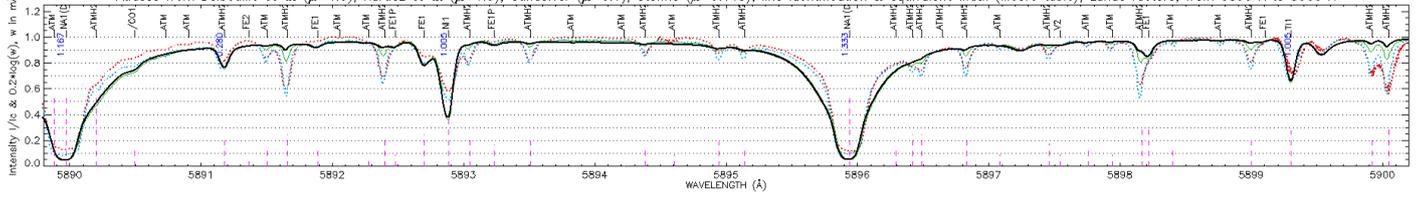

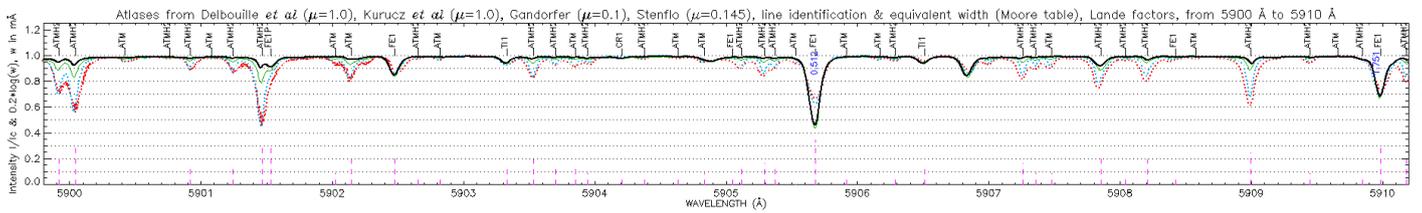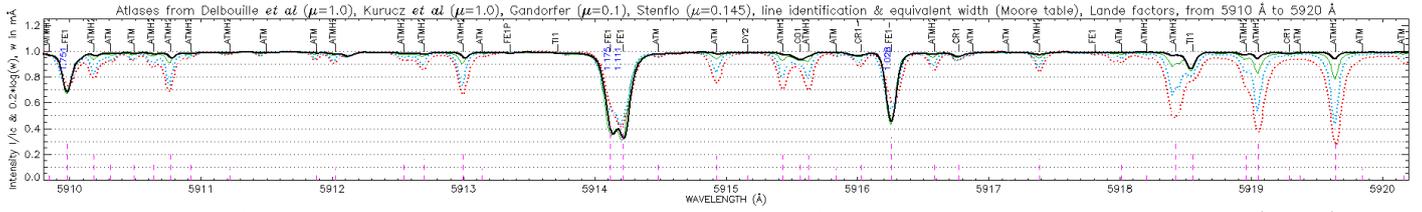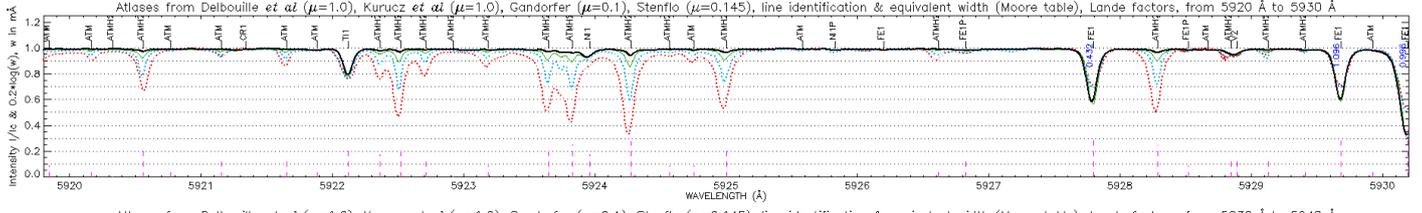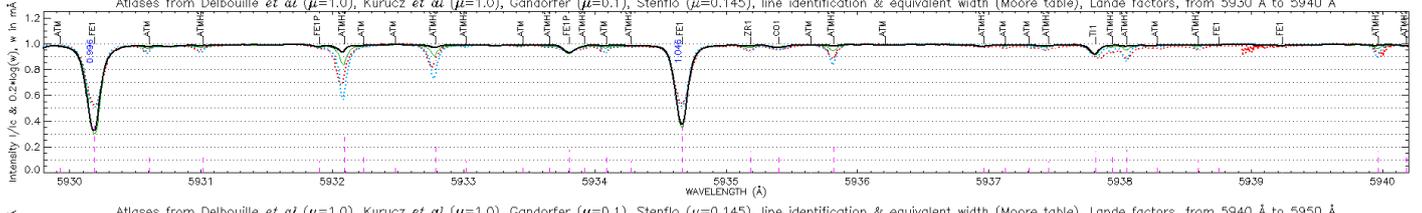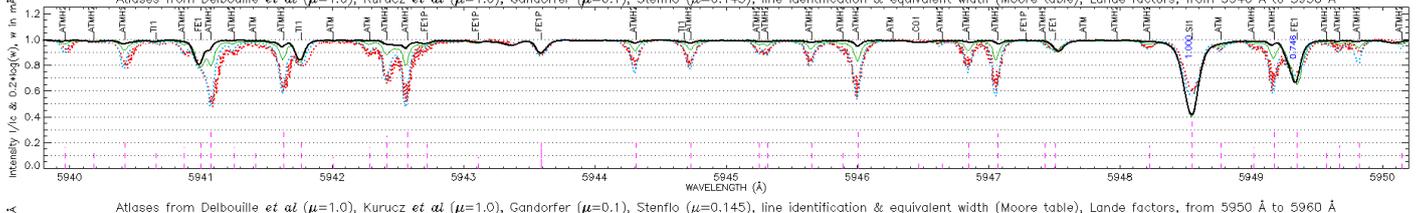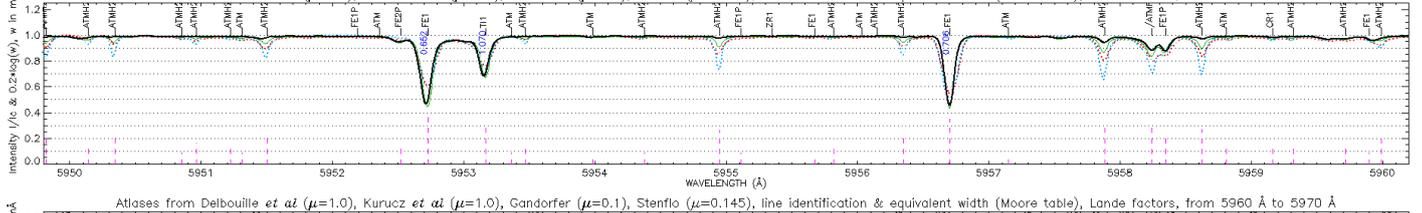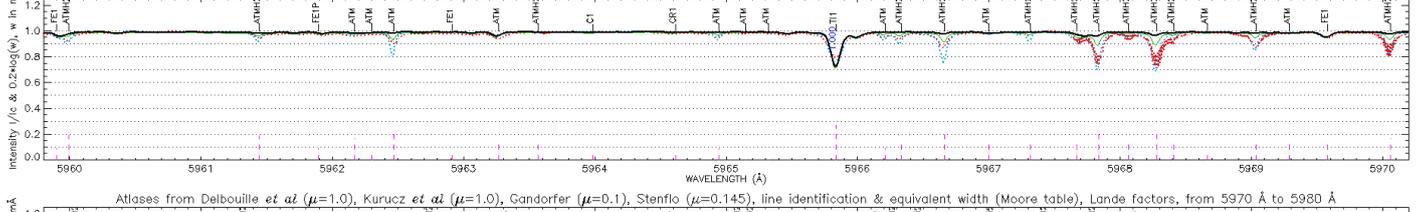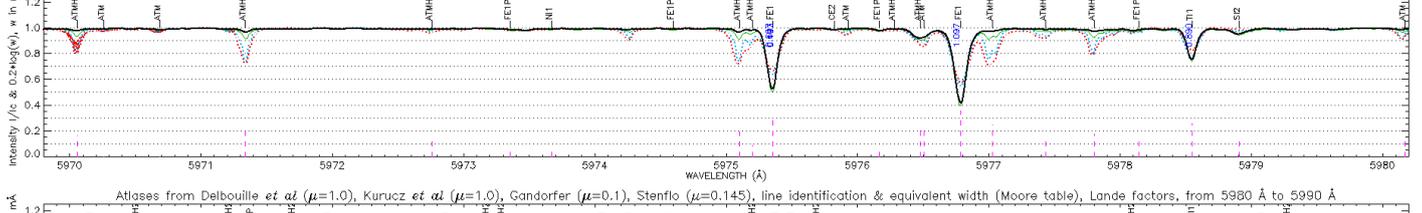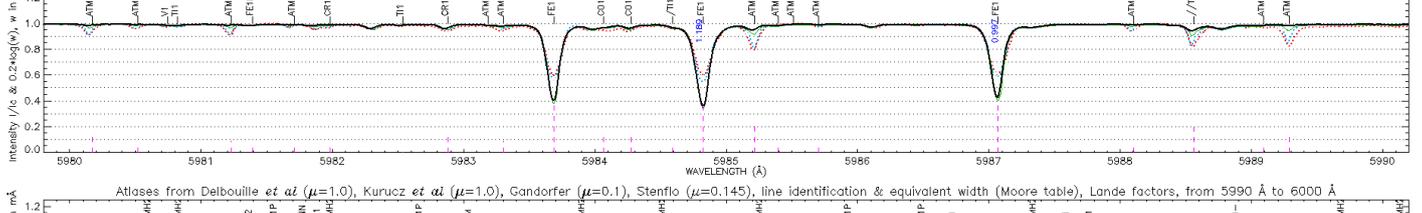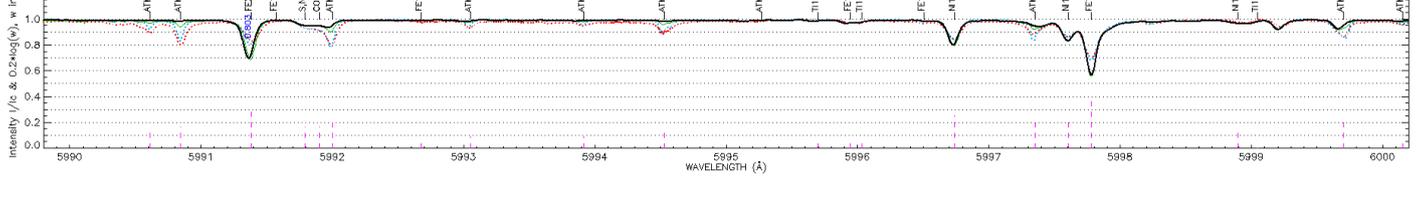

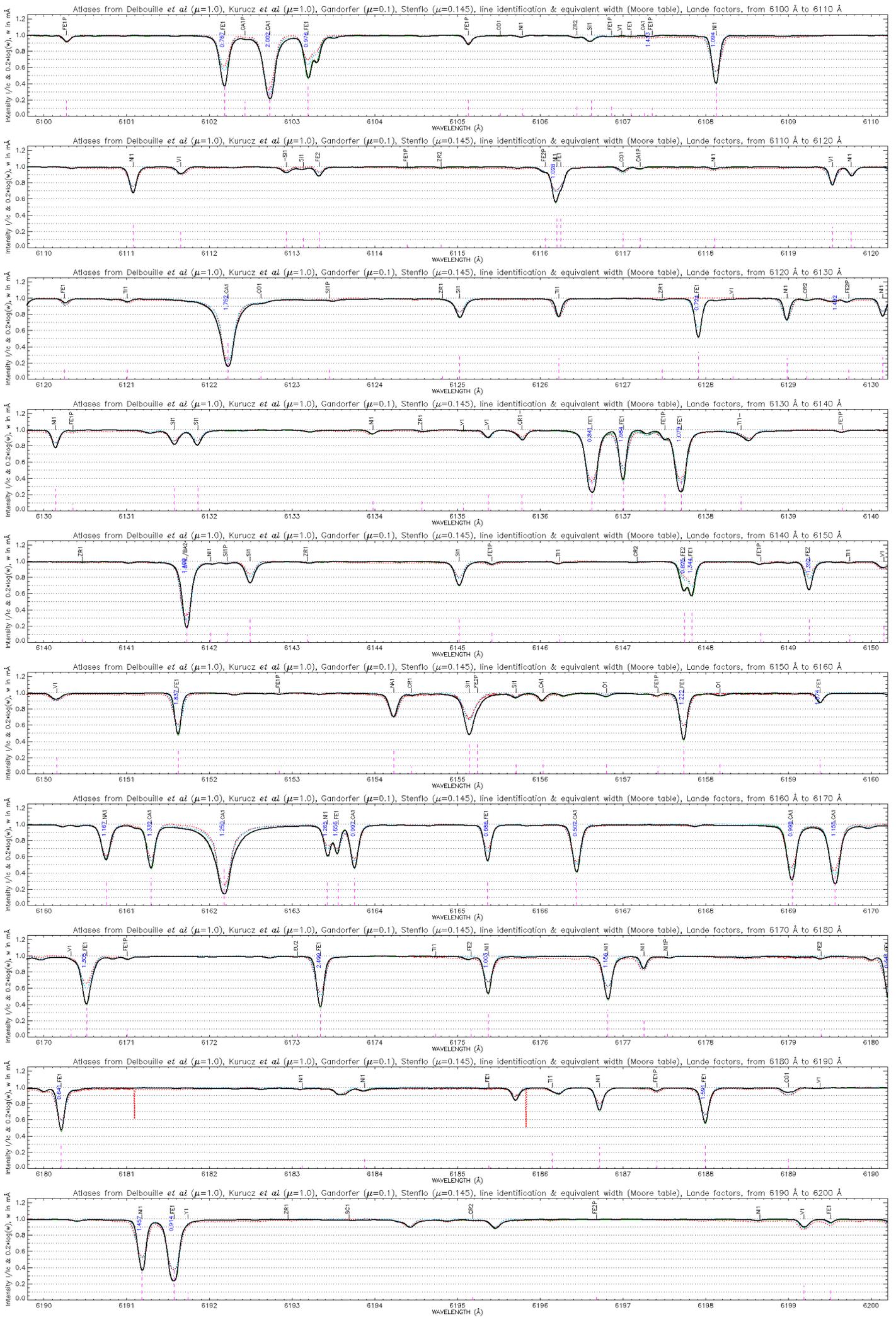

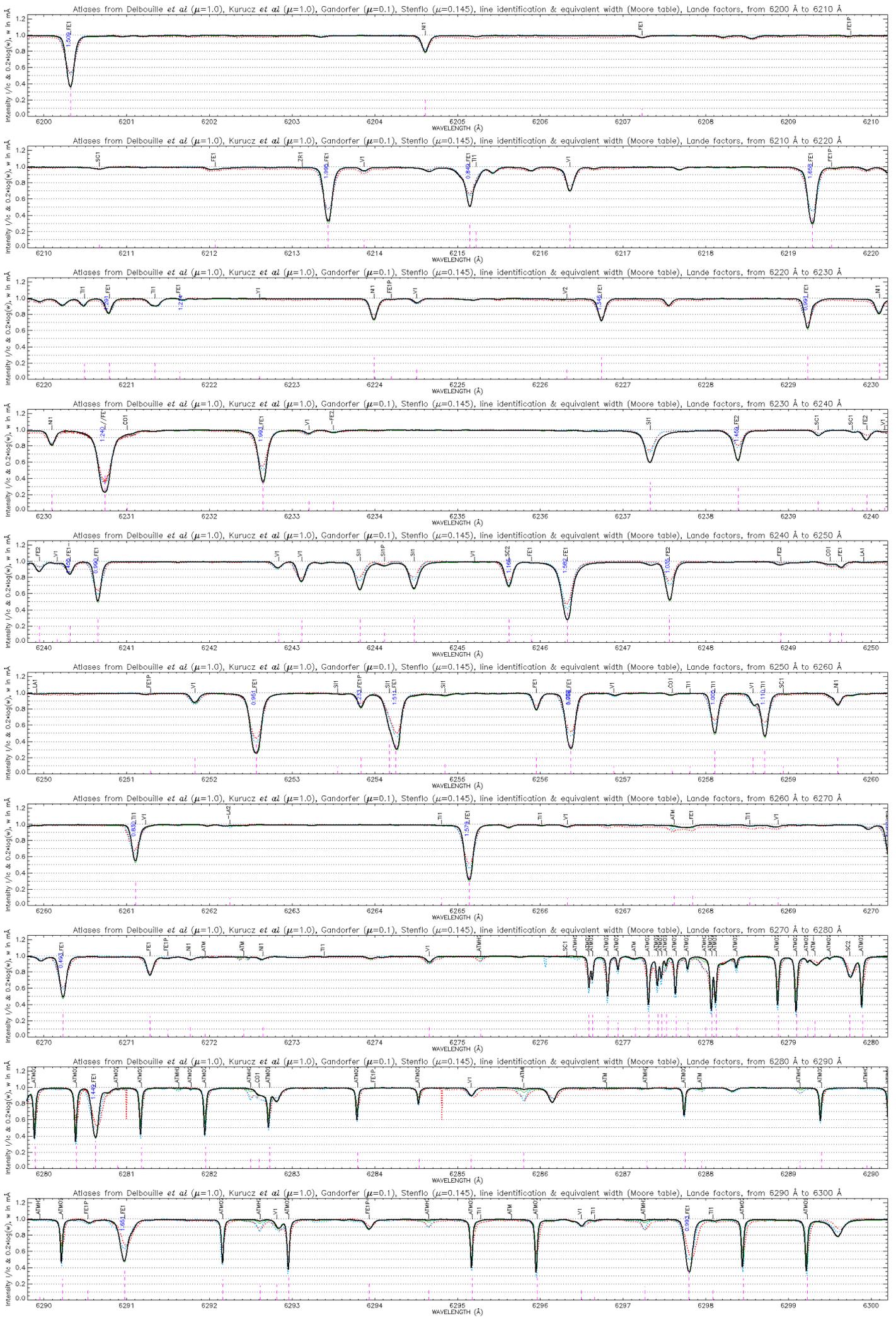

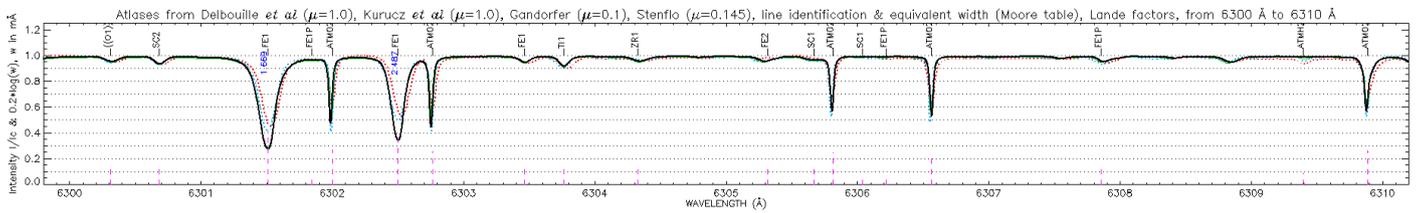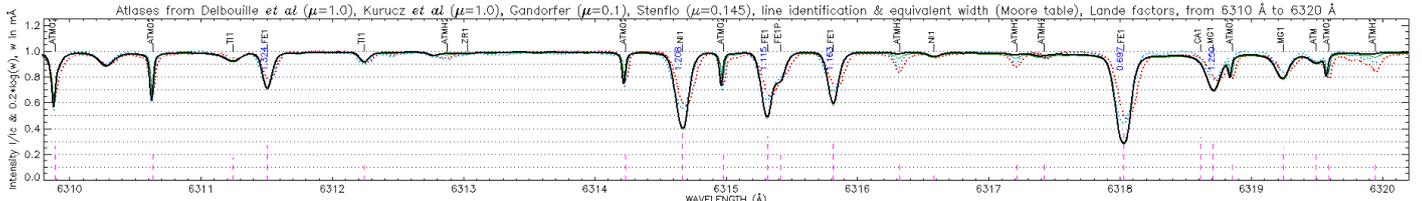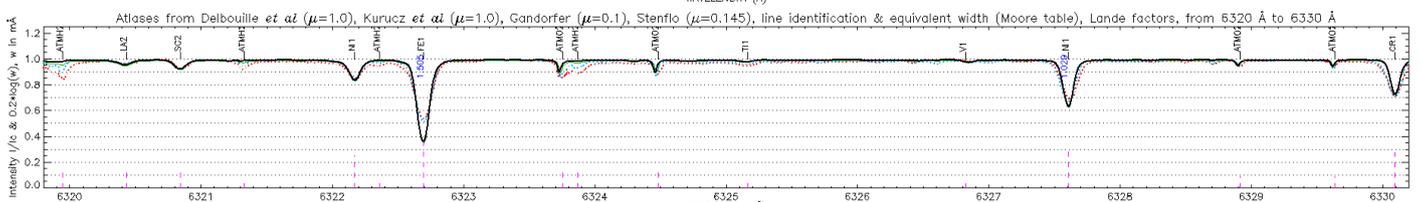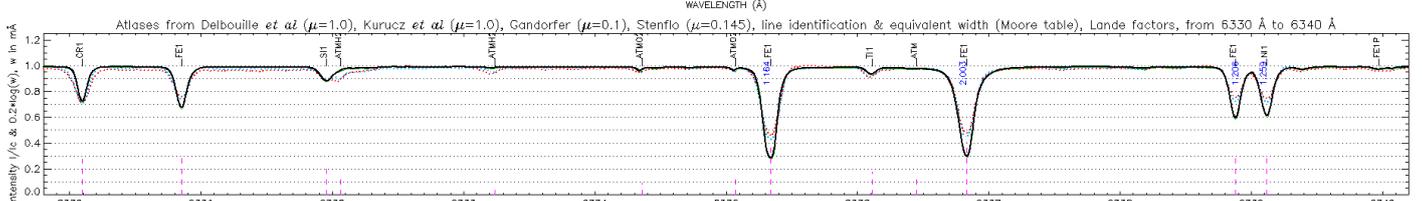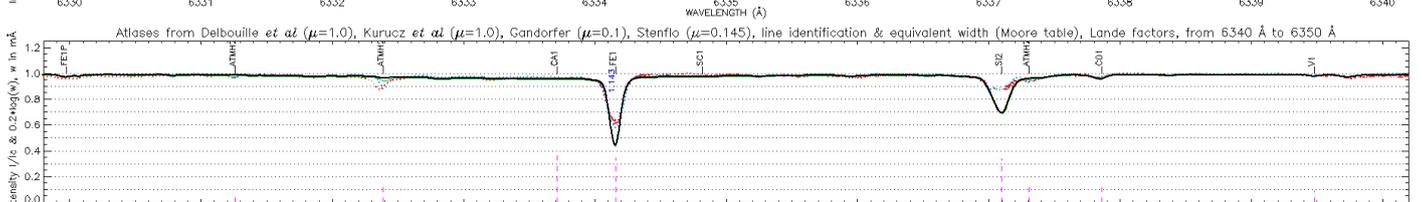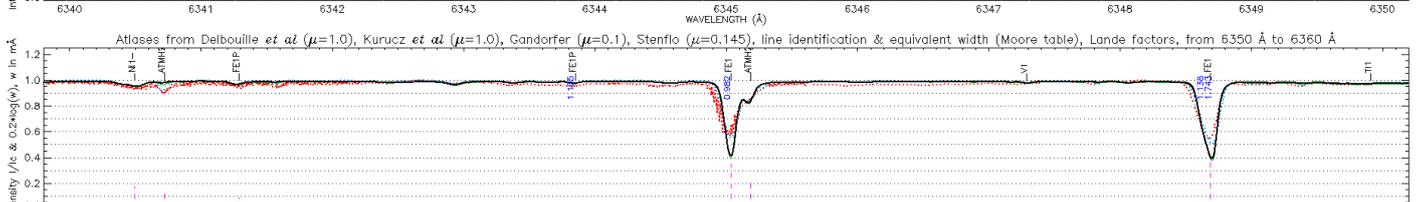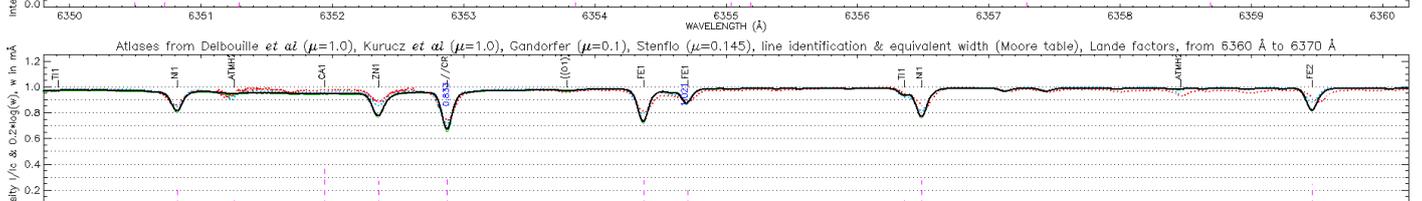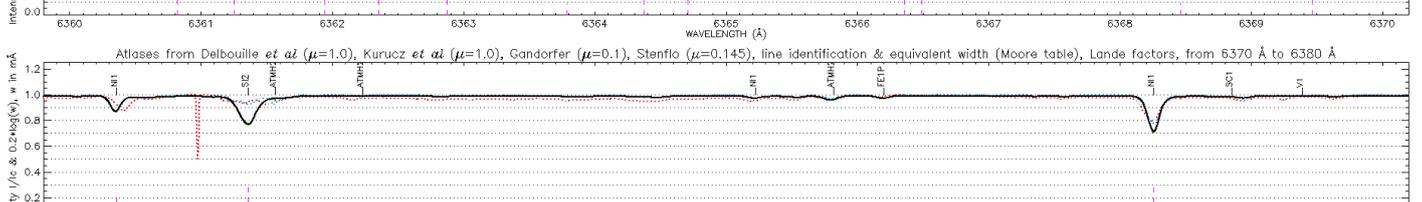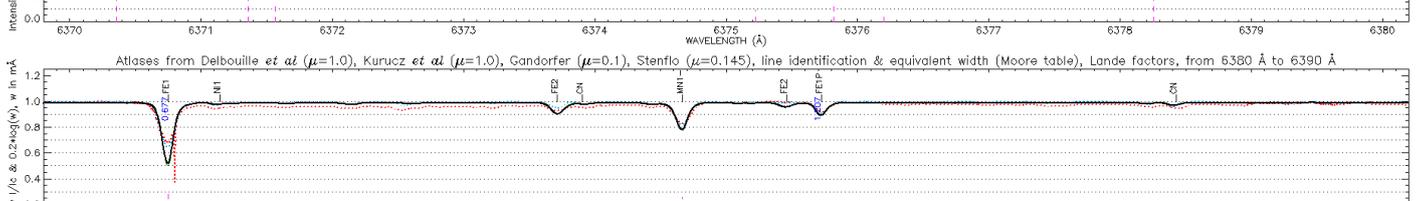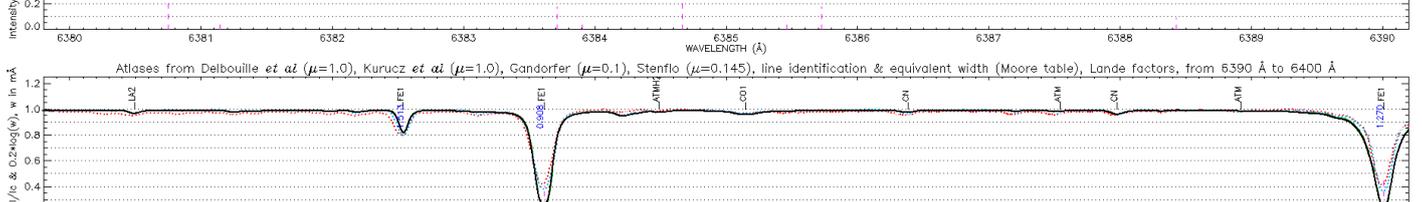

Atlases from Delbouille et al ($\mu=1.0$), Kurucz et al ($\mu=1.0$), Gandorfer ($\mu=0.1$), Stenflo ($\mu=0.145$), line identification & equivalent width (Moore table), Lande factors, from 6400 Å to 6500 Å

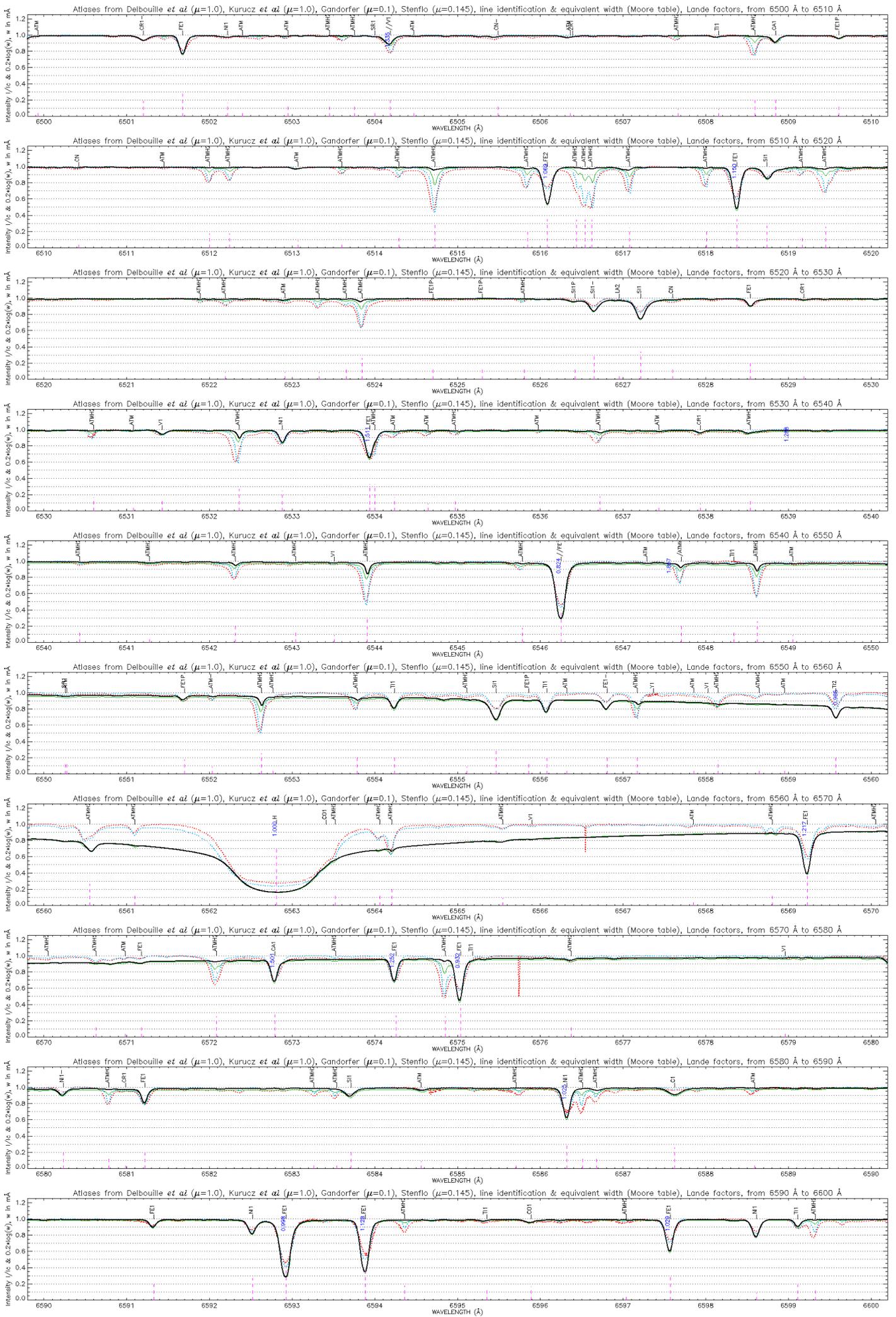

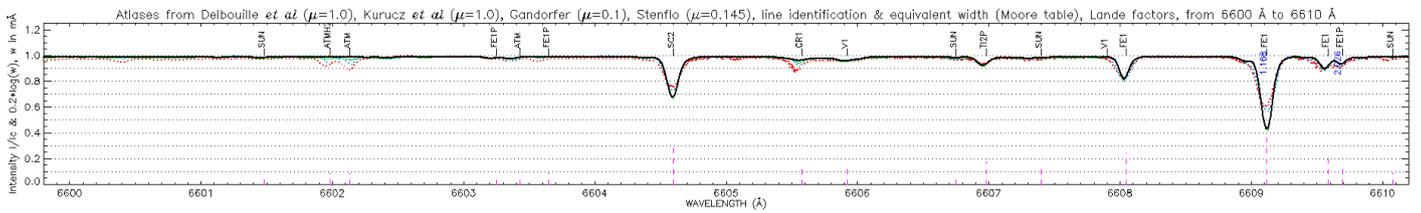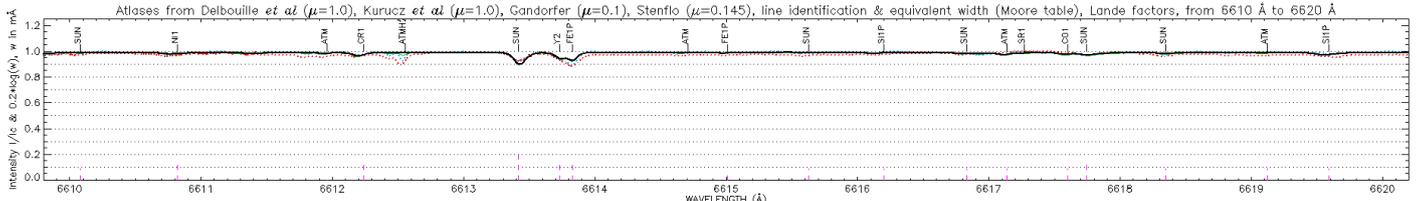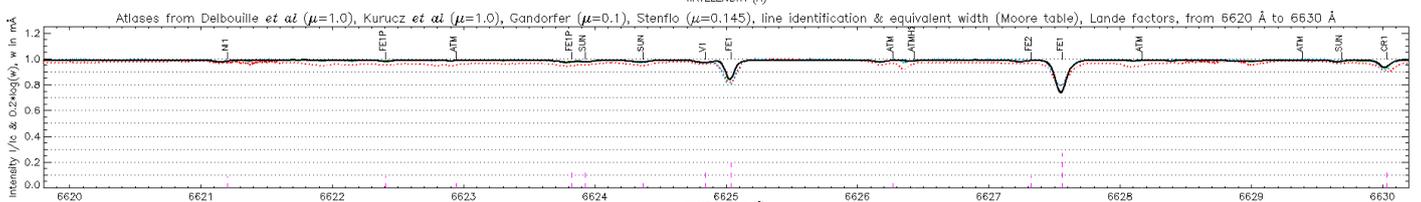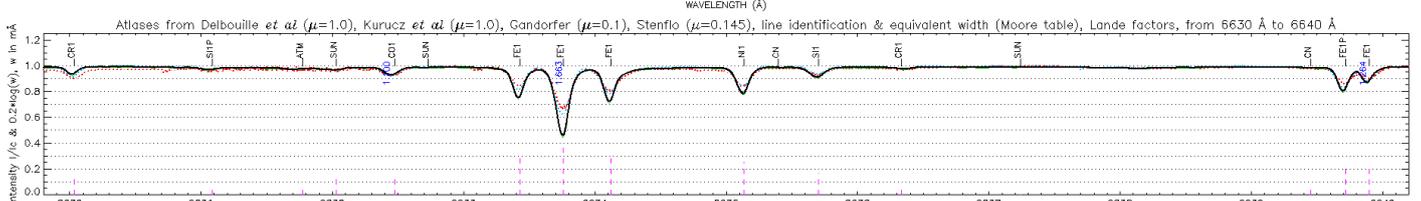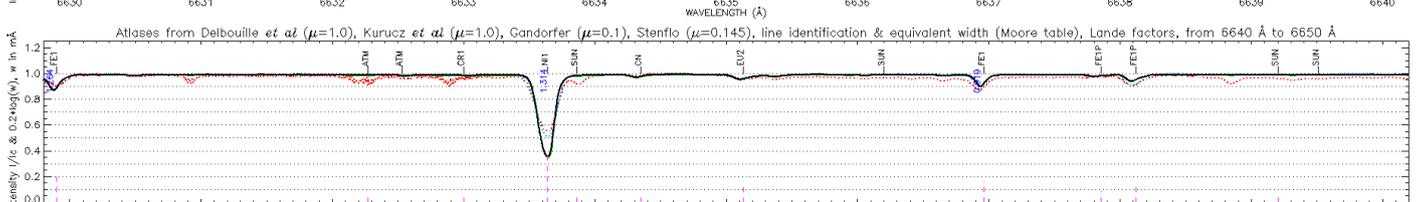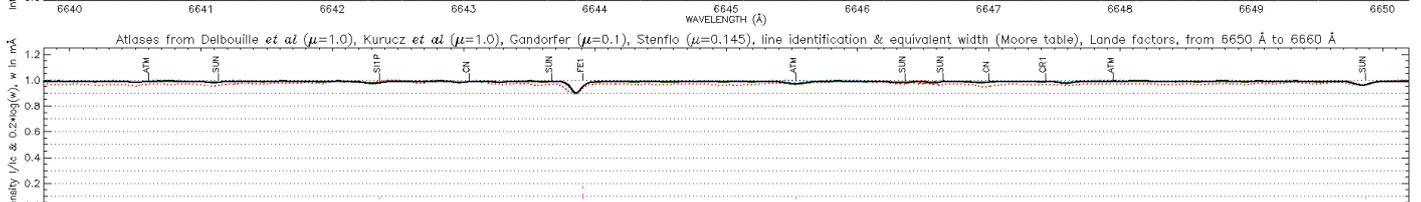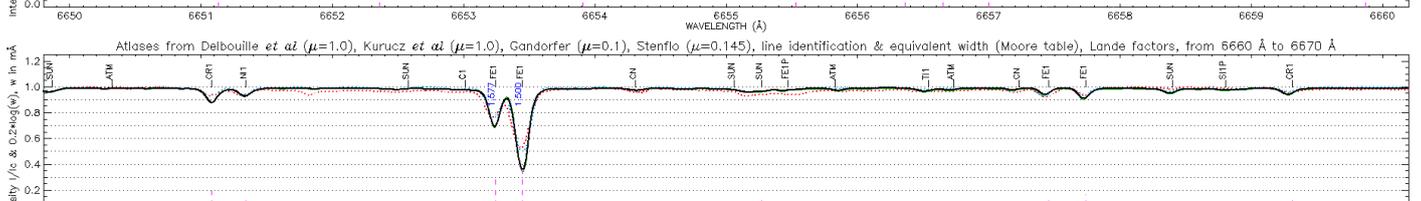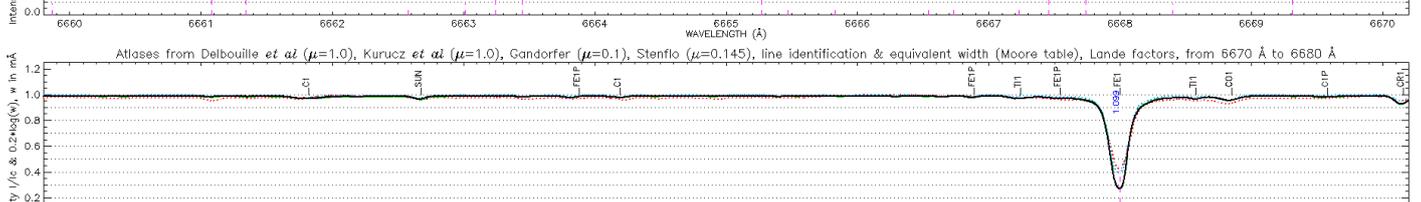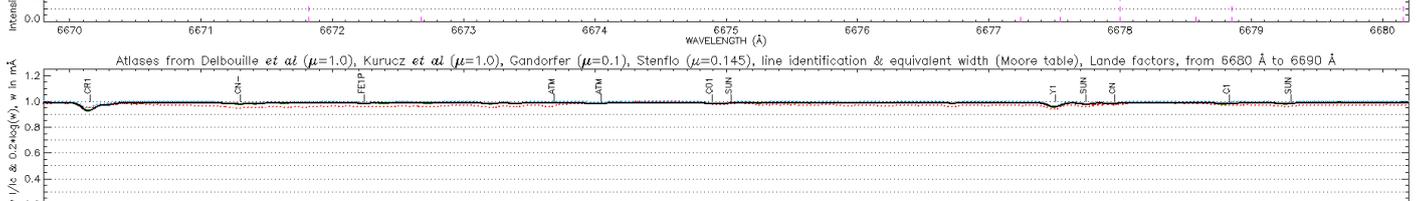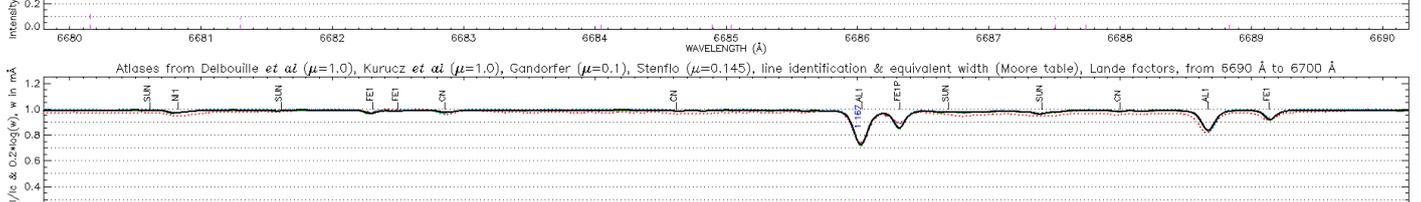

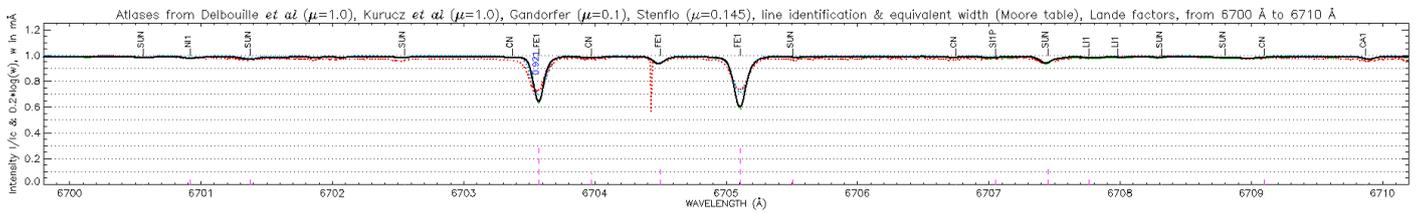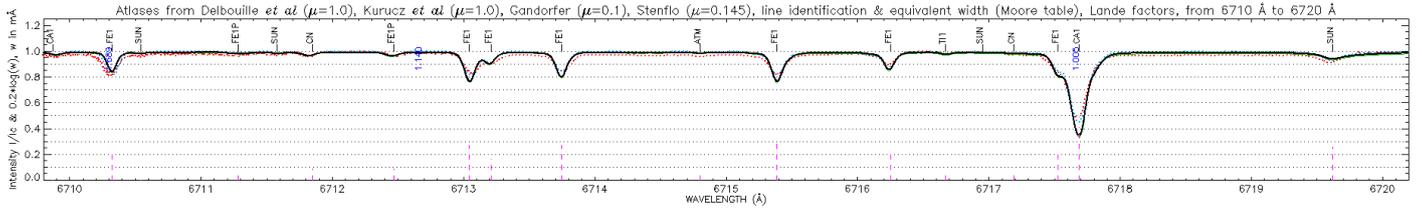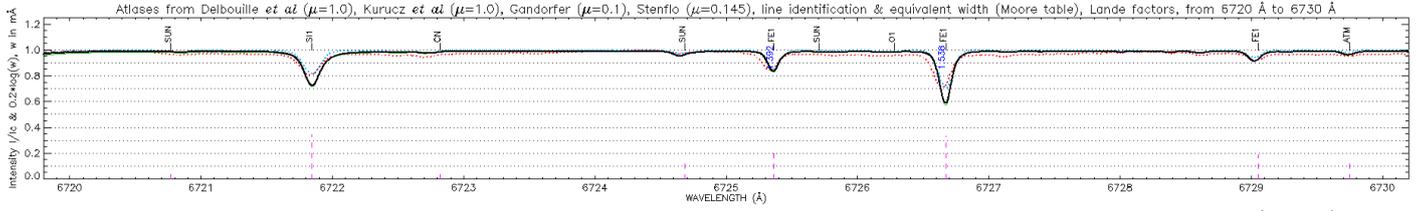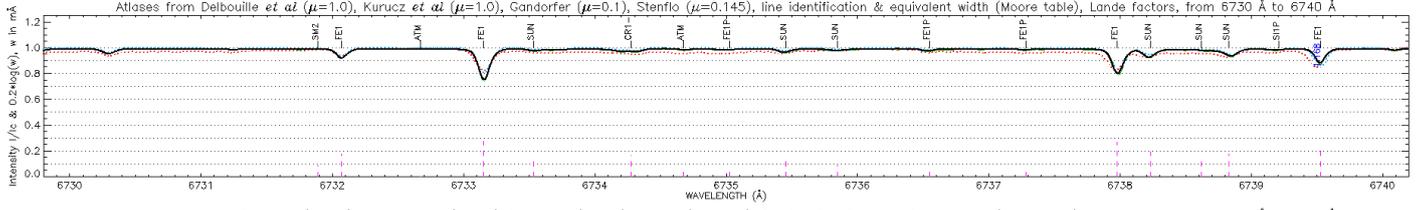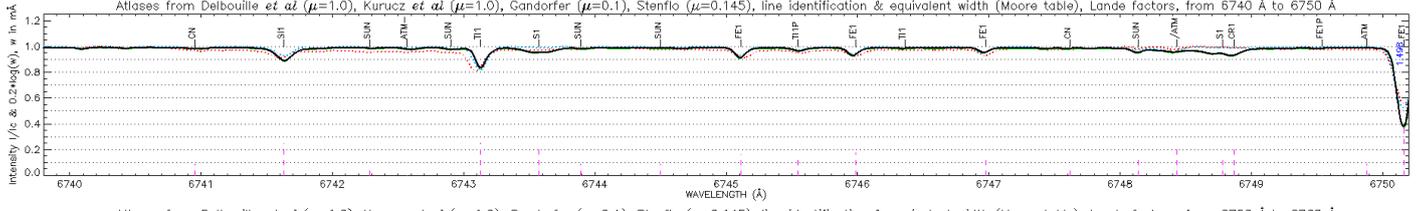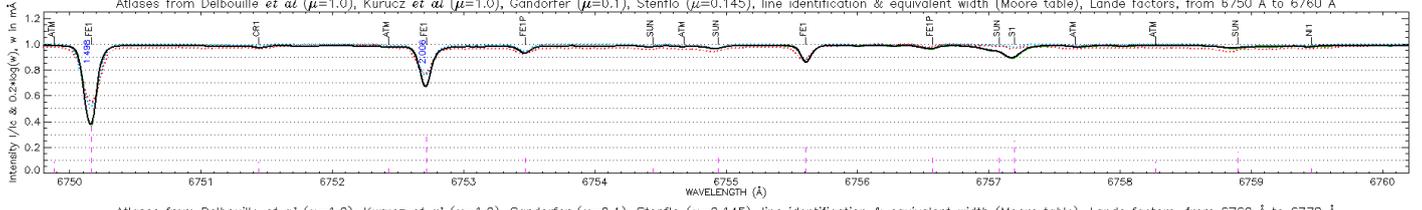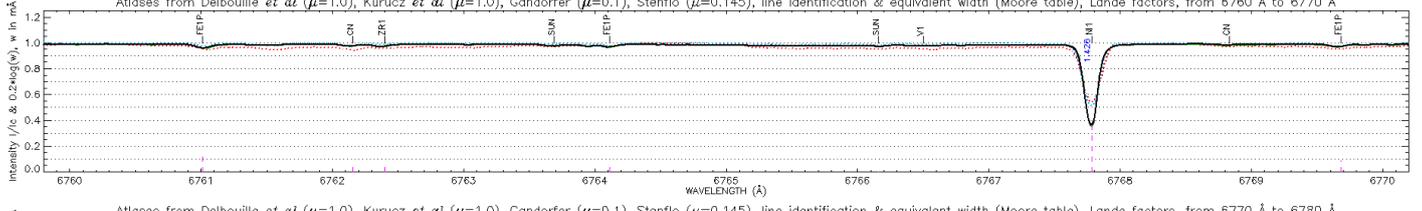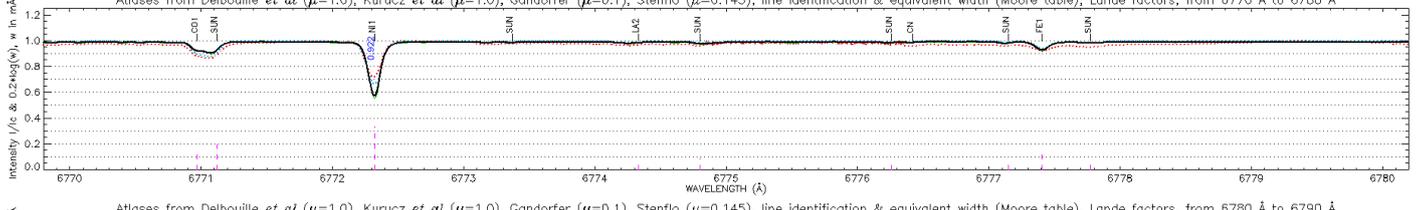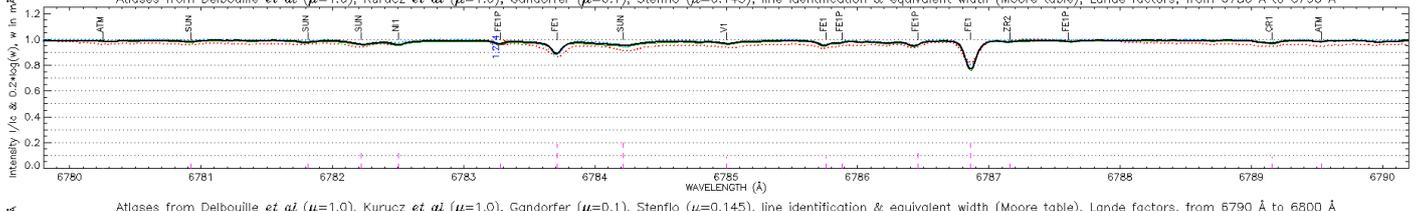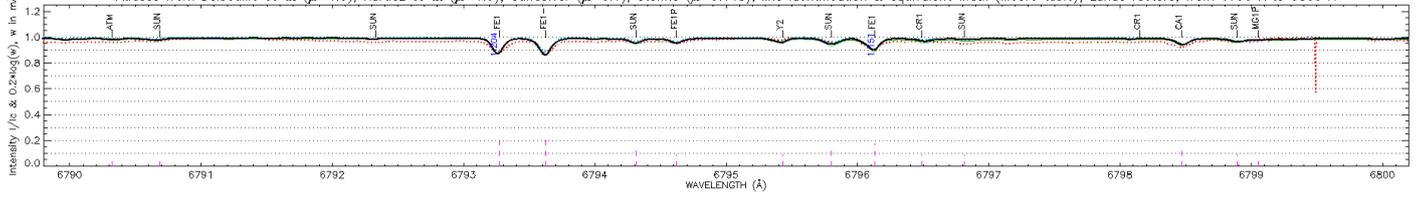

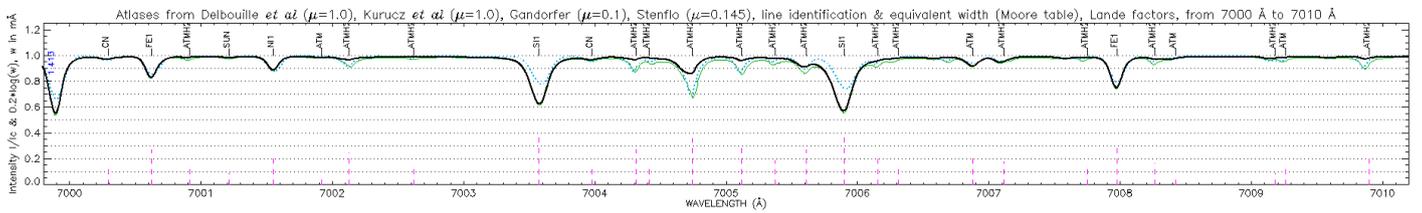
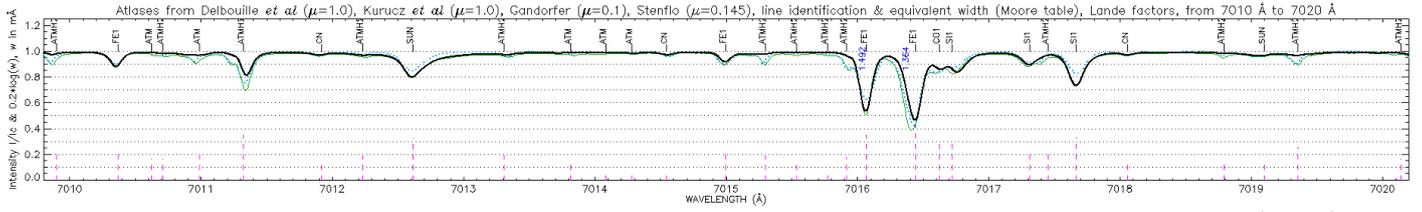
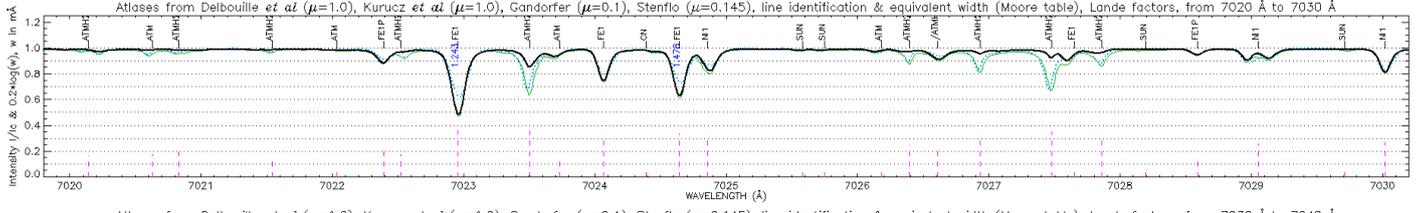
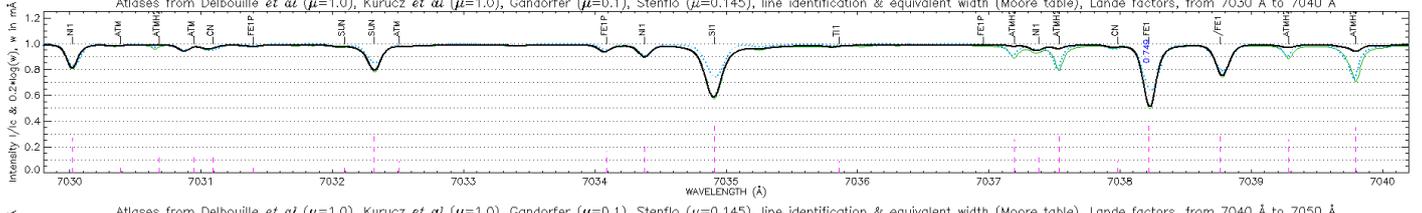
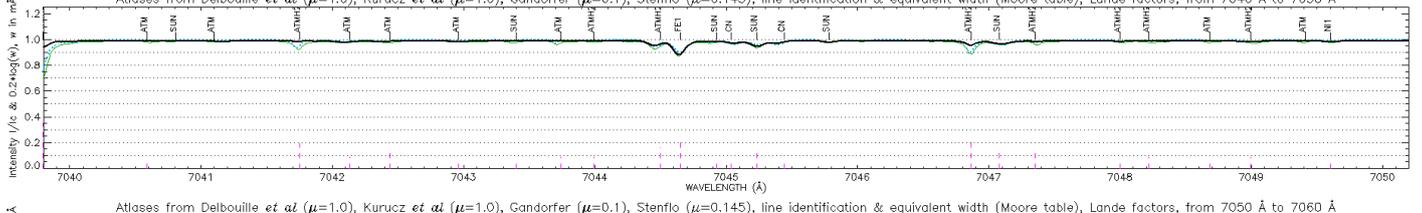
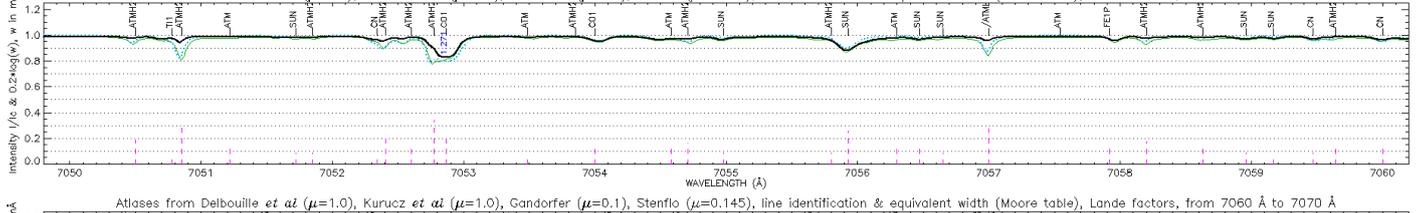
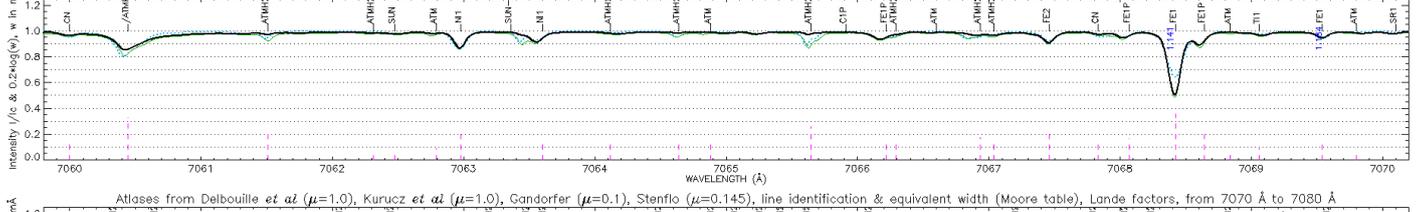
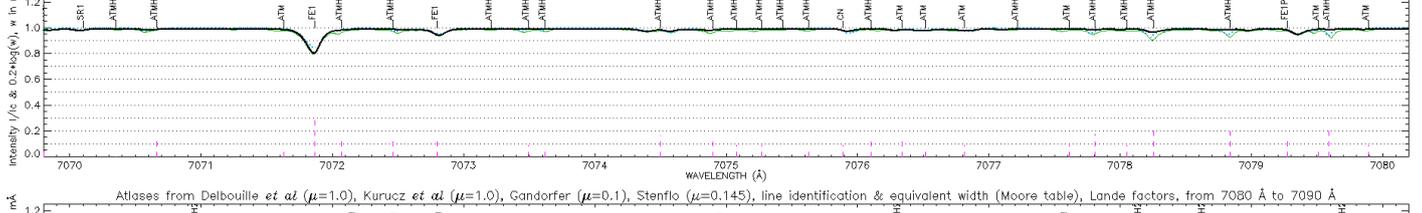
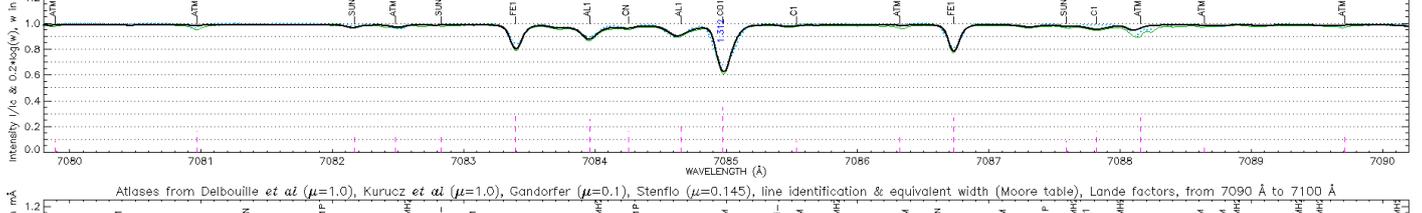
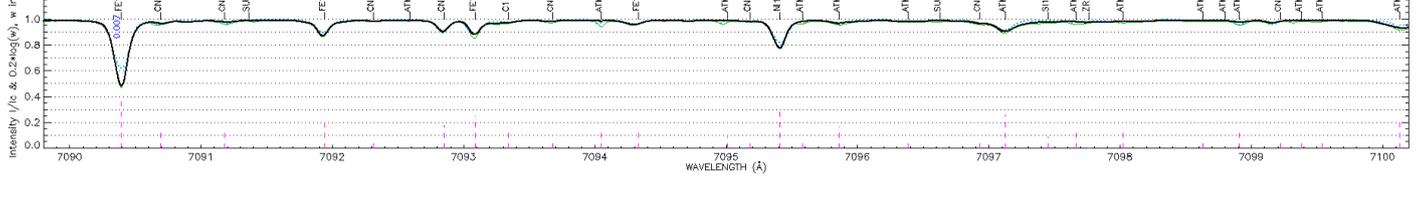

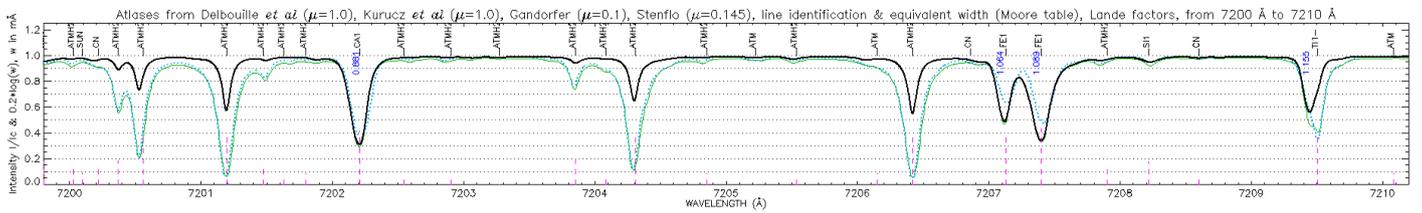
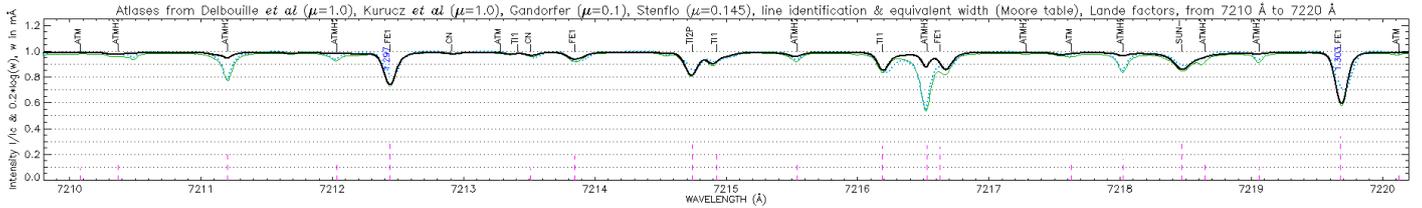
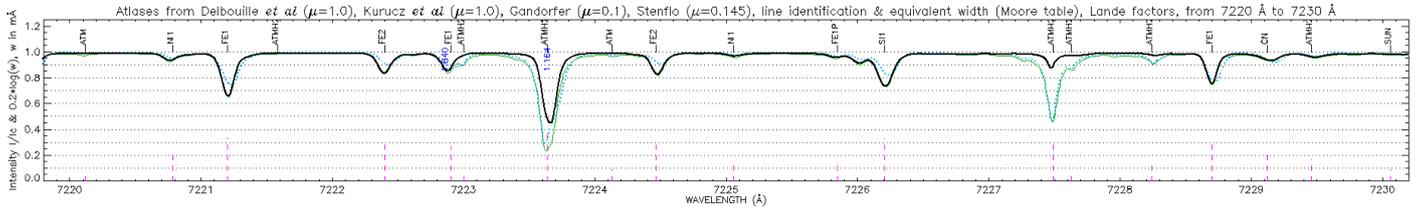
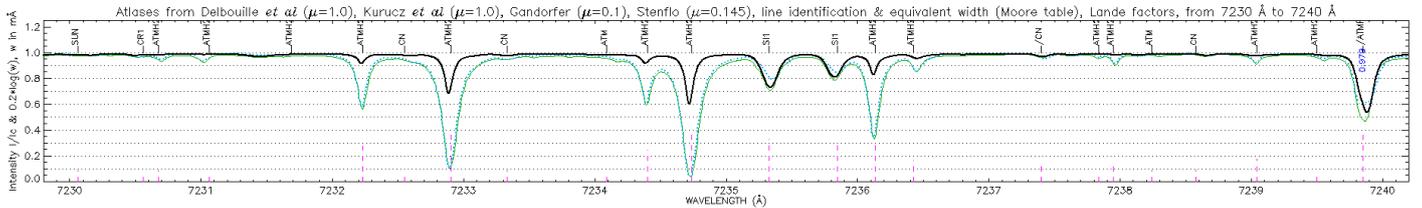
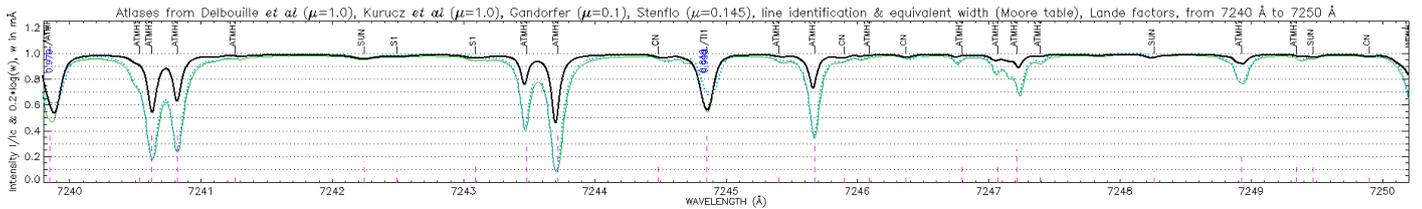
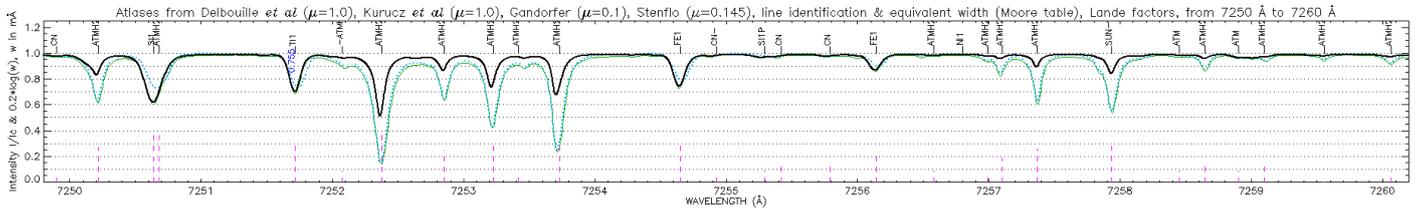
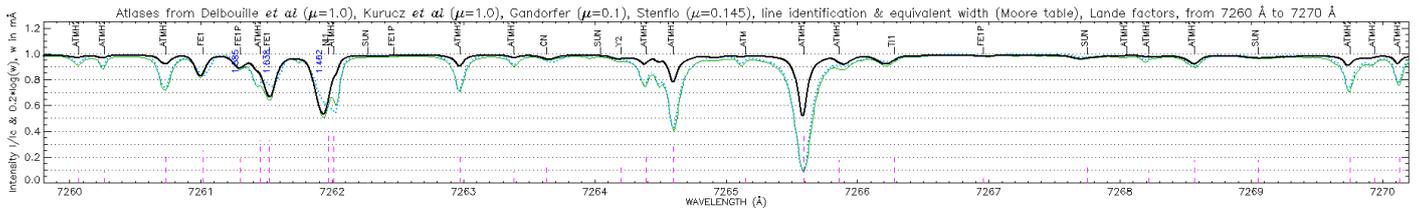
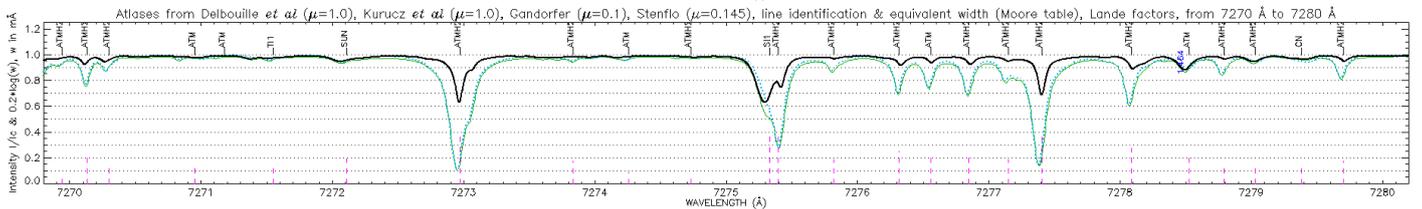
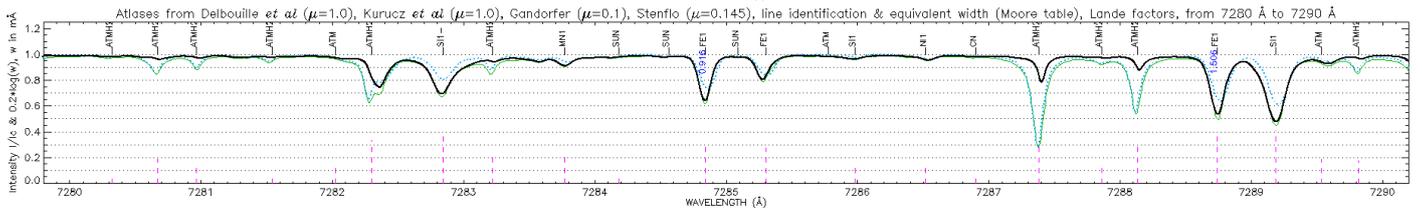
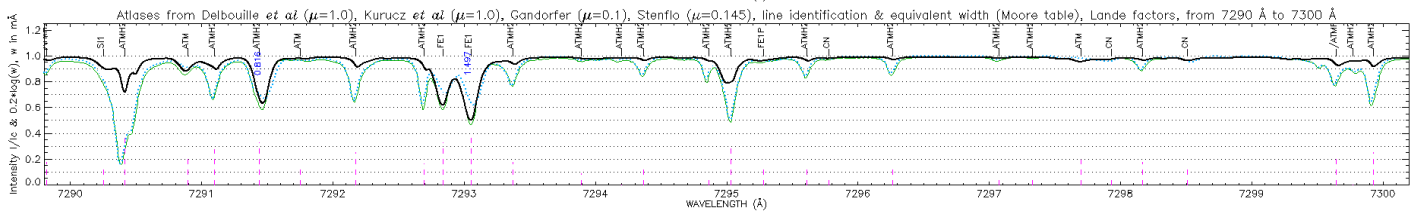

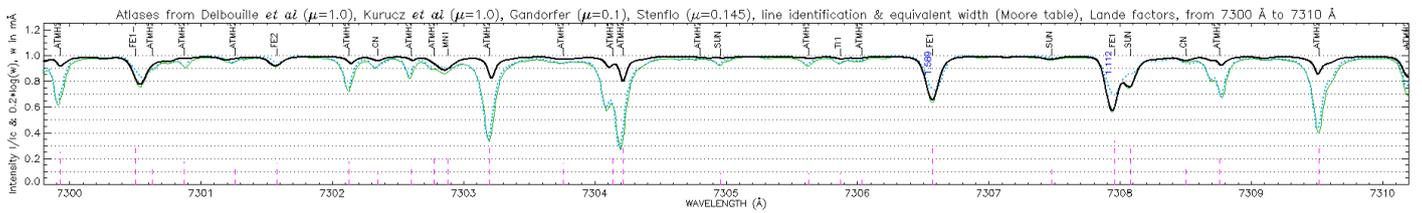
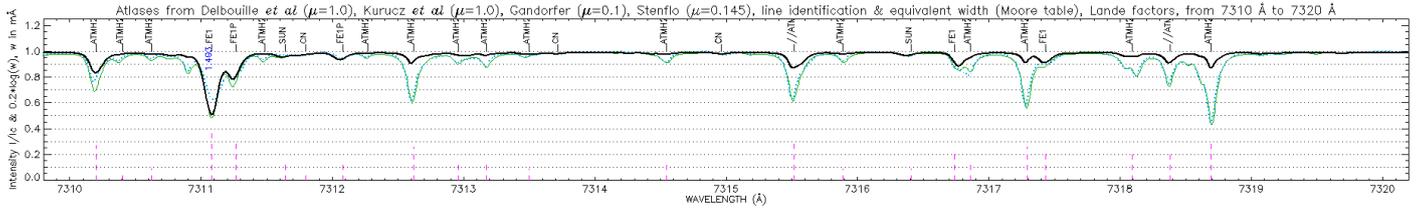
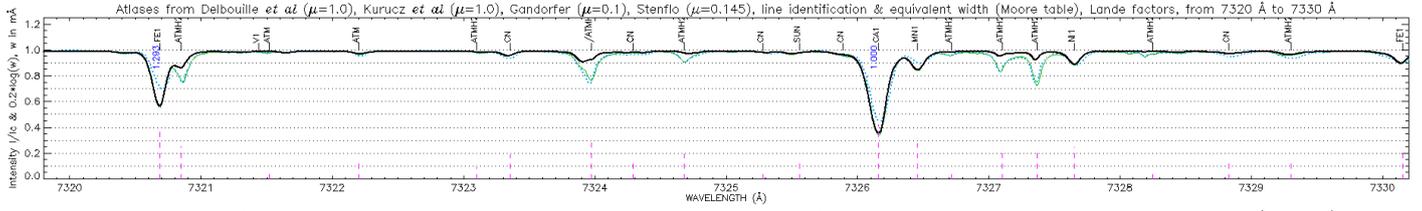
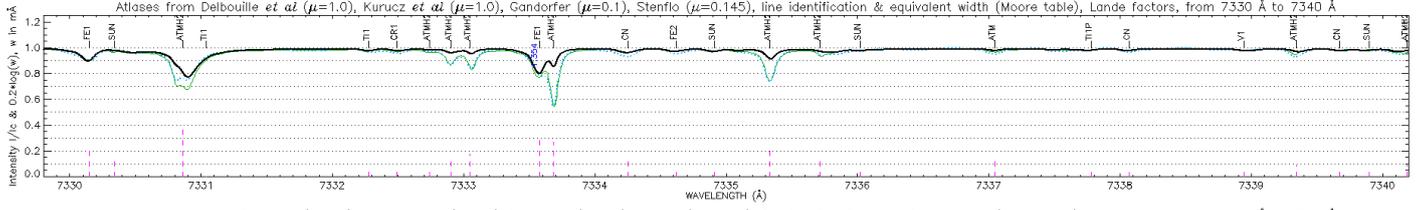
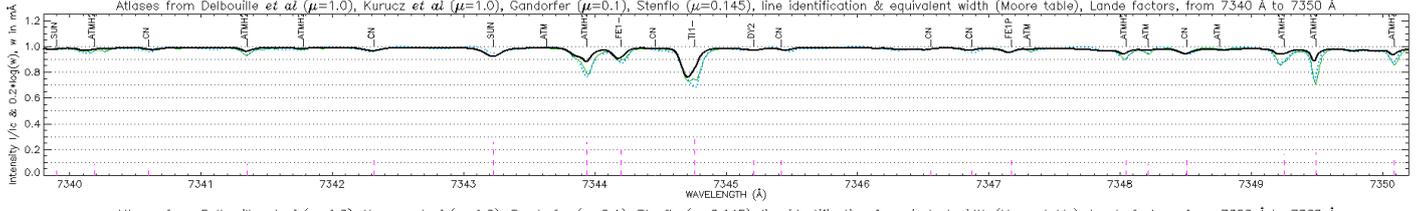
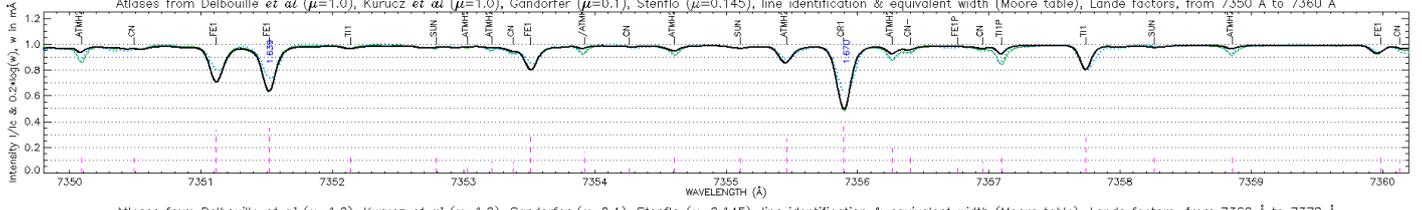
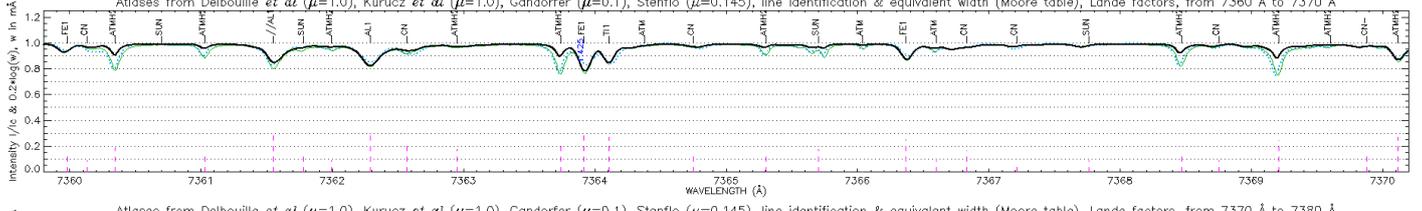
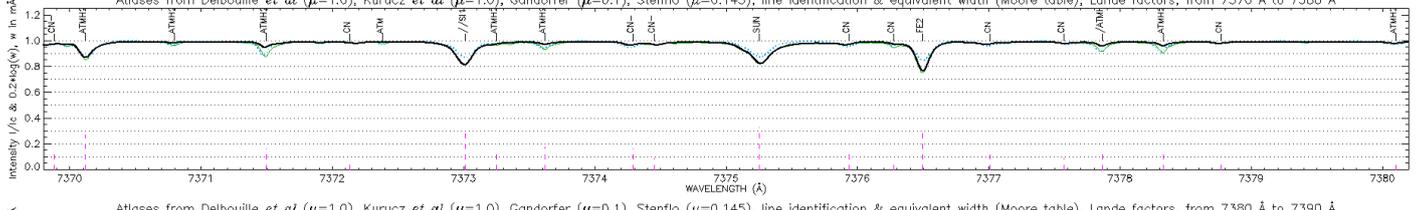
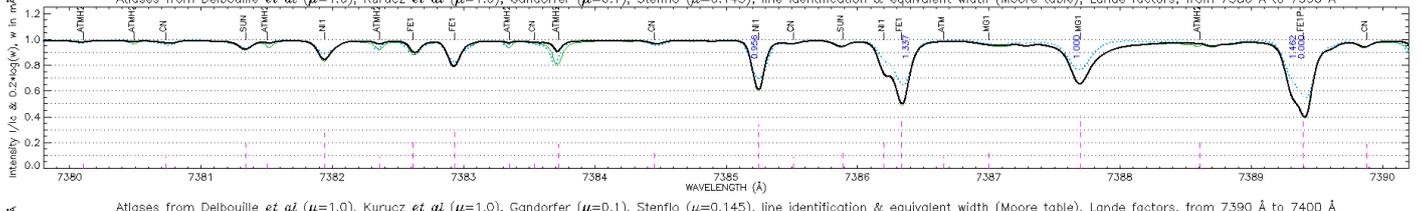
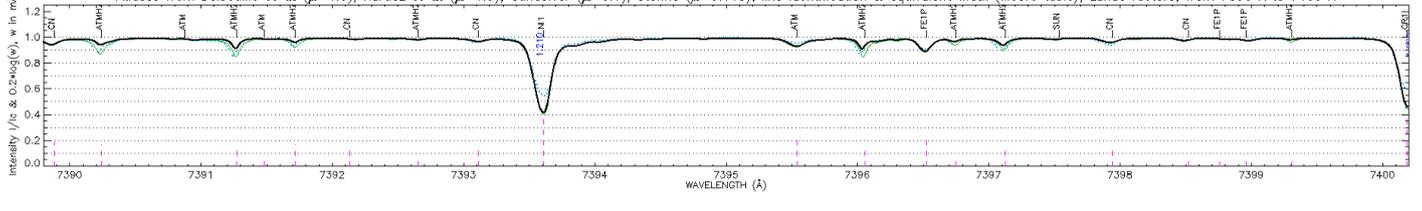

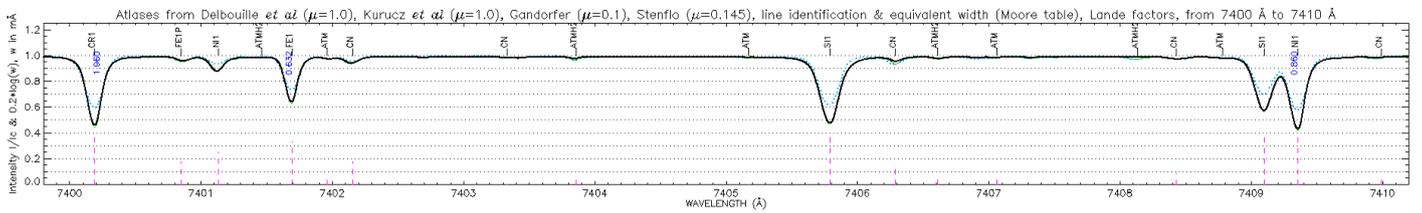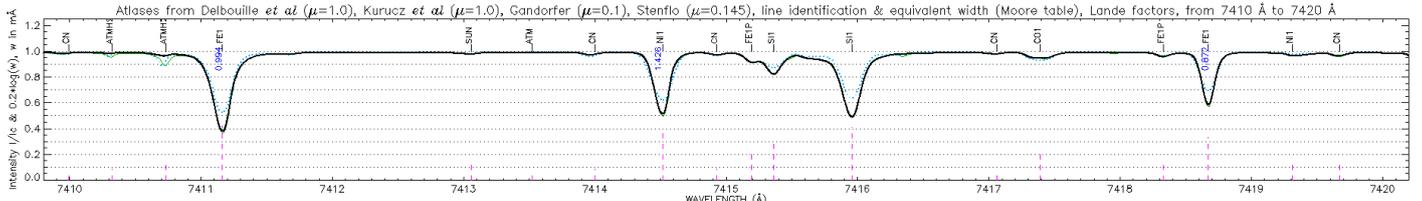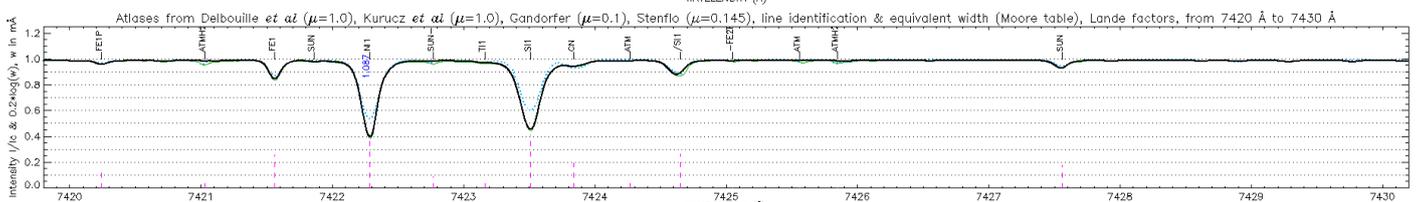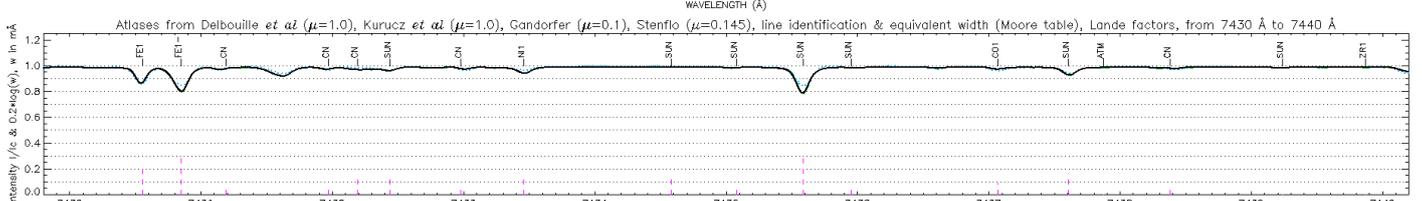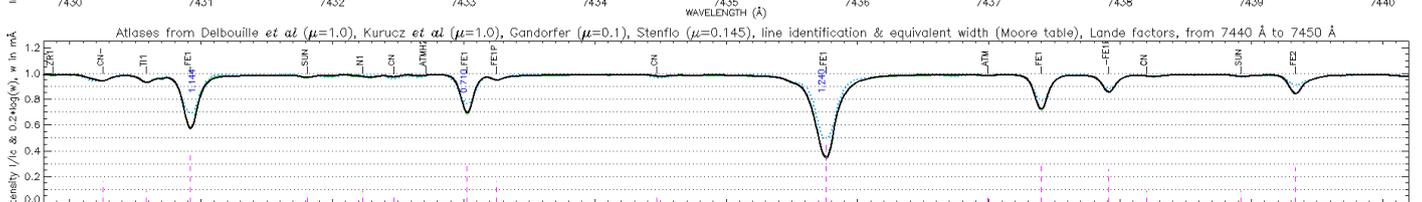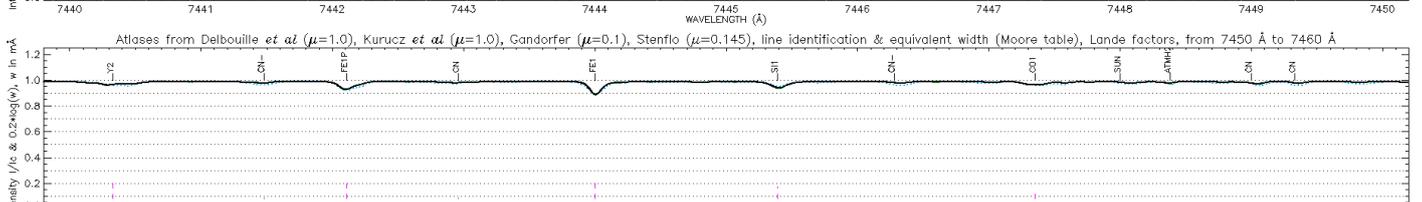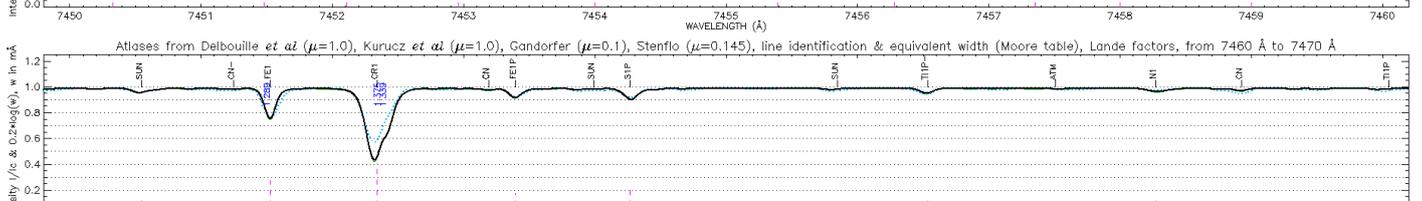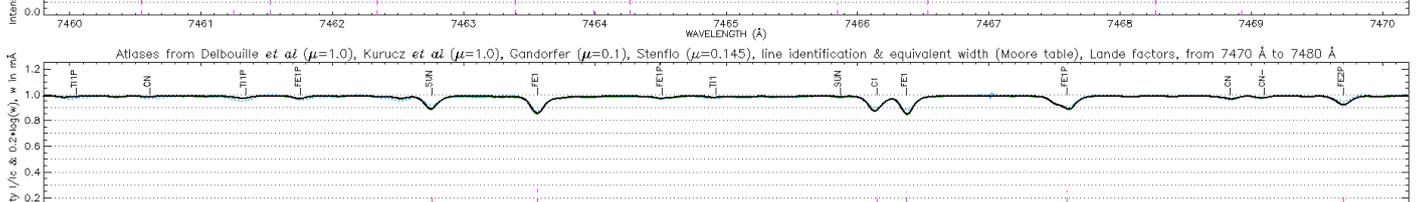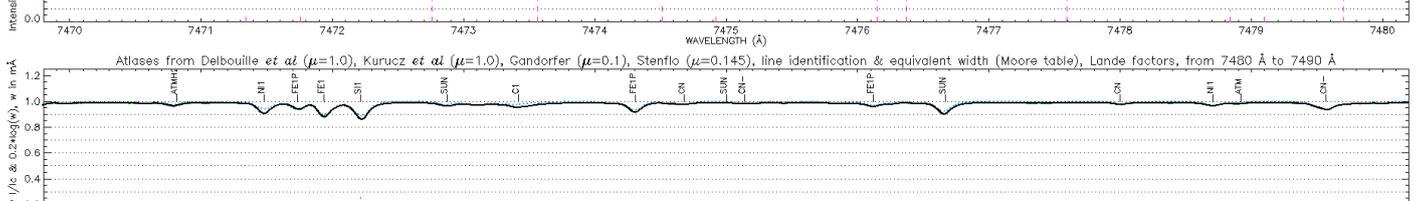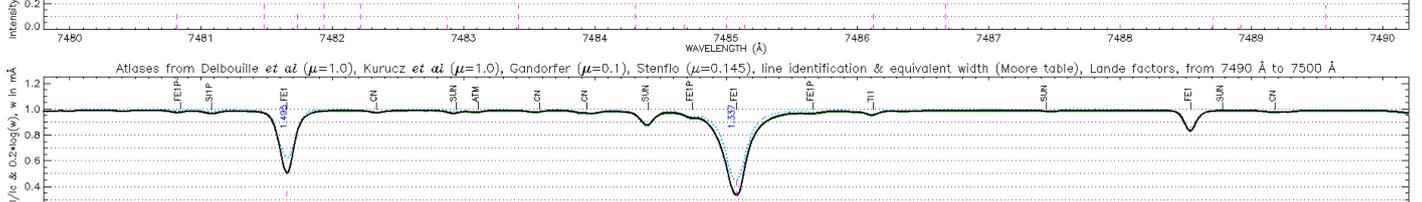

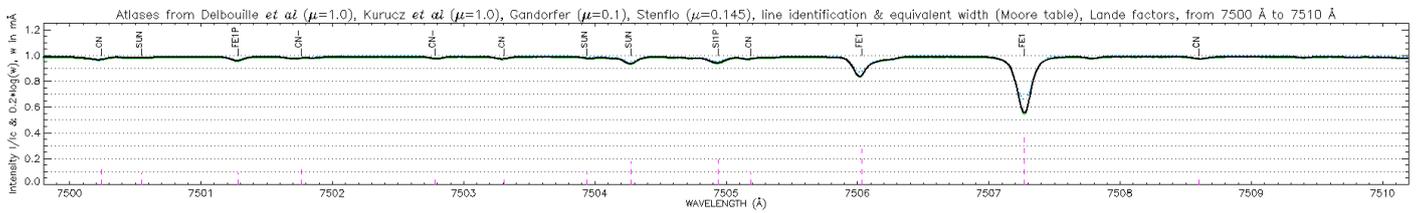
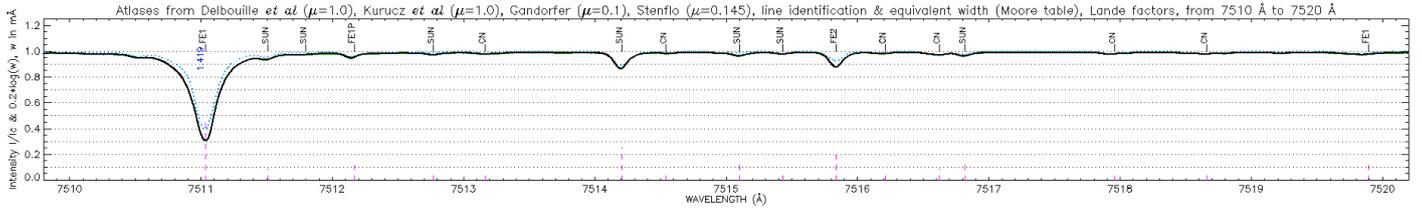
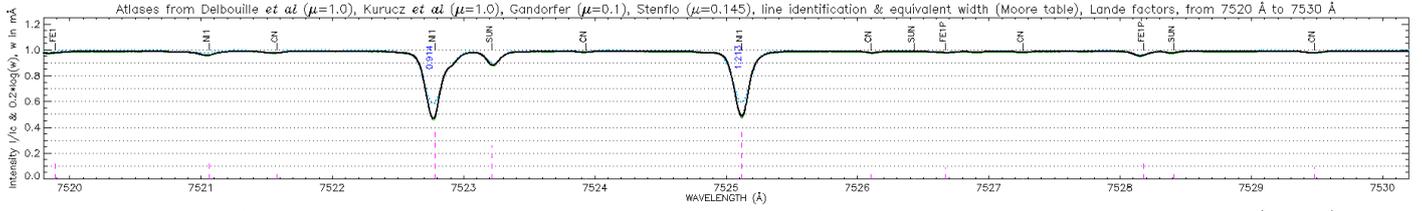
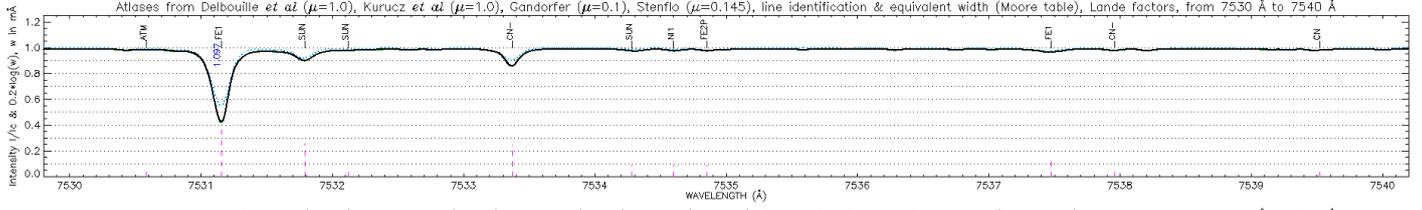
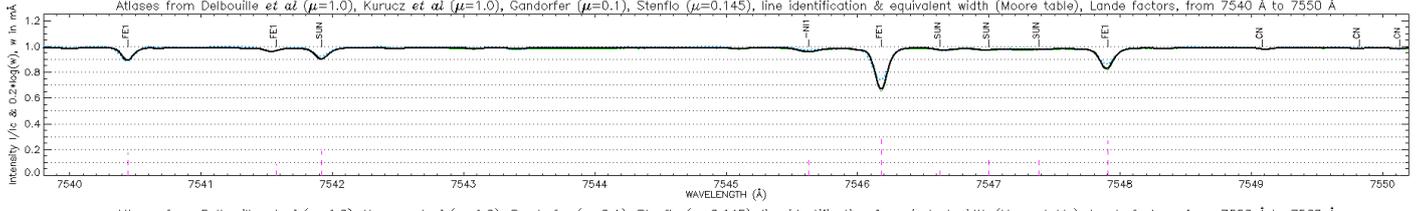
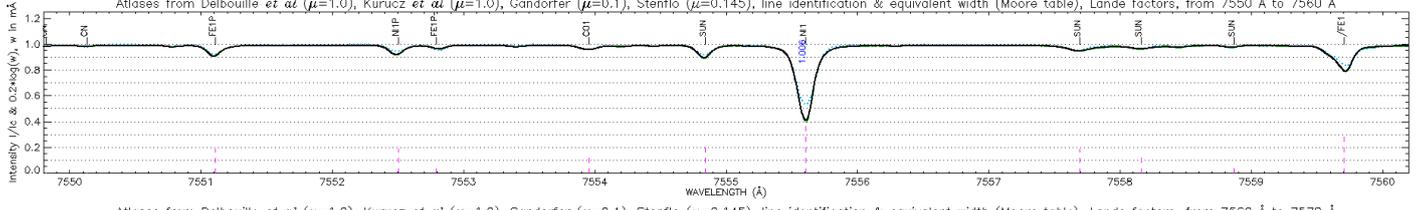
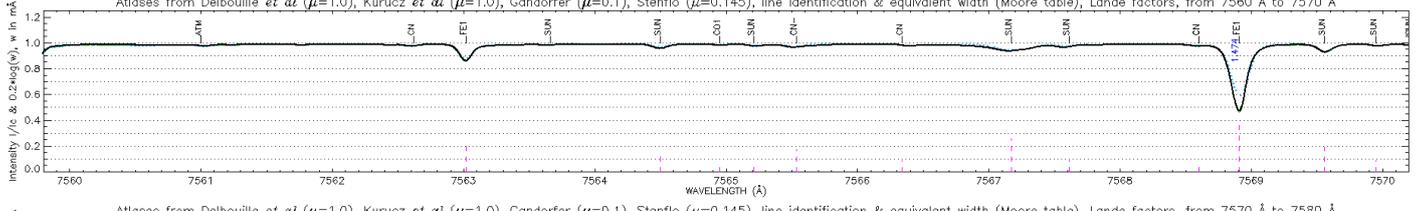
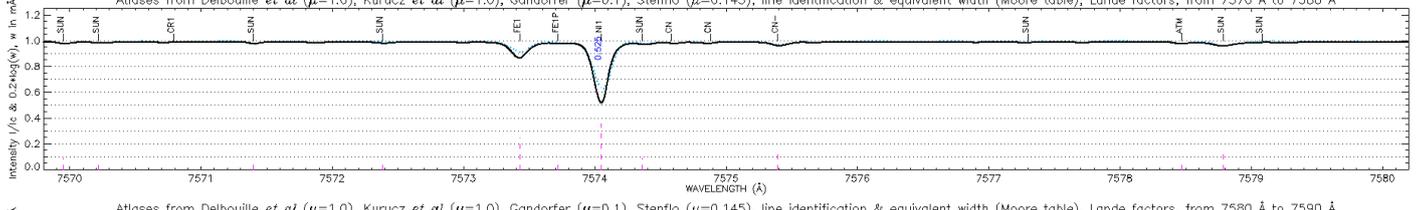
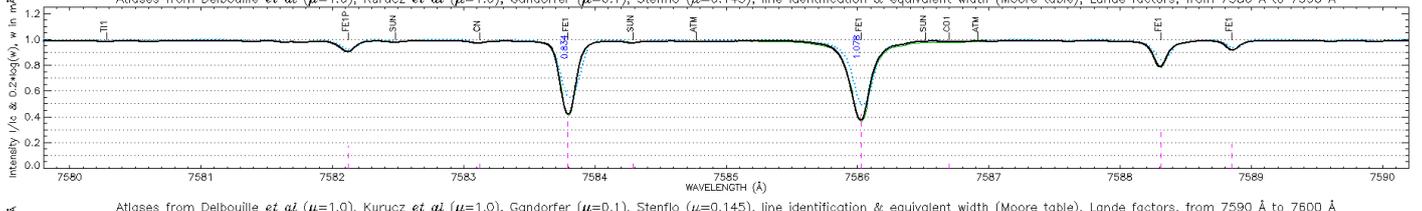
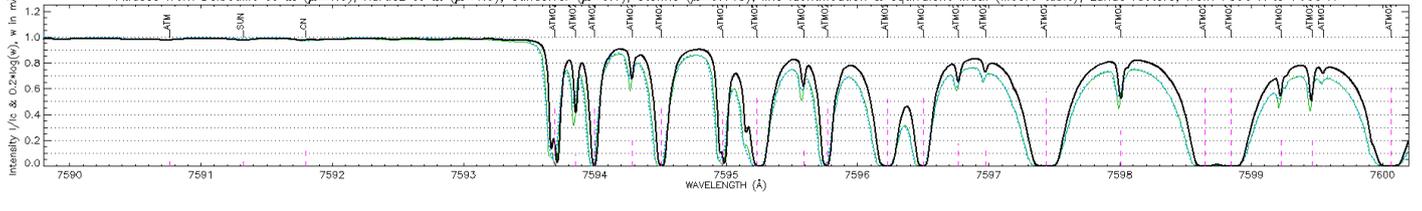

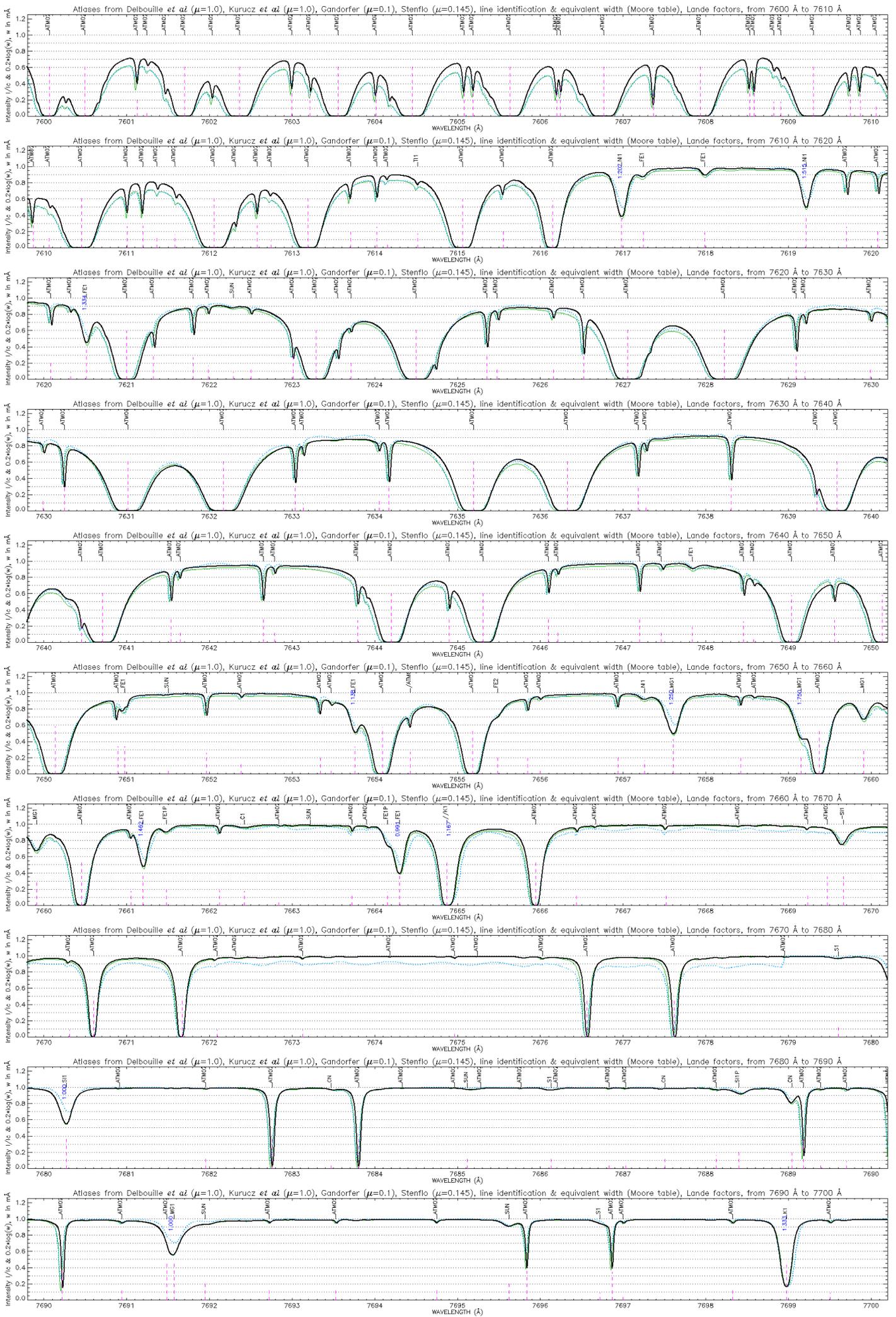

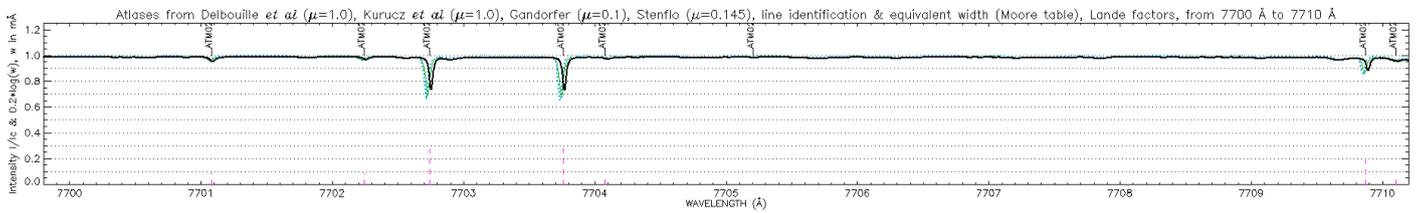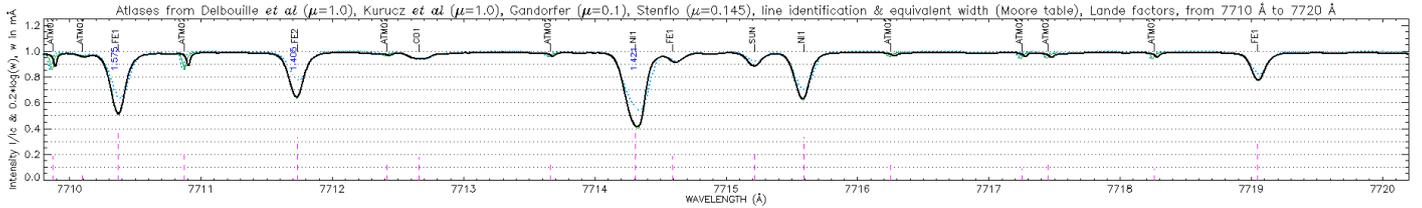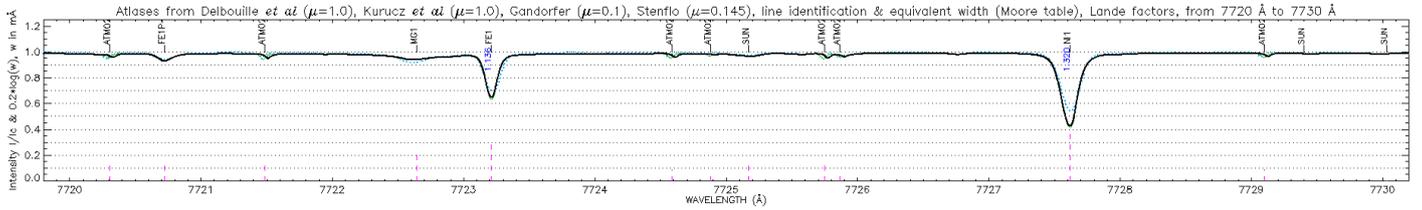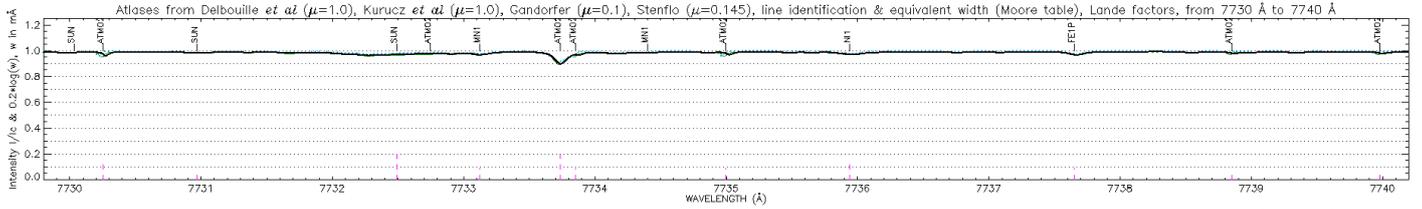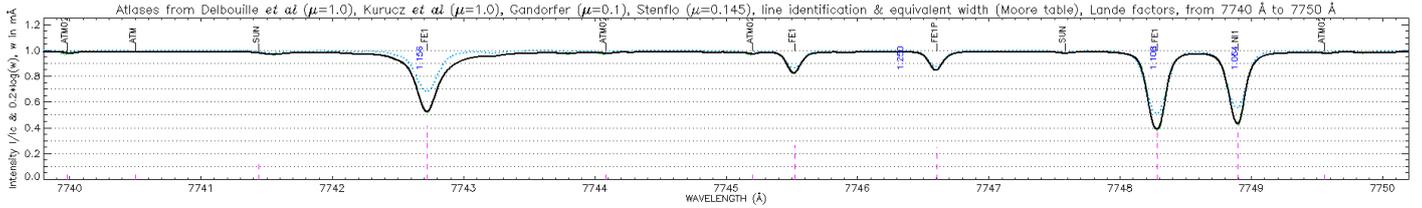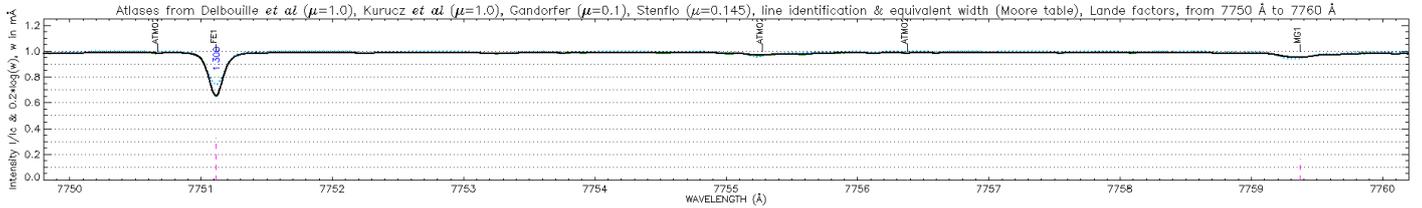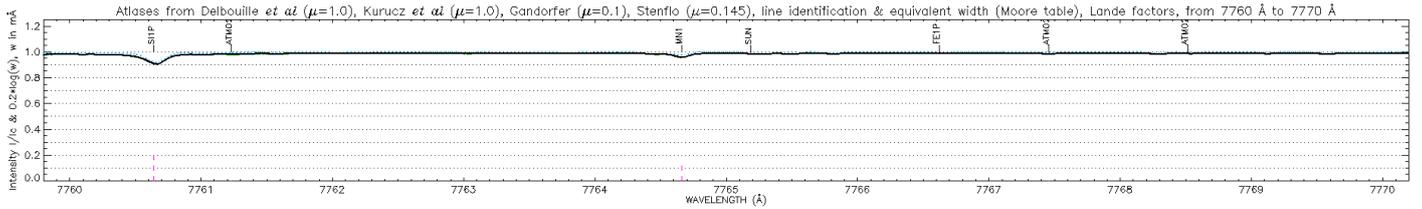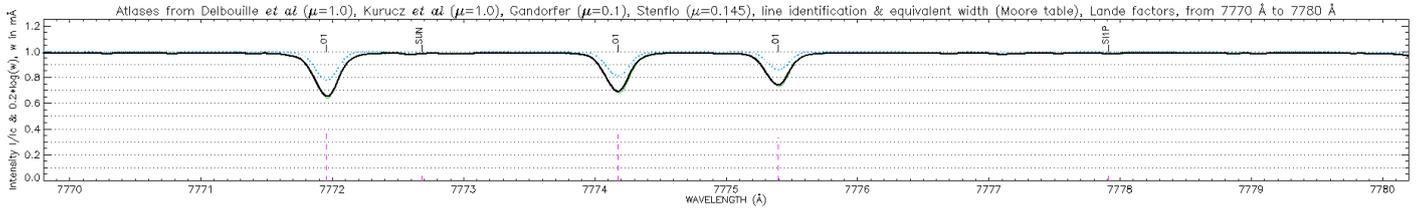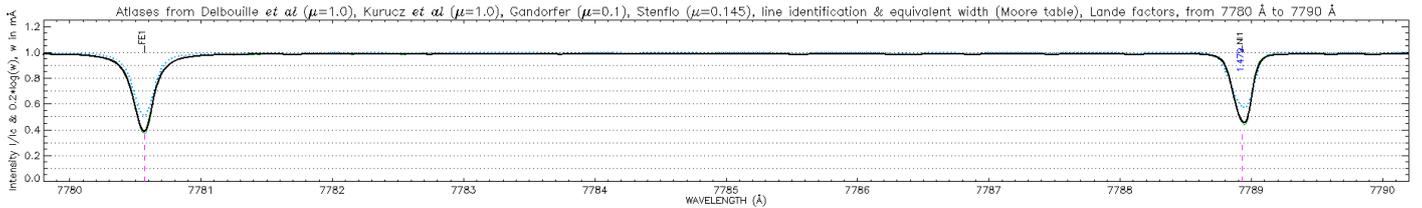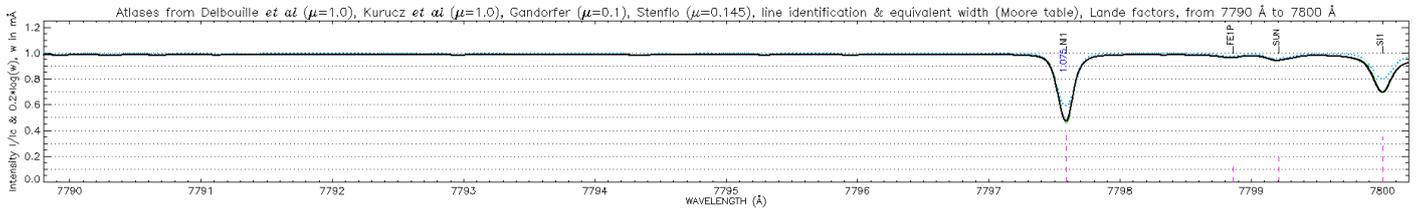

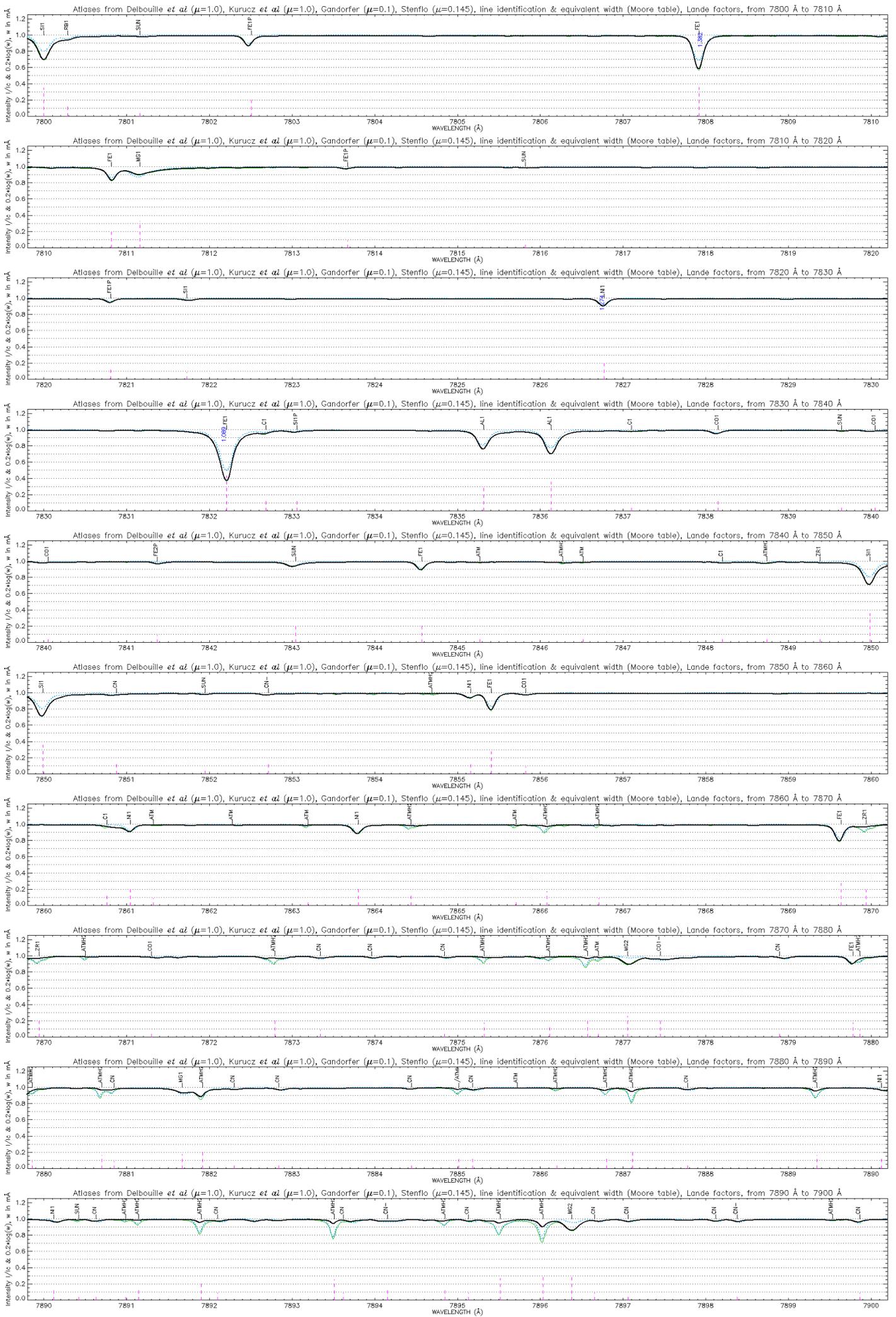

Atlases from Delbouille et al (μ=1.0), Kurucz et al (μ=1.0), Gandorfer (μ=0.1), Stenflo (μ=0.145), line identification & equivalent width (Moore table), Lande factors, from 8000 Å to 8100 Å.

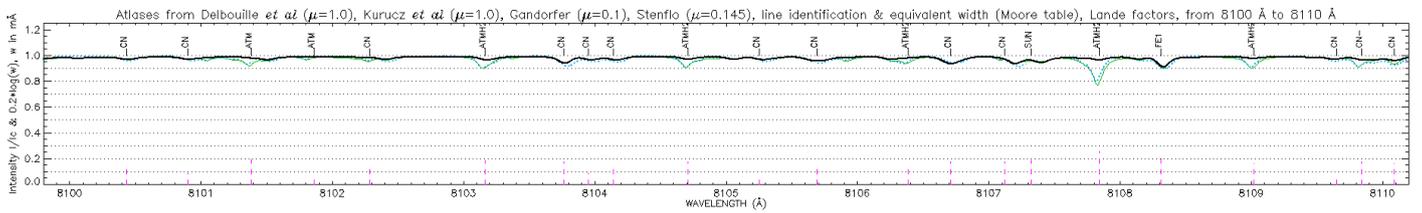
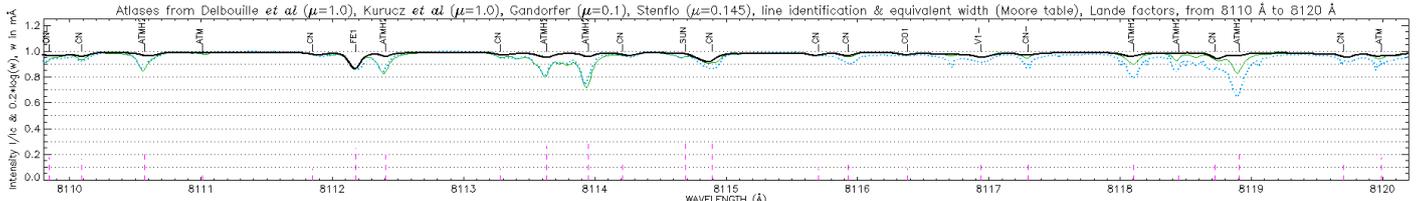
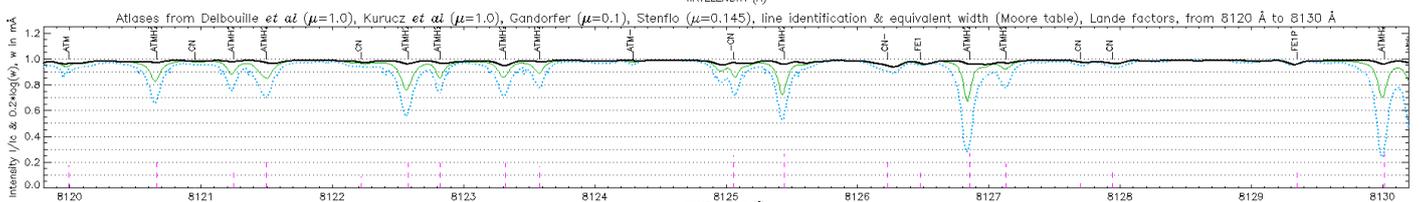
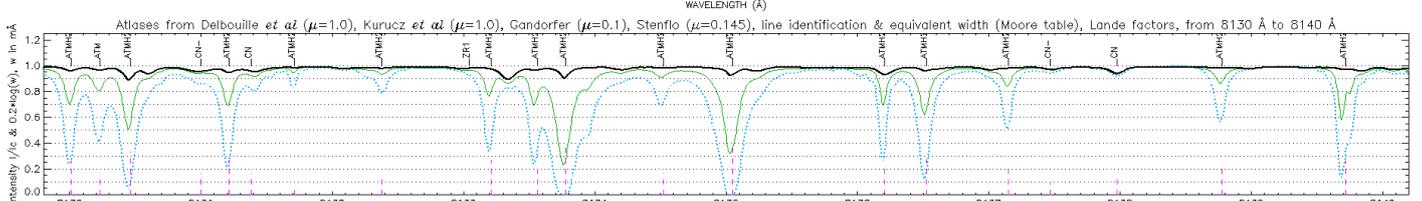
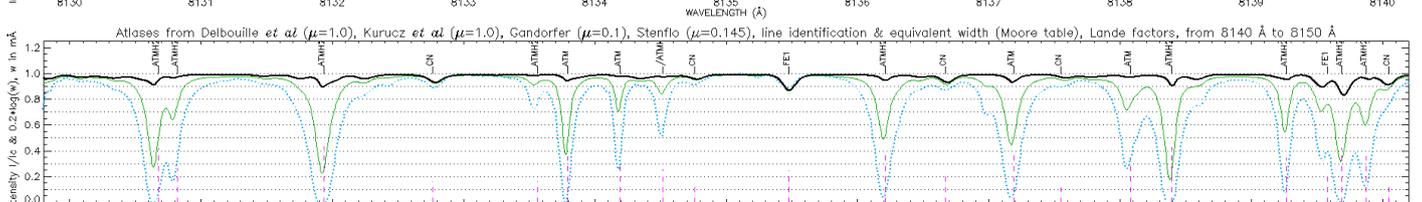
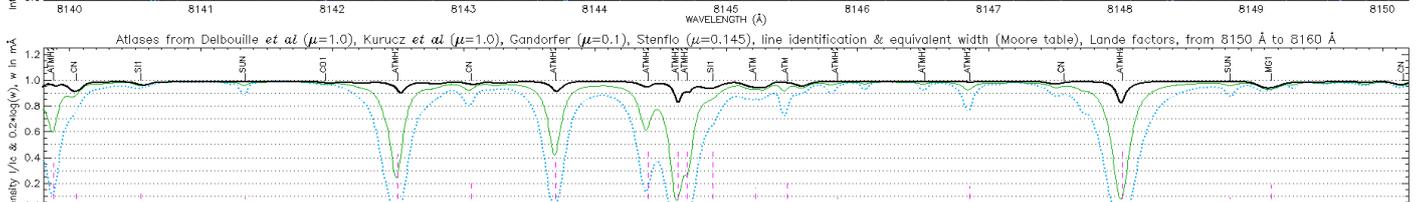
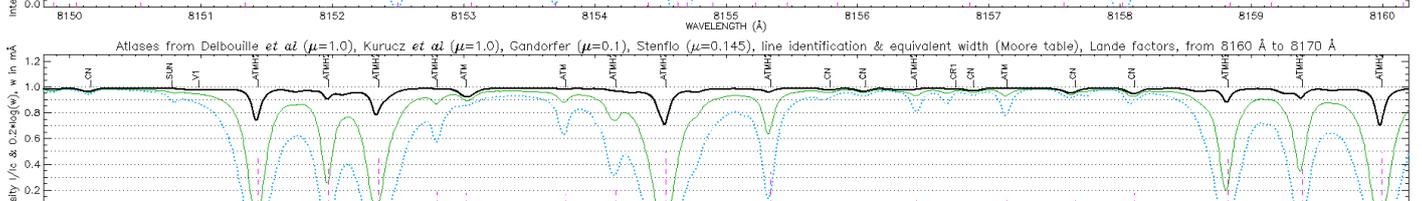
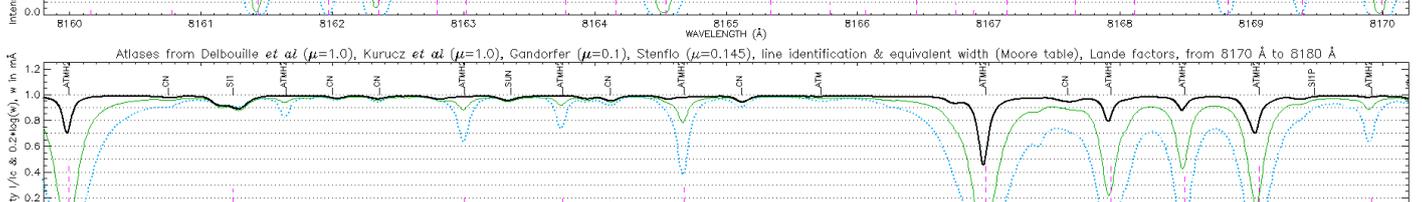
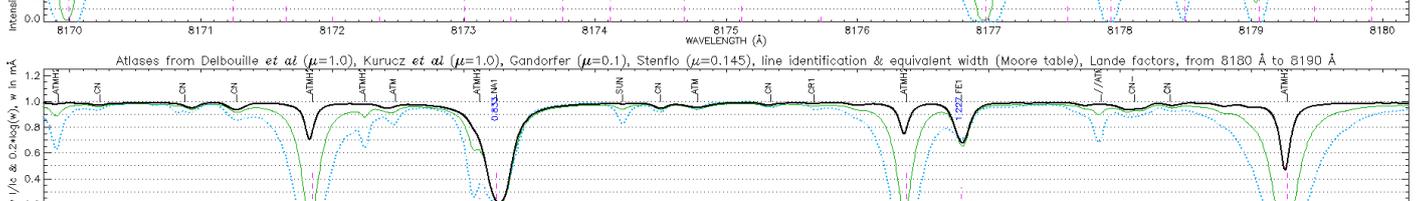
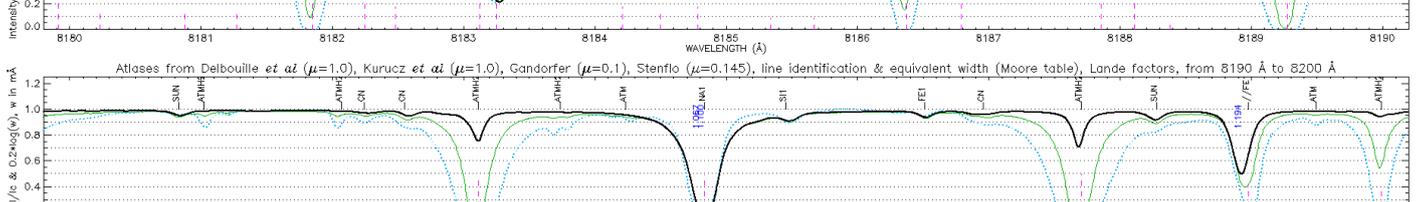

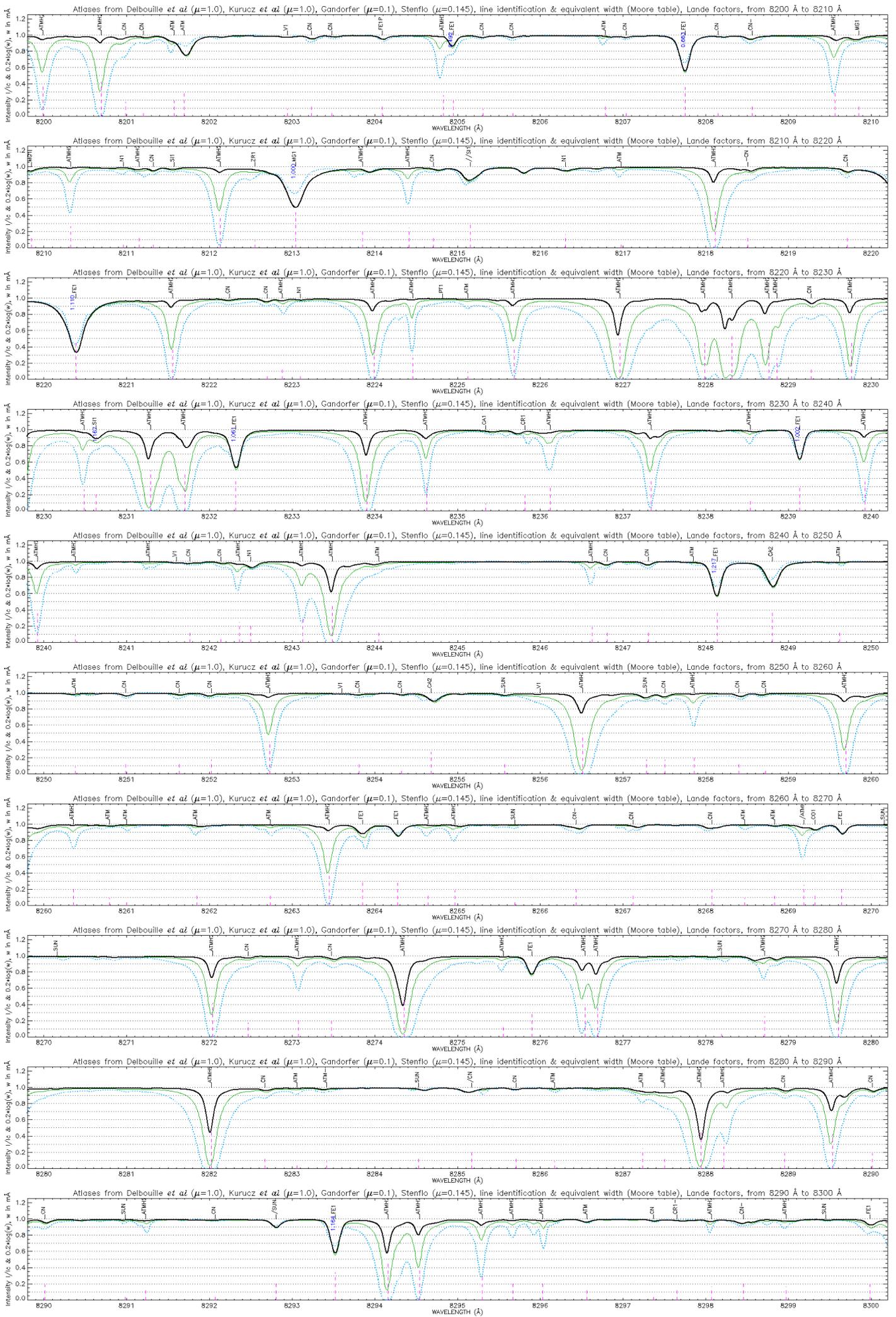

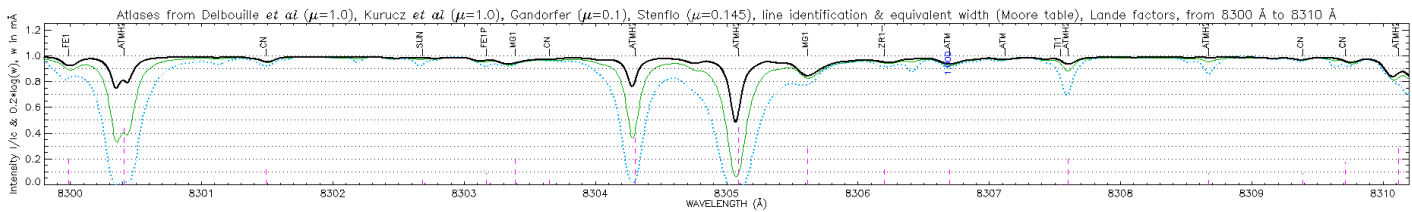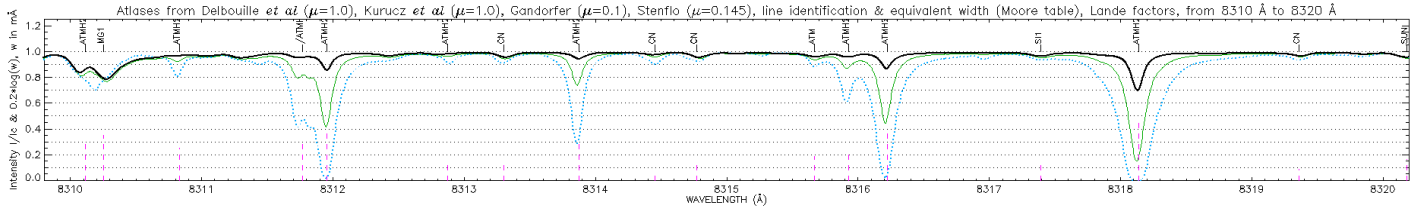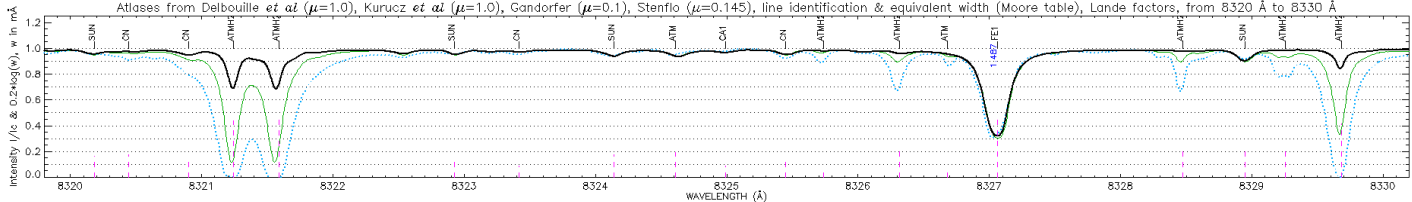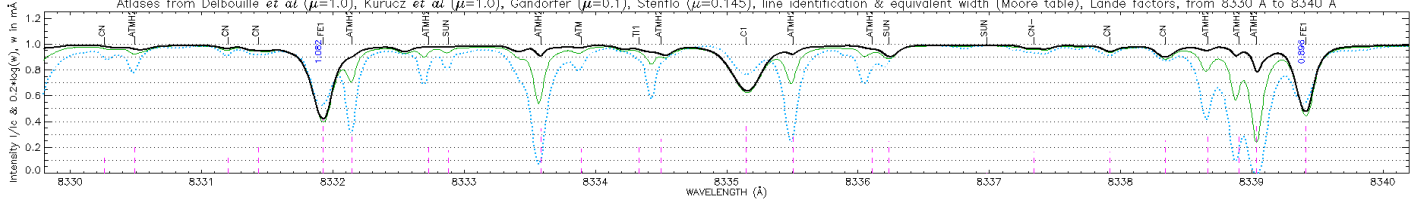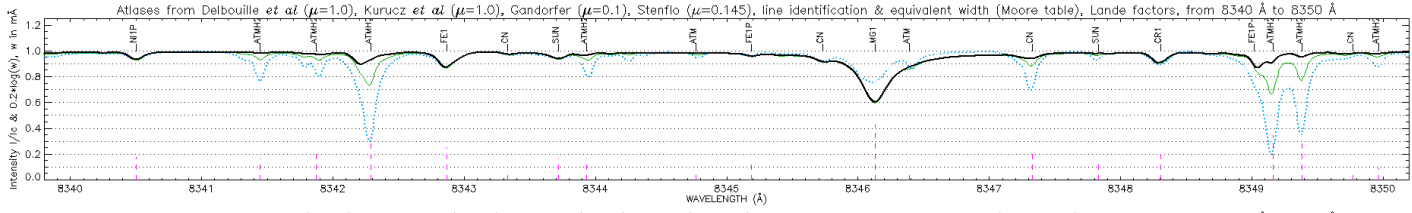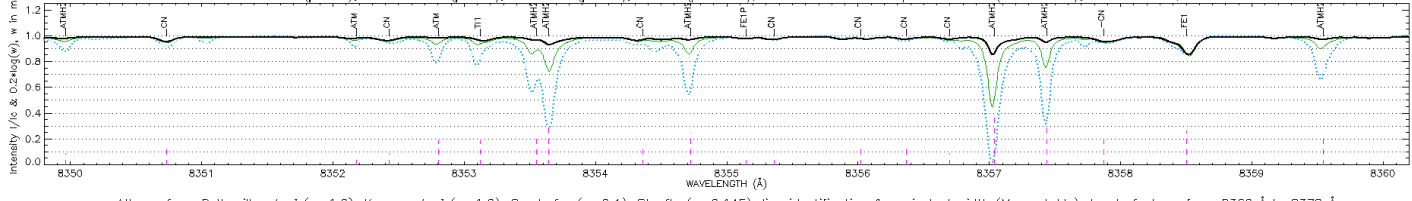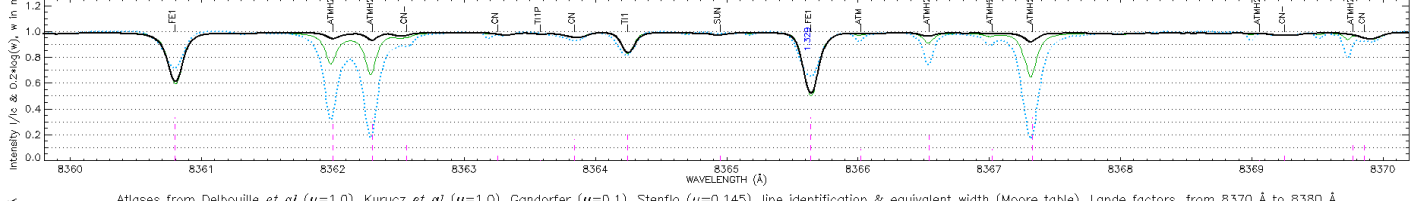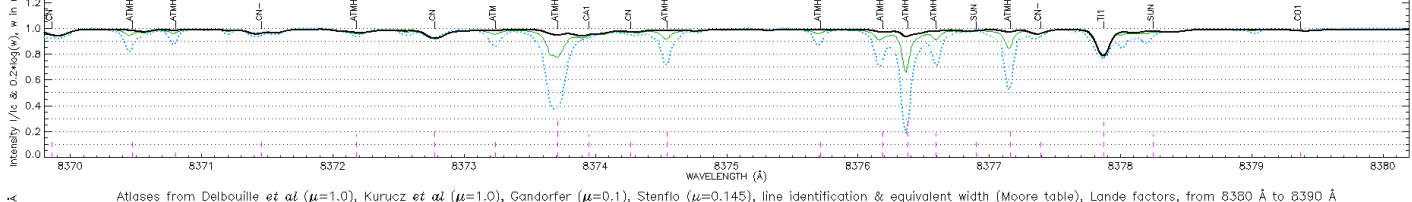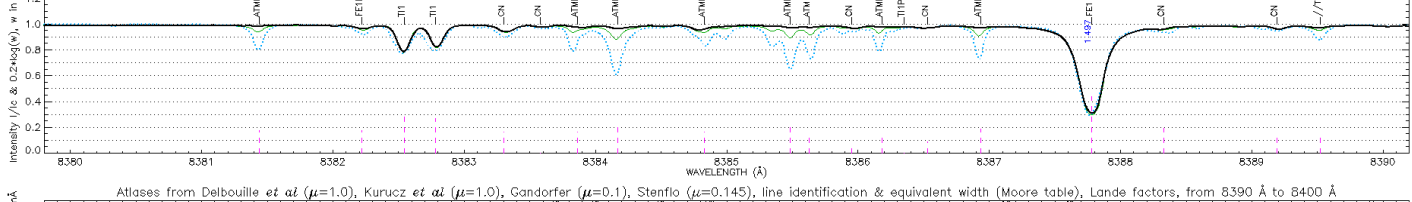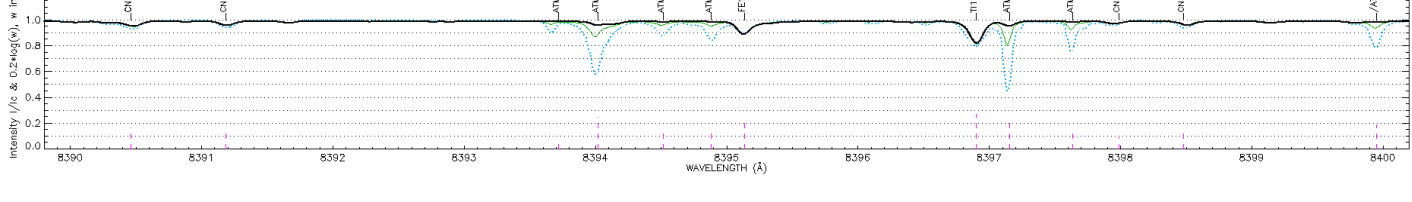

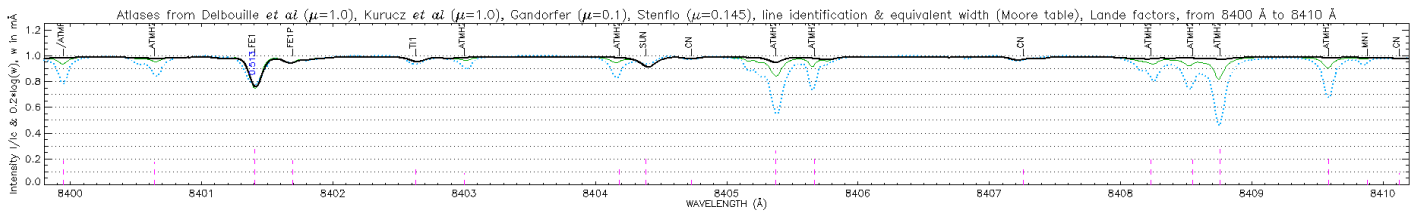
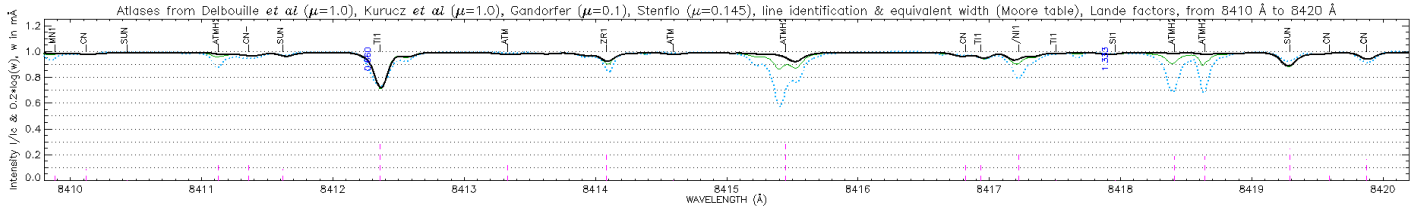
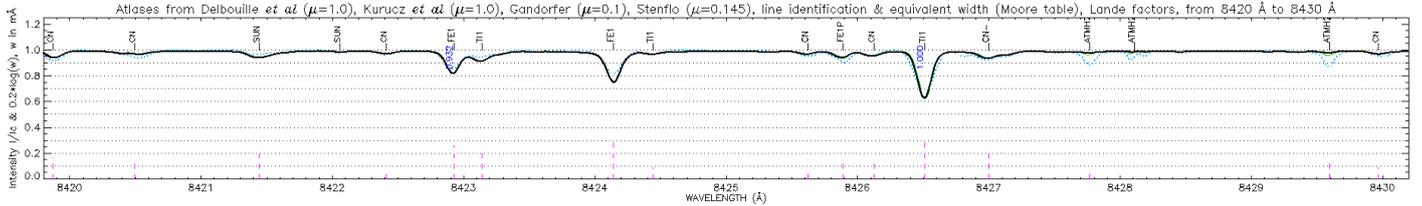
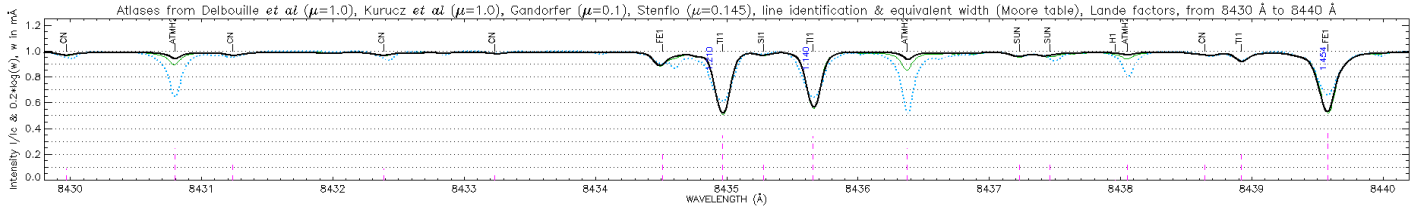
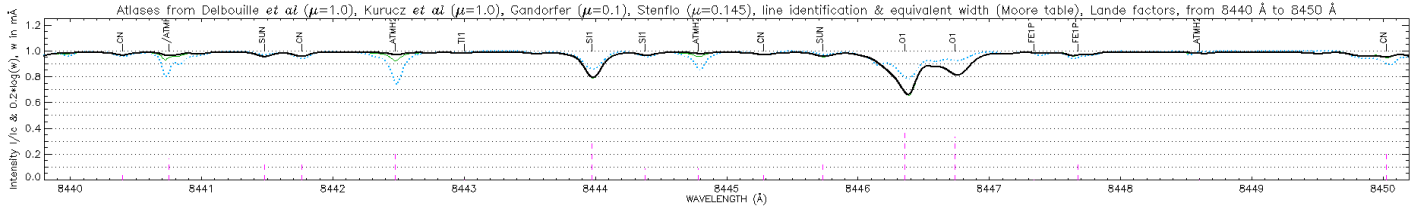
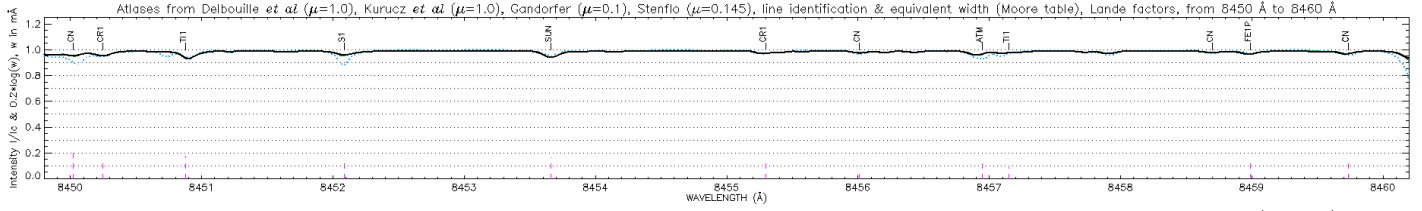
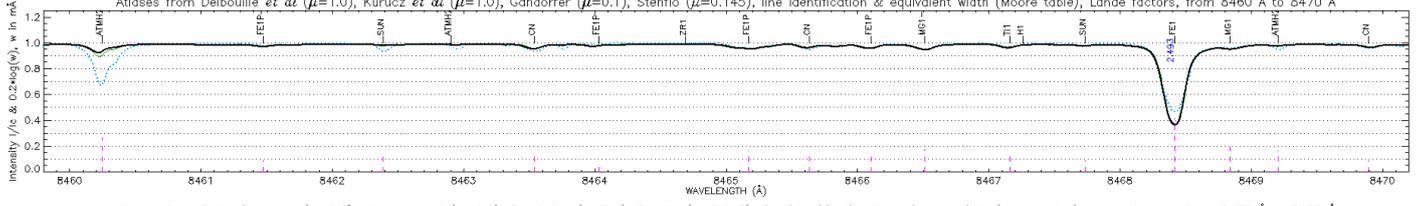
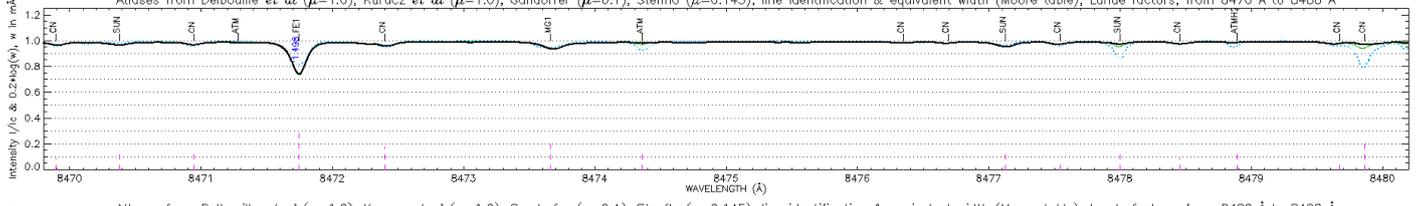
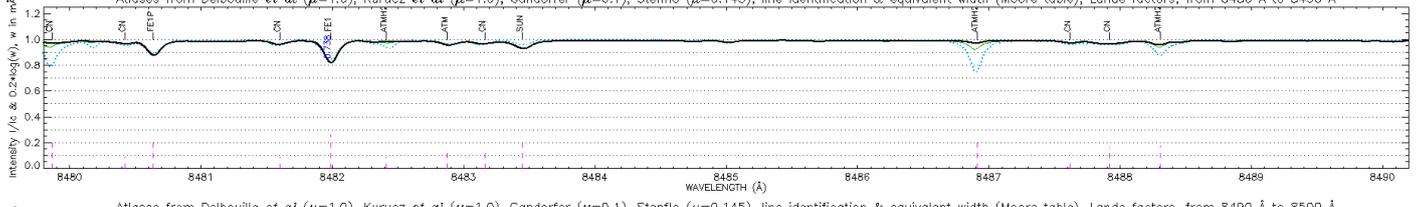
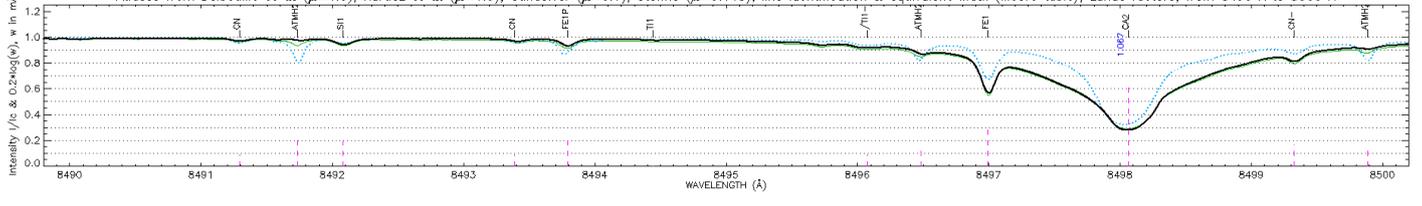

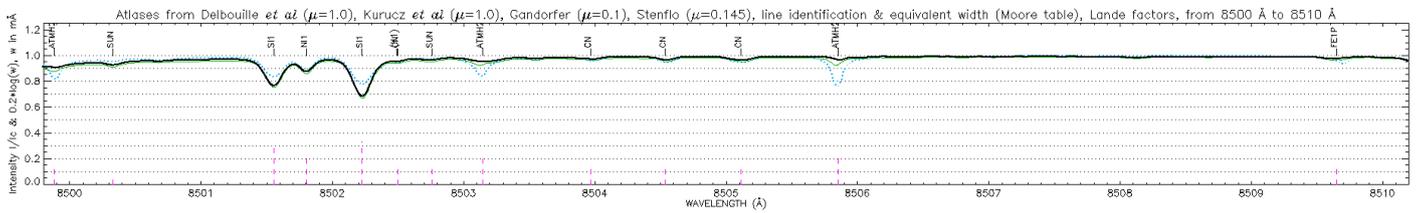
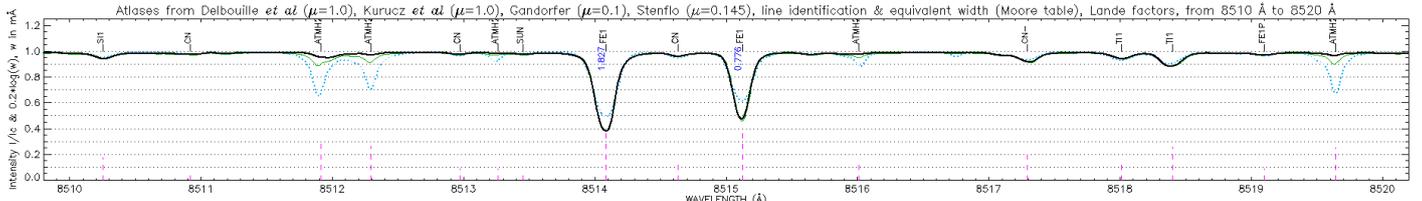
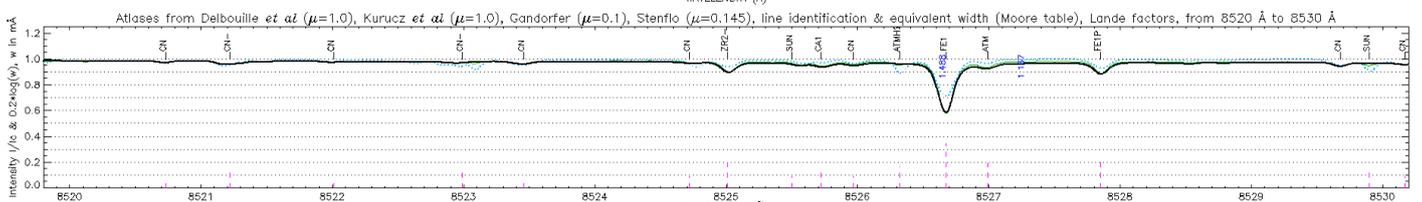
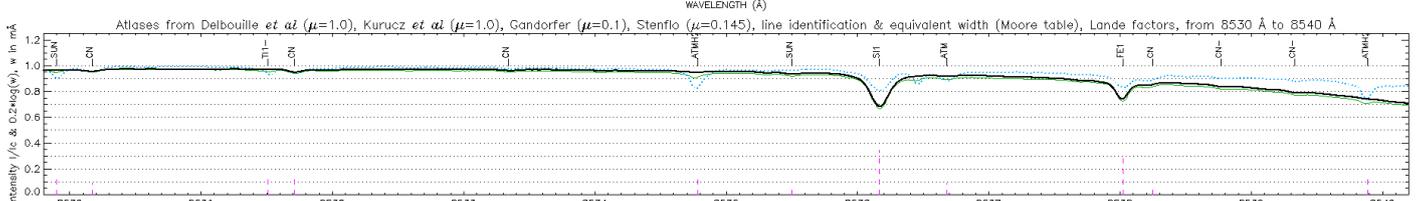
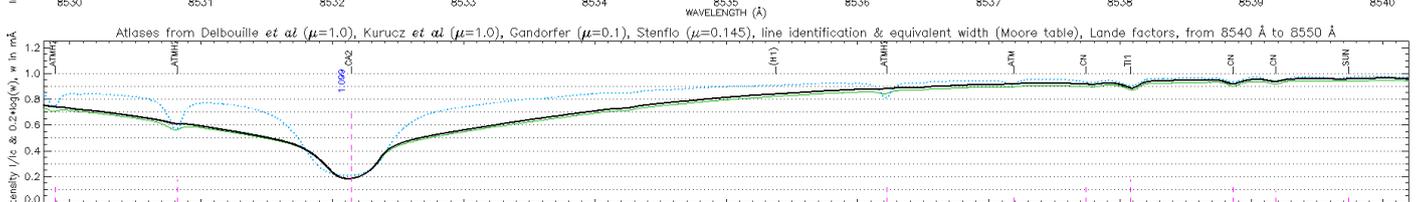
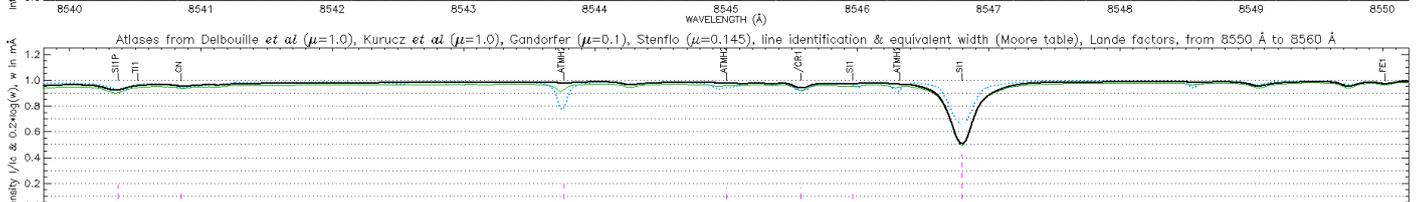
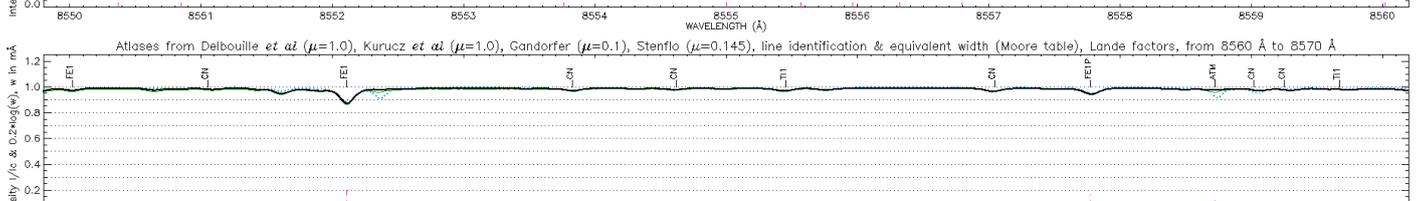
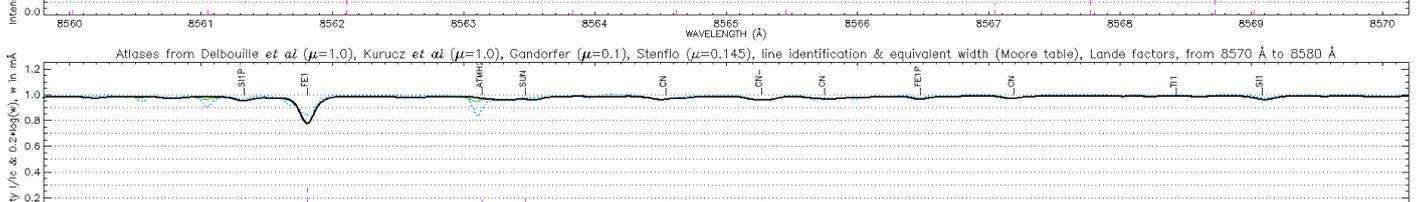
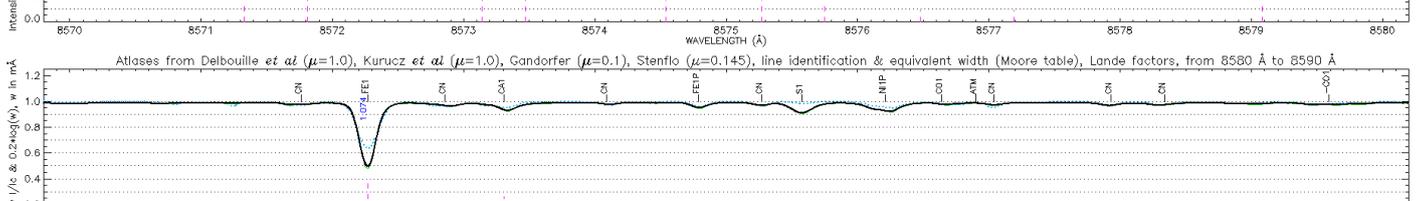
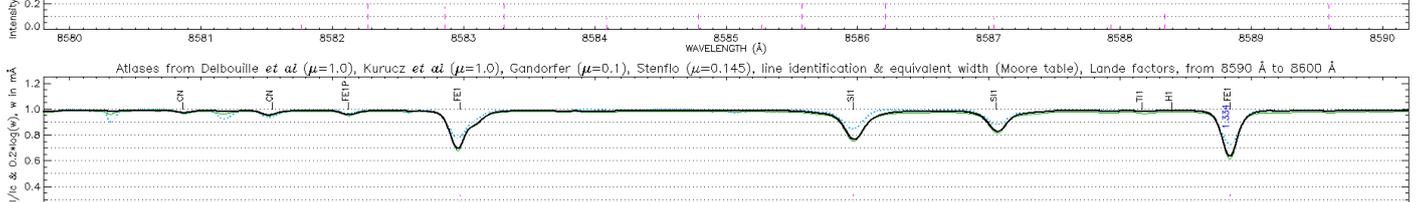

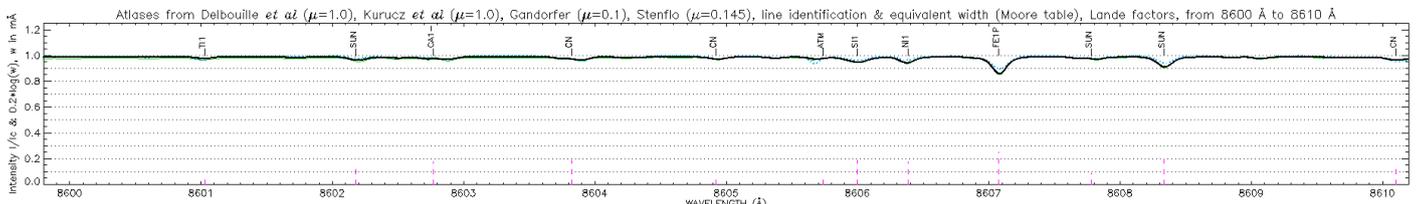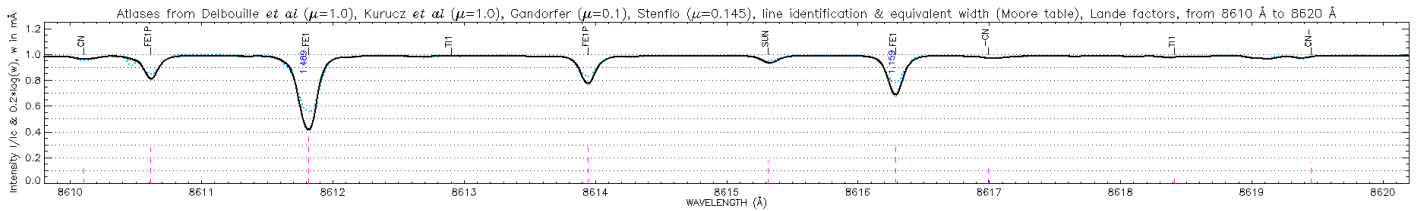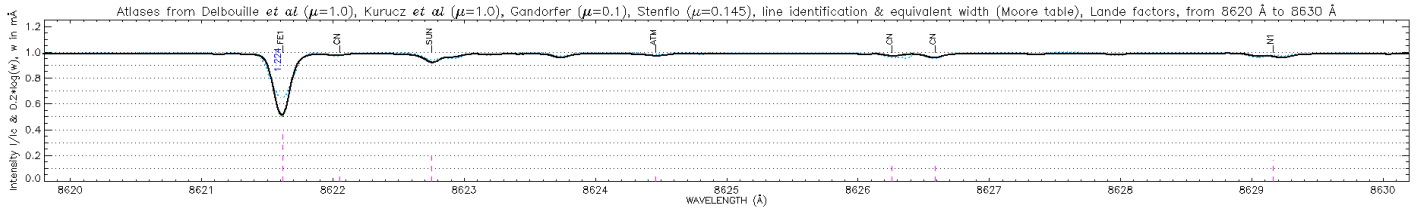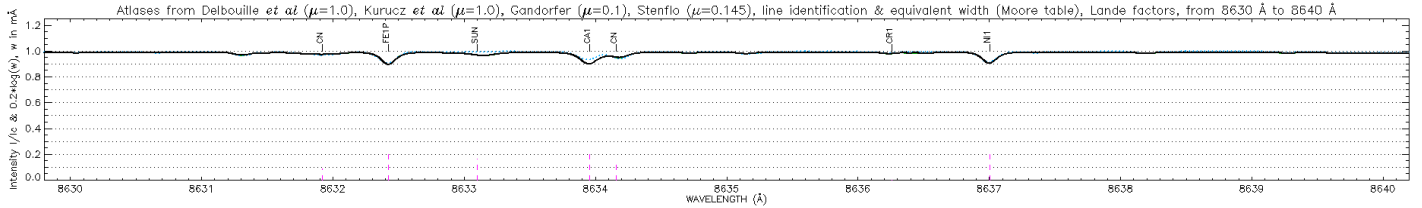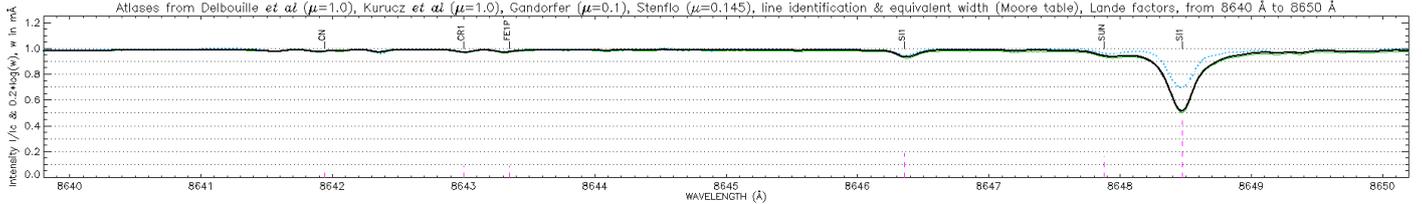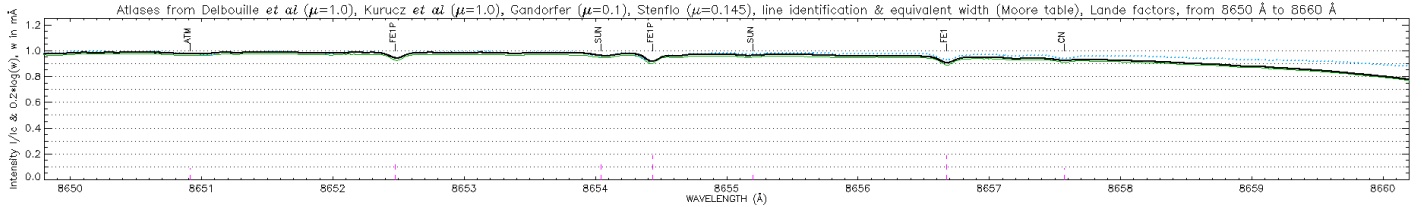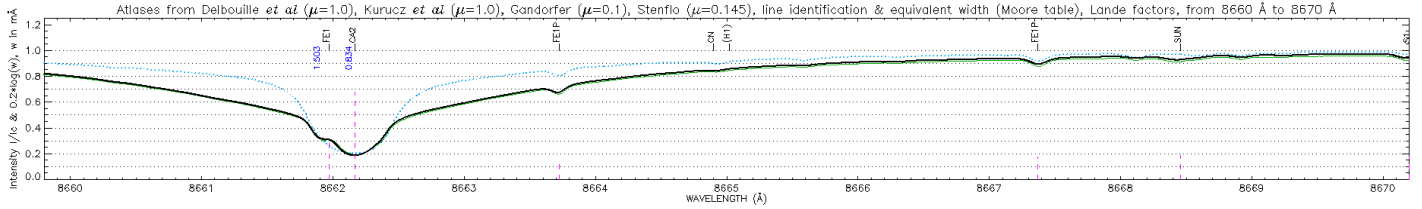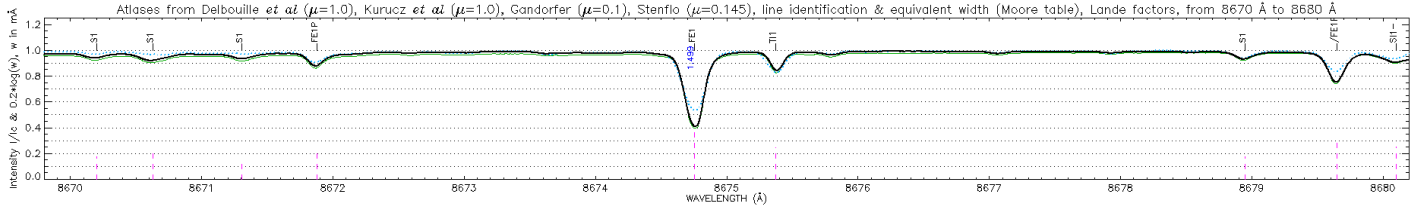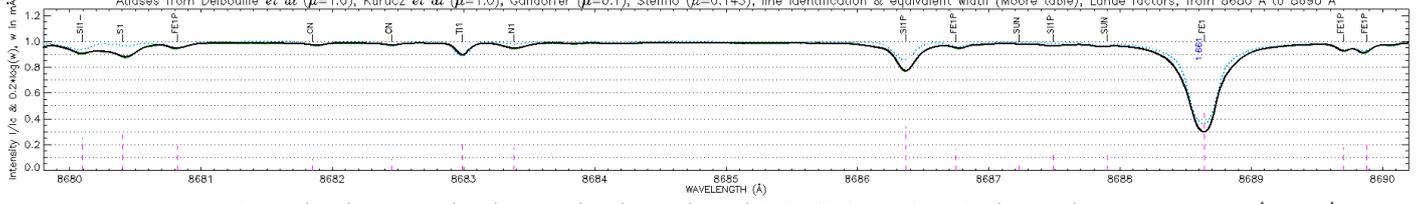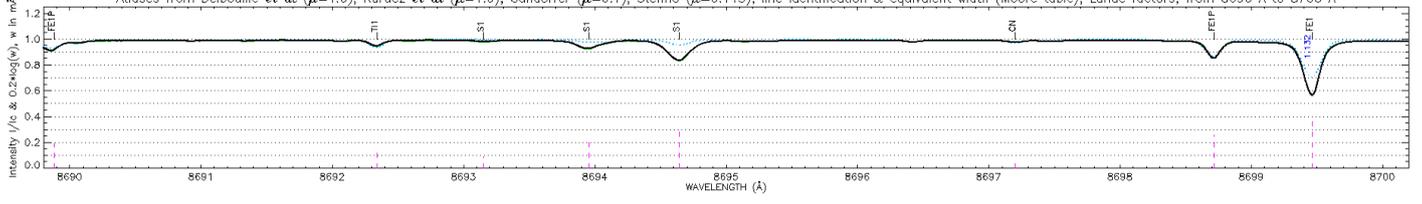

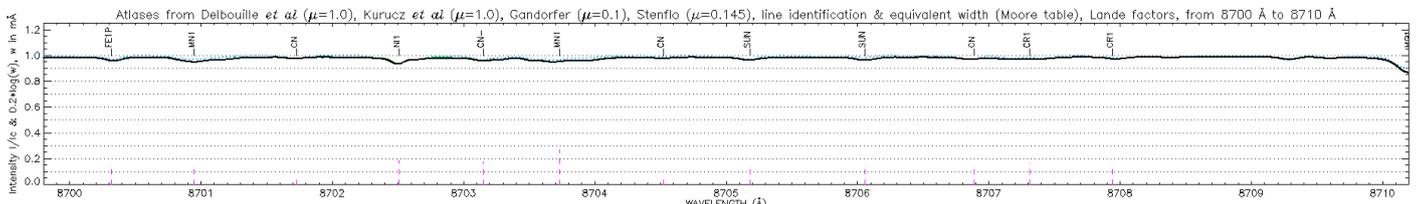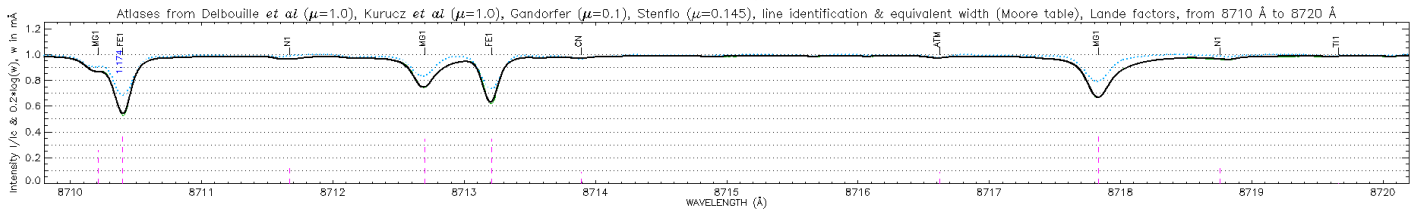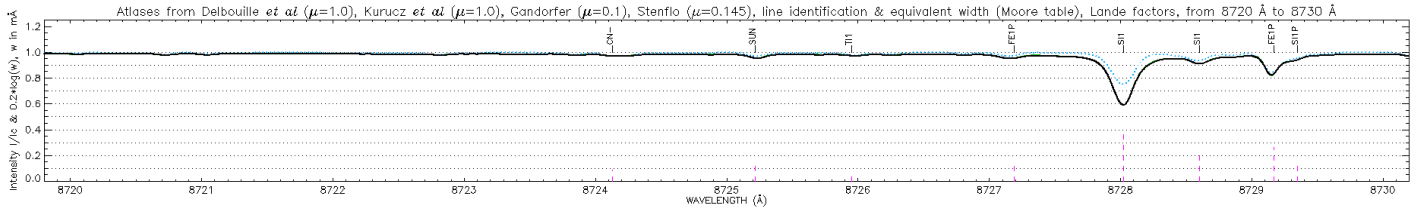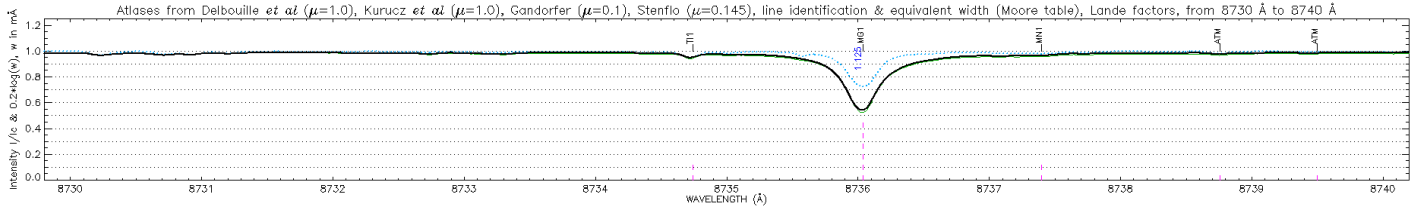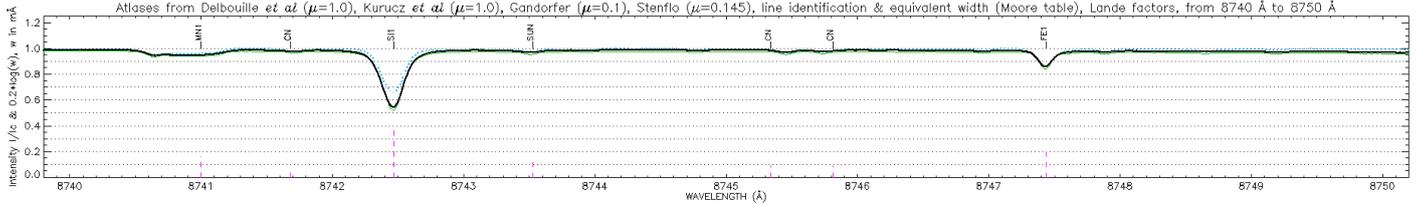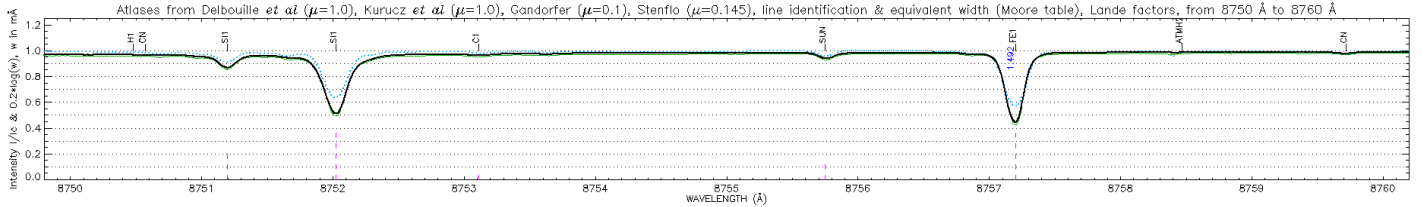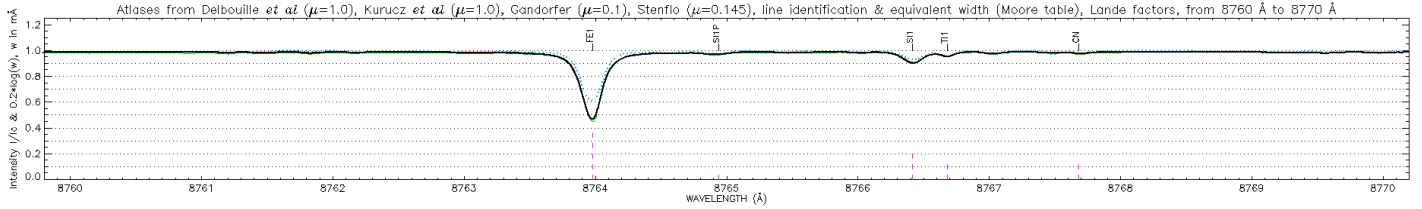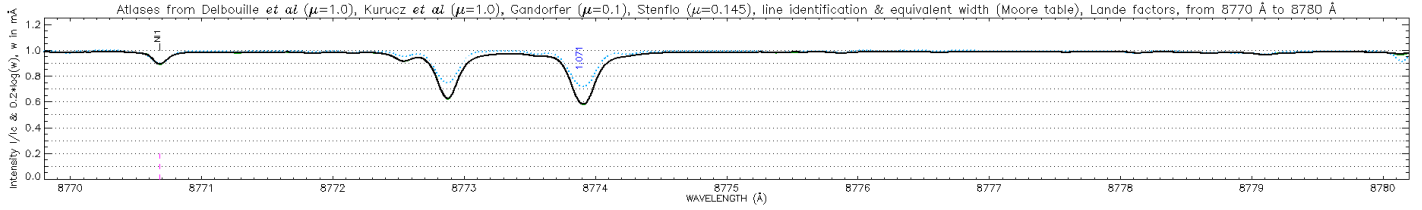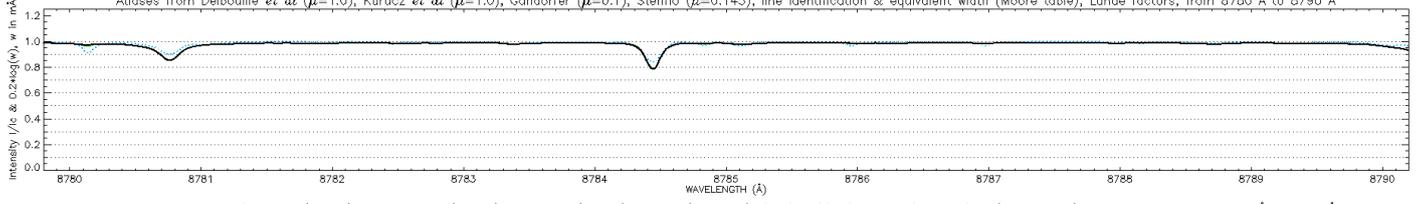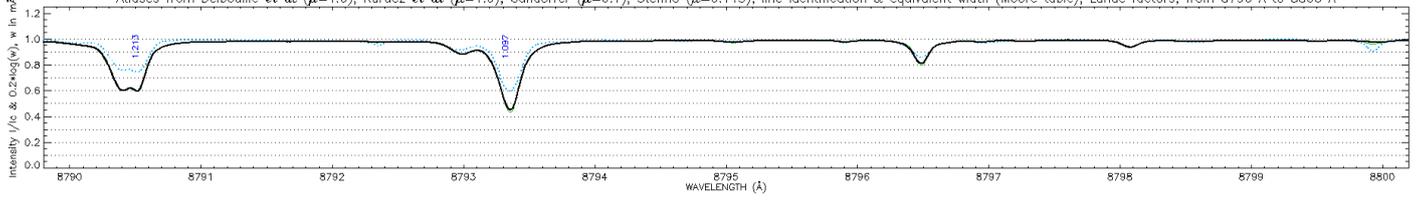